\documentclass[12pt]{article}
\usepackage{authblk}

\usepackage{float}
\usepackage{amsmath}
\usepackage{graphicx}
\usepackage{multirow}
\usepackage{mathrsfs}
\usepackage[title]{appendix}
\usepackage{textcomp}
\usepackage{booktabs}
\usepackage{algpseudocode}
\usepackage{listings}
\usepackage{subfig} 
\usepackage[normalem]{ulem}
\usepackage{geometry}
\usepackage{svg}
\usepackage{epsfig}
\usepackage{overpic}
\usepackage{dcolumn}
\usepackage{ulem}
\usepackage{bm}
\usepackage{graphics}
\usepackage{lineno}
\usepackage{xspace}
\usepackage{epstopdf}
\usepackage{xcolor}
\usepackage{soul}
\usepackage{verbatim}
\usepackage{slashed} 
\RequirePackage{hyperref}
\hypersetup{%
        colorlinks,
        breaklinks=true,
        plainpages=false,%
        citecolor=blue,
        linkcolor=blue,
        urlcolor=blue,
        bookmarksopen=true,%
        bookmarksnumbered=false,%
        bookmarksdepth=5%
}

\usepackage[textsize=tiny]{todonotes}

\usepackage{caption}
\usepackage{lineno}

\usepackage{etoolbox}
\makeatletter

\newcounter{symfn}
\newcounter{numfn}

\newtoggle{numfootmode}
\togglefalse{numfootmode}

\newcommand{\symbolfootnotes}{\togglefalse{numfootmode}}
\newcommand{\numberedfootnotes}{\toggletrue{numfootmode}}
 
\newcommand{\myfootnote}[1]{%
  \iftoggle{numfootmode}{%
    \refstepcounter{numfn}%
    \textsuperscript{\arabic{numfn}}%
    \insert\footins{%
      \noindent
      \footnotesize
      \textsuperscript{\arabic{numfn}}\;#1\par
    }%
  }{%
    \refstepcounter{symfn}%
    \textsuperscript{\@fnsymbol{\value{symfn}}}%
    \insert\footins{%
      \noindent
      \footnotesize
      \textsuperscript{\@fnsymbol{\value{symfn}}}\;#1\par     
    }%
  }%
}
\makeatother

\title{Search for Long-lived Particles at Future Lepton Colliders Using Deep Learning Techniques}
\date{}
\author[1,$\diamond$]{Yulei Zhang}
\author[1,$\diamond$]{Cen Mo}
\author[1,$\diamond$]{Xiang Chen}
\author[2]{Bingzhi Li}
\author[2]{Hongyang Chen}
\author[3,4,*]{Jifeng Hu}
\author[1,*]{Liang Li }

\affil[1]{State Key Laboratory of Dark Matter Physics, Key Laboratory for Particle Astrophysics and Cosmology (MOE), Shanghai Key Laboratory for Particle Physics and Cosmology, School of Physics and Astronomy, Shanghai Jiao Tong University, Shanghai 200240, China}

\affil[2]{Research Center for Scientific Data Hub, Zhejiang Lab, Hangzhou 310000, China}

\affil[3]{State Key Laboratory of Nuclear Physics and Technology, Institute of Quantum Matter, South China Normal University, Guangzhou 510006, China}

\affil[4]{Guangdong Basic Research Center of Excellence for Structure and Fundamental Interactions of Matter, Guangdong Provincial Key Laboratory of Nuclear Science, Guangzhou 510006, China}
\begin{document}

\maketitle

\vspace{-25pt}

{\centering
Email: \href{mailto:hujf@m.scnu.edu.cn}{hujf@m.scnu.edu.cn}, \href{mailto:liangliphy@sjtu.edu.cn}{liangliphy@sjtu.edu.cn}
\par}

\footnotetext{$\diamond$ These authors contributed equally to this work.}
\footnotetext{* Authors to whom any correspondence should be addressed. }

\begin{abstract} 
\unboldmath 
Long-lived particles (LLPs) provide an unambiguous signal for physics beyond the Standard Model (BSM). They have a distinct detector signature, with decay lengths corresponding to lifetimes of around nanoseconds or longer. Lepton colliders allow LLP searches to be conducted in a clean environment, and such searches can reach their full physics potential when combined with machine learning (ML) techniques.
This experimental study, utilizing comprehensive full simulation data samples, focuses on LLP searches resulting from Higgs decay in $e^+e^-\to ZH$. We demonstrate that, by employing deep neural network approaches the LLP signal efficiency can be improved up to 95\% for an LLP mass around 50 GeV and a lifetime of approximately 1 nanosecond, while rejecting all SM backgrounds. Furthermore, the signal sensitivity for the branching ratio of Higgs decaying into LLPs reaches a state-of-art limit of $1.0 \times 10^{-6}$ with a statistics of $4 \times 10^{6}$ Higgs.
\end{abstract}

\section{Introduction}
In 2012, the discovery of the Higgs boson completed the final piece of the Standard Model (SM)~\cite{CMS:2012qbp,CMS:2013btf}. Remarkably, SM predictions align with almost all experimental observations. However, the SM does not address several critical questions in the universe's evolution, such as the existence of dark matter~\cite{Kaplan:2009ag,Dienes:2011ja,Kim:2013ivd},  the matter-antimatter imbalance puzzle~\cite{Bouquet:1986mq,Campbell:1990fa} and the origin of the neutrino mass~\cite{Graesser:2007yj,Maiezza2015}. Consequently, the search for new physics beyond the Standard Model (BSM) is both intriguing and necessary. Despite numerous efforts, BSM signals have remained elusive, suggesting that new physics may either require a higher energy threshold than currently achievable or that the coupling strength between SM and BSM particles is too weak to produce a statistically significant number of observable signals. Amidst this scenario, the hypothesis that BSM particles could possess long lifetimes, evading detection, has transitioned from a nascent idea to a widely accepted and vigorously pursued avenue within the physics community. 
These particles, often called long-lived particles (LLPs), serve as sensitive probes into BSM physics, such as supersymmetry (SUSY) theory~\cite{Arvanitaki:2012ps,Fan:2011yu} or dark matter models~\cite{Bertone:2016nfn}. LLPs could also provide insights for the matter-antimatter asymmetry~\cite{Dine:2003ax} and the neutrino mass hierarchy, with their lifetimes varying widely. 
Recently, the search for LLPs has intensified across several experiments, such as ATLAS~\cite{ATLAS_llp_muon,ATLAS2024LLP}, 
CMS~\cite{CMS:llp2023,CMS_llp_2024}, and LHCb~\cite{Calefice:2022upw}, as well as Belle II~\cite{Ferber:2022rsf}, BESIII~\cite{2023137698}, Babar~\cite{BaBar2015}, marking a keen interest in their potential to reveal new aspects of physics, even though definitive discoveries are yet to be made.

LLPs are characterized by their extended decay lengths and distinctive detector signatures, such as displaced vertices or internal jets, allowing them to evade detection by the inner detectors after being produced. Instead, they decay into SM particles upon reaching the outer detectors, which could be electromagnetic or hadronic calorimeters, or even external detectors positioned significantly away from the interaction point. The elusive nature of LLP signals can be attributed to factors such as a low signal-over-background ratio and the complexity of event reconstruction involving displaced objects.
Consequently, specialized reconstruction techniques are required to effectively detect and measure LLPs.

Searching for LLPs using machine-learning (ML) assisted approaches at hadron colliders~\cite{Cui:2014twa, Bhattacherjee:2019fpt, CMS:2024trg} has emerged as a promising strategy to surmount these experimental challenges. Such research employs ML techniques for object-level tagging to identify LLP signatures following comprehensive event reconstruction. The efficacy of this tagging is dependent on a constrained set of high-level input variables. The demanding environment at hadron colliders complicates background control, possibly necessitating neural networks with higher granularity and more extensive input features. Furthermore, the deployment of advanced triggers powered by rapid neural networks is crucial for the efficient detection of LLP-like events at hadron colliders~\cite{Alimena2020LLP,Coccaro2023LLP}.

\begin{figure}[h]
\centering
    \includegraphics[width=0.42\linewidth]{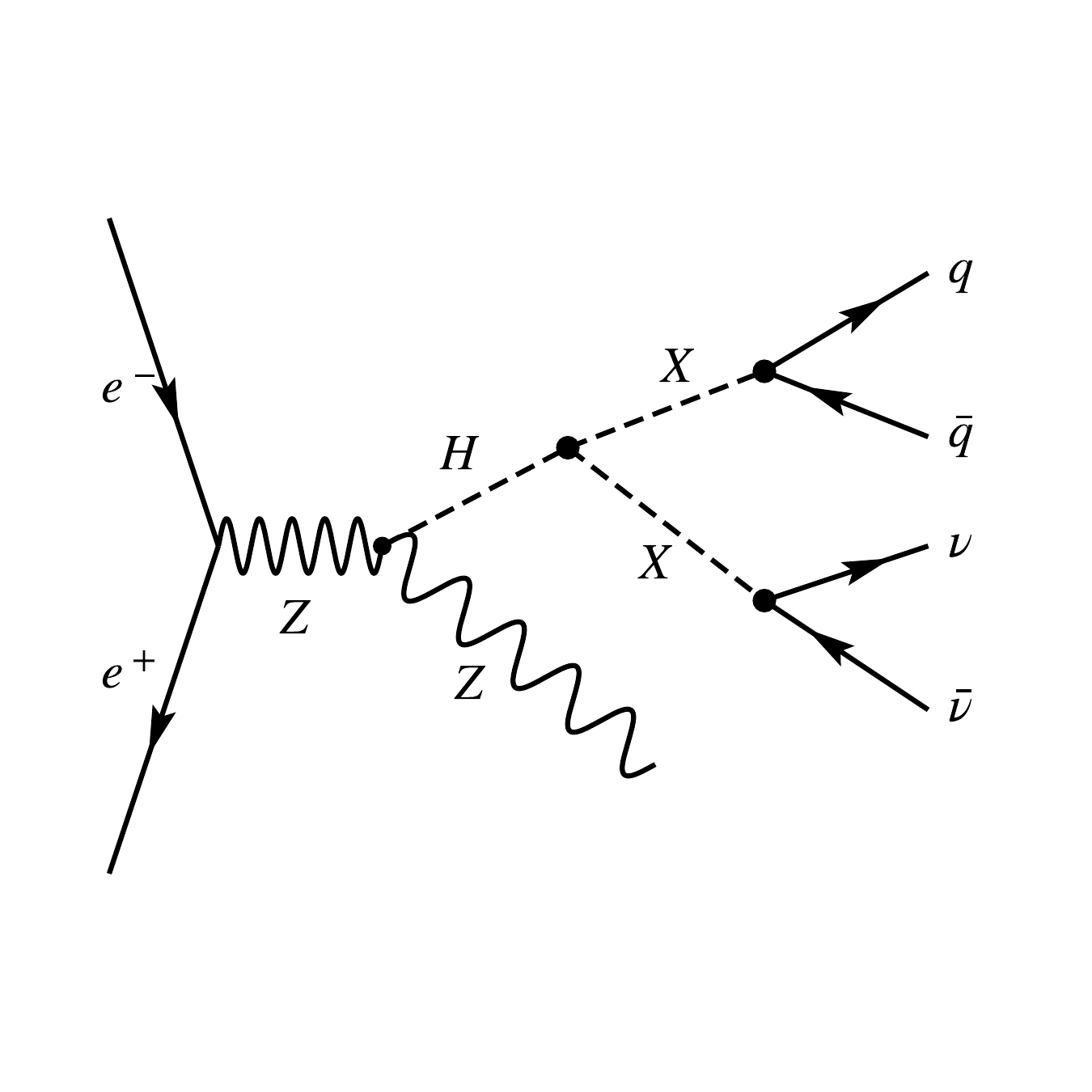}
    \includegraphics[width=0.42\linewidth]{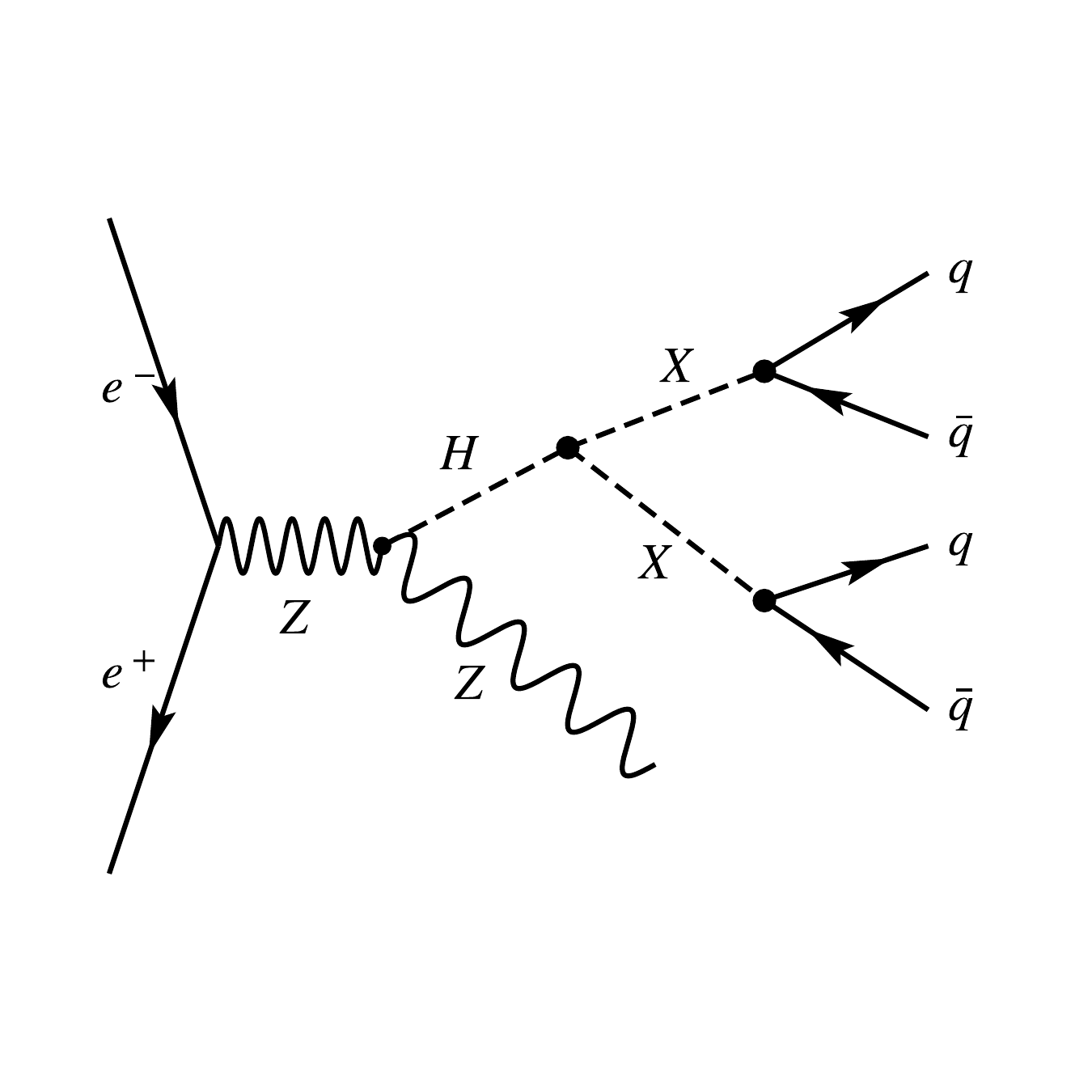}
\caption{Feynman diagrams of LLP production and decay.
Two Feynman diagrams are presented illustrating the generation of 
LLPs, denoted as \( X \), through the Higgsstrahlung mechanism.
On the left, the diagram shows the production of \( XX \) followed by their subsequent decay into a $\nu\bar{\nu}$ pair and a $q\bar{q}$ pair, respectively, resulting in two jets. On the right, both \( X \) decay into $q\bar{q}$ pairs, leading to the four jets final state.
}
  \label{fig:Sig_Fey}
\end{figure}

Electron-positron colliders, such as the Circular Electron Positron Collider (CEPC)~\cite{CEPCStudyGroup:CDR1} and the International Linear Collider (ILC)~\cite{ILC:2019gyn}, 
provide clean collision environments with precise initial-state conditions.
The Higgs boson production via $e^{+} e^{-}\rightarrow ZH$ at these facilities offers a clean channel to explore rare LLP decays, benefiting from well-defined initial kinematics and significantly reduced backgrounds compared to hadron colliders like the LHC~\cite{ATLAS:2018tup,CMS:2020iwv}.
The primary goal of this analysis is to measure the cross section of LLP production through Higgs boson decays at future lepton colliders. Given that the SM Higgs production cross section at electron-positron colliders is accurately calculable and well-known, the LLP production cross section can be directly derived by multiplying the predicted SM $ZH$ cross section by the assumed branching ratios (BRs) of Higgs into LLPs. Therefore, these branching ratios are the fundamental parameters that we aim to constrain in this analysis.

The Higgs boson can decay into a pair of long-lived particles (%\glspl{LLP}
 LLP, denoted as \(X\)), each of which subsequently decays into final state objects such as jets, charged leptons, or neutrinos.
In the scope of our study, the search strategy is inspired by the glueball and Gauge Mediated Supersymmetry Breaking (GMSB) models~\cite{Giudice:1998bp, Meade:2010ji}, where such LLPs naturally emerge and have accessible decay modes as defined by the process of \(H \rightarrow XX \rightarrow \textrm{jets}\). This focus stems from the distinguishable detector signatures that the jet decay products can offer.

Figure~\ref{fig:Sig_Fey} shows Feynman diagrams for the LLP production process from the Higgs decay in lepton colliders with two jets (type I signal) or four jets (type II signal) in the final state.
 We focus on scenarios where the mixing between the new scalar particle \( X \) and the Higgs boson is negligible. This assumption allows us to isolate and analyze specific decay channels of \( X \) without the complexities introduced by universal couplings to all SM particles that would result from significant \( X \)-Higgs mixing.
 While \( X \) could, in principle, decay into gauge bosons, our study focuses on the hadronic and invisible decay channels. This focus is motivated by scenarios such as effective \( Z' \) models, where the new scalar \( X \) couples preferentially to quarks and neutrinos over charged leptons and gauge bosons~\cite{Fox:2011qd, Ilten:2016tkc, Ilten:2018crw}. The invisible decay channel of \( X \) is defined to include any final states that result in missing energy signatures, such as neutrinos or potential dark matter candidates. This approach aligns with various BSM scenarios where \( X \) serves as a mediator between the SM and the dark sector, leading to missing energy signatures in the detector. Additionally, hadronic and invisible final states pose significant experimental challenges due to large QCD backgrounds and missing energy signatures, respectively, making them compelling channels for detailed study.

Other potential decay channels such as \(H \rightarrow XX \rightarrow \textrm{invisible}\) and \(H \rightarrow XX \rightarrow \textrm{leptons}\) are not considered in this study.
 Particularly, the \(H \rightarrow XX \rightarrow \textrm{invisible}\) channel, classified as a type III signal, results in decay products that leave no detectable signature within the detector's sensitivity range. Such invisibly decaying LLPs do not contribute to observable events that can be analyzed within our current experimental framework. Consequently, these decays are more appropriately addressed in analyses specifically dedicated to \(H \rightarrow \textrm{invisible}\) decays~\cite{tan_2020}.
Similarly, while \(H \rightarrow XX \rightarrow \textrm{leptons}\) presents an interesting decay channel, the current study prioritizes jet signatures due to their robust detection capabilities and the analytical focus on hadronic final states. Future research may expand to encompass these additional decay modes, taking into account different event topologies and analysis methodologies.

In this study, we investigate the feasibility and sensitivity of employing advanced machine learning methods, particularly deep learning techniques such as Convolutional Neural Networks (CNNs) and Graph Neural Networks (GNNs), to learn the intricate topologies and kinematic features of LLPs. 
We conduct a comparative analysis of the two distinct machine learning architectures to evaluate their respective performances in reconstructing LLP signals. Each model is independently trained and assessed, providing insights into their individual capabilities and limitations within the context of LLP detection.
By applying these methods directly to low-level detector data, we achieve remarkable improvements in LLPs signal reconstruction efficiency and background suppression. 
The search sensitivity for LLPs at future \(e^+e^-\) colliders achieved using deep learning techniques significantly surpasses that of traditional methods, establishing state-of-the-art limits on the Higgs branching ratio into LLPs.
This approach enables us to derive stringent constraints on the branching ratios, thus enhancing our ability to identify potential LLP signals and probe BSM physics effectively.
Importantly, the ML-based methodology we present offers broad applicability, promising an easy adaptation for future lepton collider experiments, including the CEPC~\cite{CEPCStudyGroup:CDR1}, ILC~\cite{ILC:2019gyn}, FCC~\cite{FCC:2018byv}, and CLIC~\cite{CLIC:2018fvx}. 

The remainder of this paper is structured as follows: details on the samples used are provided in Section ``Signal and Background Samples''. In Section ``Method", we describe the methodology of our analysis, including analysis strategy, reconstruction algorithms, and neural network architectures. Sections ``Results" and ``Discussion" present detailed numerical results and performance evaluations. Section ``Conclusion" summarizes our findings and discusses implications for future LLP searches.

\section{Signal and Background Samples}

In this study, we utilize fully simulated Monte Carlo (MC) samples for both the signal and background analyses. To generate the long-lived particle (LLP) signal samples, \texttt{MadGraph 3.0.1}~\cite{Alwall:2014hca} is employed to cover both type I and type II signals across various mass points (50 GeV, 10 GeV, and 1 GeV) and lifetimes (0.001 ns, 0.1 ns, 1 ns, 10 ns, 100 ns). A total of $1.5 \times 10^7$ events are produced, distributing $10^6$ events for each combination of mass point and lifetime.

The backgrounds predominantly consist of SM processes resulting in jets in the final state, notably \(e^+e^- \rightarrow \bar{q}q\) and \(e^+e^- \rightarrow VV\), where \(VV\) includes both \(WW\) and \(ZZ\). Additionally, the SM process \(e^+e^- \rightarrow ZH \rightarrow \textrm{inclusive}\) is also considered, given its significance as the main production mode for SM Higgs. These SM processes are simulated using \texttt{Whizard 1.95} \cite{Kilian:2007gr, Moretti:2001zz}. A total of \(10^7\) events is simulated for both the \(e^+e^- \rightarrow \bar{q}q\) and \(e^+e^- \rightarrow VV\) processes, while the \(e^+e^- \rightarrow ZH \rightarrow \textrm{inclusive}\) process has \(10^6\) events simulated. 
The top quark and tri-boson related backgrounds have negligible cross sections at a center-of-mass energy of 240 GeV and therefore are not considered.
Minor backgrounds such as pileup background and cosmic ray background are also neglected in this analysis~(Appendix~\ref{pile_up_and_cosmic_ray}). 
Details on the cross-sections and simulated statistics for the LLP signal and SM background are provided in Table~\ref{tab:mc_samples}.

For each MC sample, a full detector simulation is performed following the CEPC conceptual design~\cite{CEPCStudyGroup:CDR2}. 
The CEPC detector comprises a silicon-based vertexing and tracking system, a Time Projection Chamber (TPC) tracker, a high-granularity calorimetry system, a 3~Tesla superconducting solenoid magnet, and a muon detection system integrated within the magnet's iron return yoke. The calorimetry system consists of an electromagnetic calorimeter (ECAL) and an iron-scintillator hadronic calorimeter (HCAL). 
Simulations are carried out using
MOKKAC software~\cite{CEPC-SIMU-2017-001}, a dedicated Geant4-based tool customized for the CEPC framework, which accounts for comprehensive particle showering effects and calibrated detector responses. 
The calorimeter resolutions are modeled to achieve an energy resolution of \(16\%/\sqrt{E/\textrm{GeV}} \oplus 1\%\) for the ECAL and \(60\%/\sqrt{E/\textrm{GeV}} \oplus 1\%\) for the HCAL and muon detector. In addition, detector hits are assigned a time resolution of 1 nanosecond. Following full detector simulation, no further event or object reconstruction is applied; rather, the digitized detector hits are directly utilized as inputs for the subsequent ML-based analyses.

\begin{table}[!htbp]
    \centering
    \caption{Cross sections and simulated statistics of the signal and background samples.}
    \label{tab:mc_samples}
    \resizebox{0.8\textwidth}{!}{
    \begin{tabular}{ccccccc}
        \toprule 
Process                 & LLP signal          & SM ZH            & \(q\bar{q}\)           & \(Z Z\)               & \(W W\)     \\
        \midrule
\(\sigma[\mathrm{fb}]\) & -                     & 203.66                & 54106.86               & 1110.37               & 16721.77          \\
Statistics              & \(1.5 \times 10^7 \) & \(1.0 \times 10^6\) & \(1.0 \times 10^7\) & \( 1.1 \times 10^6\) & \( 8.9 \times 10^6\) \\
        \bottomrule
    \end{tabular}
    }
\end{table}

\section{Method}

\subsection{General Analysis Strategy}

We develop a novel analysis strategy leveraging deep learning techniques based on low-level detector information, circumventing the need for event or object reconstruction. This approach marks a significant departure from traditional selection-based methods.

Initially, a signal acceptance selection is applied to ensure that at least one LLP is within the geometric boundaries of the CEPC detector~\cite{CEPCStudyGroup:CDR2}, with acceptance factors detailed in Table~\ref{tab:signal_acceptance}. Subsequently, signal events are classified based on the number of LLP decays detectable within the detector and their consequent final states. This classification yields three categories: $\mathcal{D}_0$, $\mathcal{D}_1$, and $\mathcal{D}_2$, denoting events with 0, 1, and 2 detectable LLPs leading to jet final states, respectively. 

\begin{table}[!htbp]
  \centering
  \caption{Signal acceptances for different LLPs masses and lifetimes, including statistical uncertainties.}
  \resizebox{1.0\textwidth}{!}{
  \begin{tabular}{cccccc}
  \toprule
   Acceptance (\%) & \multicolumn{5}{c}{Lifetime [ns]} \\
  \cmidrule{2-6}
   Mass [GeV] & 0.001 & 0.1 & 1 & 10 & 100 \\ 
   \midrule
    1 & 100.00 $\pm$ 0.00 & 99.86 $\pm$ 0.01 & 48.76 $\pm$ 0.18 & 6.49 $\pm$ 0.09 & 0.67 $\pm$ 0.03 \\
    10 & 100.00 $\pm$ 0.00 & 100.00 $\pm$ 0.00 & 99.78 $\pm$ 0.01 & 46.80 $\pm$ 0.16 & 6.22 $\pm$ 0.08 \\
    50 & 100.00 $\pm$ 0.00 & 100.00 $\pm$ 0.00 & 100.00 $\pm$ 0.00 & 99.31 $\pm$ 0.03 & 40.37 $\pm$ 0.16 \\
  \bottomrule
  \end{tabular}
  }
\label{tab:signal_acceptance}
\end{table}

For the background analysis, events are categorized based on their final-state jet multiplicities into two categories: \textit{2-jet} and \textit{4-jet}. The dominant background for the \textit{2-jet} category arises from the process $e^+e^- \rightarrow q\bar{q}$ , while the \textit{4-jet} background primarily originates from the diboson production process $e^+e^- \rightarrow VV$. Additionally, the inclusive SM Higgs production process ($e^+e^- \rightarrow ZH \rightarrow$ inclusive) contributes to both the \textit{2-jet} and \textit{4-jet} categories.

\begin{figure*}[!htb]
    \centering
    \includegraphics[width=0.99\linewidth]{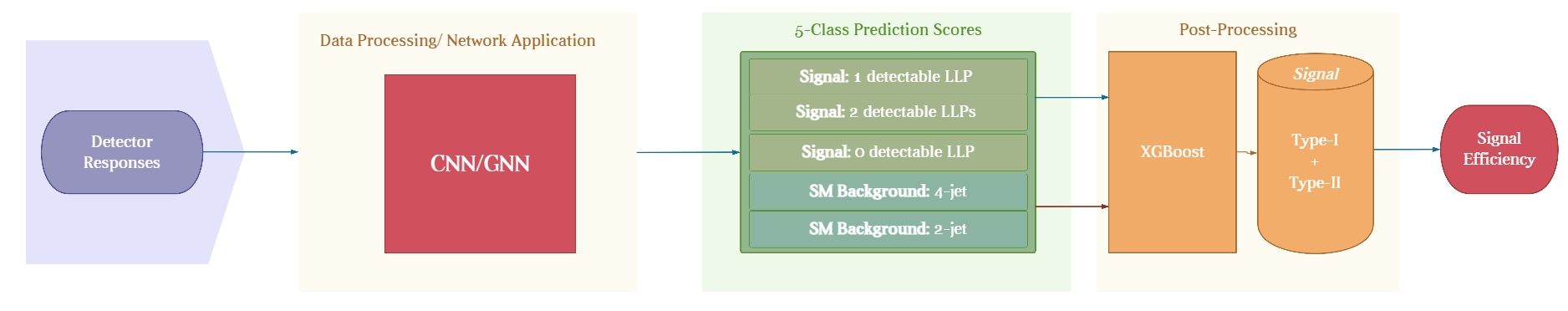}
    \caption{
        Workflow of event classification for LLP signals.
        This chart outlines the event classification process utilizing CNN and GNN models, followed by XGBoost analysis. 
        The process starts with formatting detector hits for NN input, progresses through 5-class classification, 
        and concludes with an XGBoost-enhanced selection to differentiate the signal from the SM background, finally determining signal efficiency.
    }
    \label{fig:llp_workflow}
\end{figure*}

Unlike the conventional method of event and object reconstruction followed by complex kinematic and object selection, our study adopts a direct application of two neural network (NN) models: a Convolutional Neural Network (CNN) and a Graph Neural Network (GNN), for the purpose of event classification and performance evaluation. Our method involves transforming low-level detector information on the energy, position, and time of detector hits, into formats suitable for these neural models, leading to a comprehensive classification of events into five distinct categories, as detailed in Figure~\ref{fig:llp_workflow}:
\begin{itemize}
\item LLPs signals:  $\mathcal{D}_0$, $\mathcal{D}_1$, $\mathcal{D}_2$ 
\item SM backgrounds: \textit{2-jet}, \textit{4-jet}
%\item \red{Higgs $\rightarrow$ invisible (including $\mathcal{D}_0$)}
\end{itemize}

The classification output is a set of prediction scores, indicating the probability of an event belonging to each of the five categories. Following this NN-based classification, an XGBoost algorithm~\cite{xgboost} is implemented to use these prediction scores as inputs to generate a single output score. This score discriminates between signal events and SM background.

To achieve a background-free analysis, we apply stringent event selection criteria based on XGBoost output scores to reject all SM background events. Events resulting in 0 detectable LLPs ($\mathcal{D}_0$), despite being classified as signal events during the neural network training stage to capture the full range of LLP behaviors, are subsequently excluded from the final signal extraction and limit calculation, since such events produce no experimentally distinguishable signatures. The fractions of events with exactly one ($\mathcal{D}_1$) or two ($\mathcal{D}_2$) detectable LLPs are derived for both type I and type II signals using Monte Carlo samples. The overall signal efficiency reported in this analysis thus reflects the combined selection efficiency of events containing at least one detectable LLP. Finally, based on the achieved signal efficiencies, we estimate the signal sensitivity and derive exclusion limits on the Higgs branching ratio into LLPs.
We employ two independent neural networks, CNN and GNN, to conduct a comparative study, evaluating their respective strengths in LLP reconstruction and signal identification.

\subsection{Convolutional Neural Network}

\begin{figure}[!htb]
    \subfloat{\includegraphics[width=0.52\linewidth]{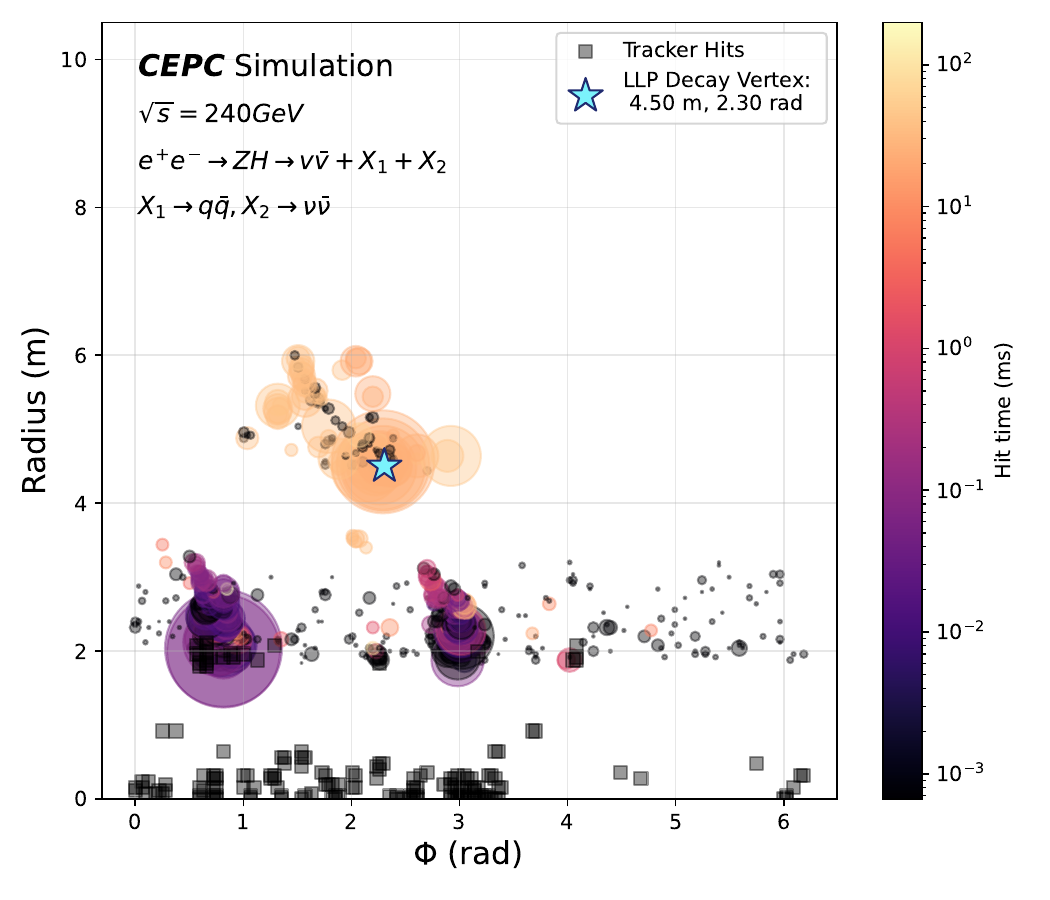}}
    \subfloat{\includegraphics[width=0.52\linewidth]{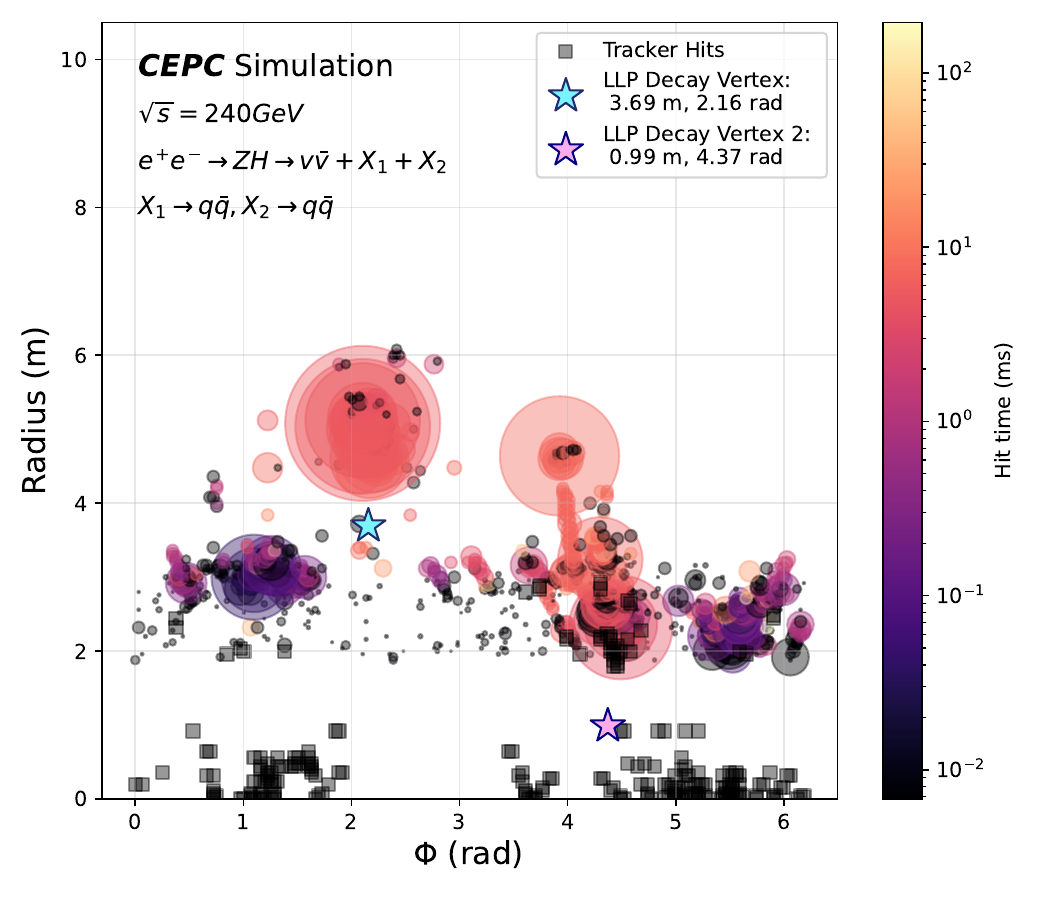} }
\caption{
Event visualization of CNN. 
\textbf{Left}: an event with one detectable LLP ($\mathcal{D}_1$, Type-I signal) featuring a decay vertex at [4.5 m, 2.3 rad]; \textbf{Right}: an event with two detectable LLPs ($\mathcal{D}_2$, Type-II signal) with decay vertices at [3.69 m, 2.16 rad] and [0.99 m, 4.37 rad].
Circles represent detector hits in the calorimeter and squares represent detector hits inside the tracker. Darker pixels represent hits with smaller time differences, and bigger pixels represent hits with larger energy.
Note that the varying circle sizes are solely for visualization purposes; in the analysis, the pixel size remains fixed.
The decay vertex of LLPs is marked with a star symbol. 
}
  \label{fig:event_display}
\end{figure}

\begin{figure*}[!htbp] 
\centering
{\includegraphics[width=0.95\linewidth]{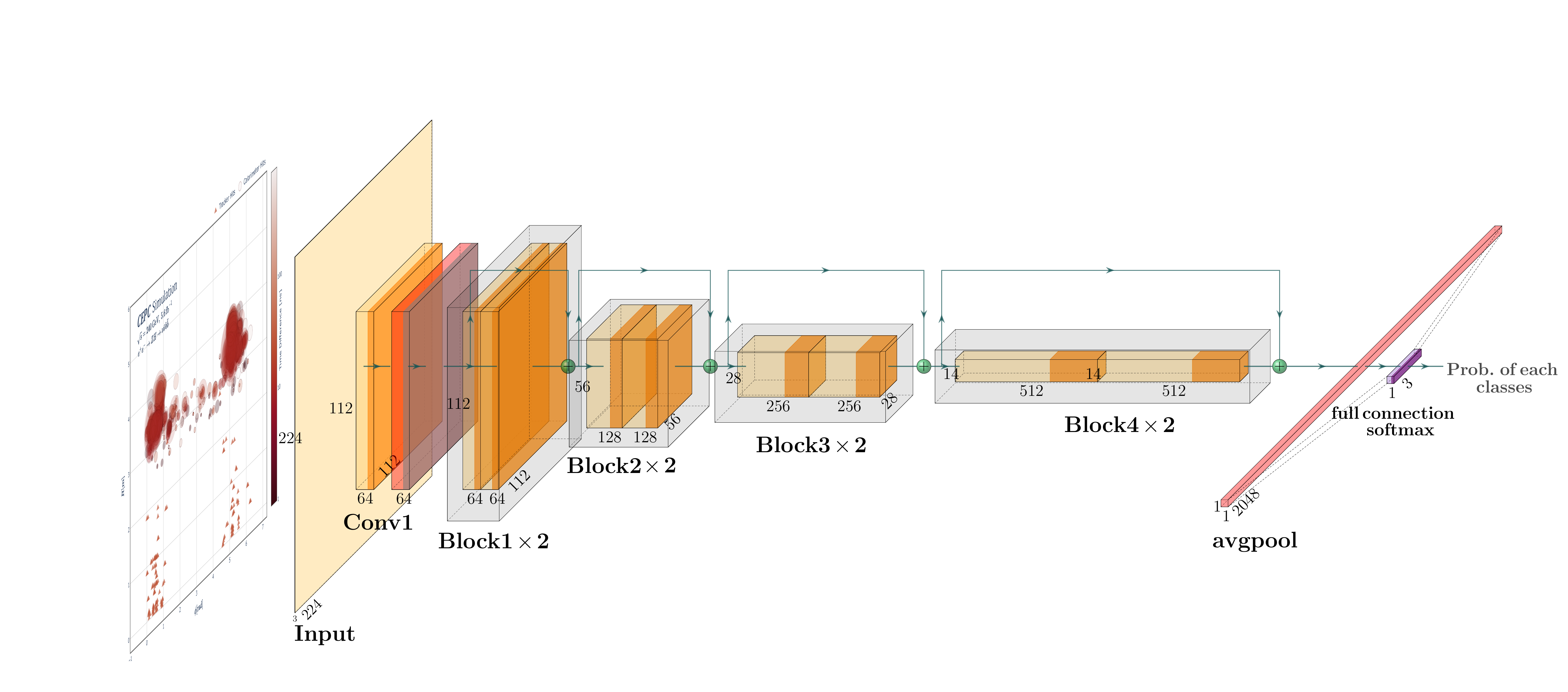}}
\caption{
Network structure of CNN-based classification.
The ResNet18 neural network architecture, consisting of two convolutional layers, followed by four ResNet blocks, an average pooling layer, and an output layer for classifications. 
}
\label{fig:ResNet}
\end{figure*}

For the CNN approach, each event with low-level detector information is transformed into a two-channel image with $200 \times 200$ pixels in $(R, \phi)$,
where $R$ represents the distance of the position of the hits to the interaction point ($0 - 6$~m) and $\phi$ represents the azimuthal angle of the hits ($0 - 2\pi$).
The first channel (energy channel) represents the sum of all hits energies in a pixel. 
The second channel (time channel) represents the time difference ($\Delta t$) associated with the most energetic hit in a pixel as shown in Eq.~\ref{eq:deltaT}.
\begin{equation}
\label{eq:deltaT}
 \Delta t = t_{hit, maxE} - {r_{hit, maxE}}/c
\end{equation}
$t_{hit, maxE}$ is the time of the hit with the maximum energy in a pixel, $r_{hit, maxE}$ is the Euclidean distance from the interaction point to the hit location, and \(c\) is the speed of light in vacuum.
A selection of hits energy larger than $0.1$~GeV is placed to suppress contributions from extremely low energy hits of secondary particles.
Figure~\ref{fig:event_display} shows images of hits for signal and background events.
The event images illustrate the discriminating power of CNN to separate LLP signals with displaced energies or objects from SM backgrounds.
By feeding event images to ResNet18 neural networks~\cite{DBLP:journals/corr/HeZRS15} shown in Figure~\ref{fig:ResNet}, a multi-label classification is performed.
The CNN model is trained using 5 GPUs with a batch size of 256 per GPU. The learning rate is adaptively set, beginning at $10^{-4}$ and progressively decreasing to $10^{-6.5}$. The training employs the Adam optimizer with a weight decay of 0.01. The training duration is set at 10 epochs, and the model's final configuration is determined by the epoch that achieves the lowest validation loss.
The distributions of the training and testing losses for CNN are shown in Appendix~\ref{app:training_loss}.

\subsection{Graph Neural Network}
In the GNN approach~\cite{Shlomi_2021}, low-level calorimeter and tracker hits from an event are clustered and then integrated into a heterogeneous event graph, encompassing both calorimeter-type and tracker-type nodes. 
    We implement a simple clustering method based on spatial proximity, grouping hits around the most energetic ones within a predefined region. This method is intentionally chosen for its computational simplicity, aiming to quickly and effectively reduce data dimensionality without requiring extensive event reconstruction. Given that lepton collider environments are relatively clean compared to hadron colliders, more complex clustering algorithms offer limited improvements in performance but at increased computational cost. Sensitivity studies confirmed that varying clustering parameters, such as minimum-hit thresholds and cluster radii, had negligible impact on signal efficiency and background rejection efficiency, thereby validating the robustness and suitability of our approach.
The clustering process helps to minimize the complexity of the graph and significantly reduces memory usage during training.
Furthermore, nodes of the same detector type are comprehensively interconnected, facilitating the formation of edges within the graph. This structure optimizes the analysis by ensuring efficient data processing and interaction modeling between different components of the detector hits.
For the clustering of hits in the calorimeter, a point cloud composition method based on energy screening and distance thresholding is introduced. The most energetic calorimeter hit is first identified as the core hit. Subsequently, hits within a predefined distance threshold of $50~mm$ from the core hit are merged, constructing a graphical node structure. It is required that the number of merged hits exceeds 3 and the total number of hits in a cluster exceeds 10. This procedure is then iterated for all hits in an event to construct the calorimeter-type nodes. Hits that fail to meet these requirements are not clustered and are excluded from the GNN.
The momentum \( p \) of each calorimeter hit is defined as parallel to its position: \( p_{i} = \frac{i}{r}E \), where \( i \) represents the position along the \( x \), \( y \), or \( z \) axis, with \( z \) as the beam line direction, \( x \) and \( y \) in the transverse plane, and \( y \) pointing towards the ground. \( r \) is the radial distance \( r = \sqrt{x^{2} + y^{2} + z^{2}} \), and \( E \) is the hit's energy.
The tracker hits are binned into $5~\times~6$ blocks based on their R-$\phi$ positions.
In each block, the tracker hits are clustered into a tracker-type node taking an arithmetic average of all hit positions.
The definitions of nodes and edges for the GNN event graph are summarized in Table~\ref{tab:gnn-fea}.

\begin{figure*}[!htb]
    \centering
    \includegraphics[width=0.95\linewidth]{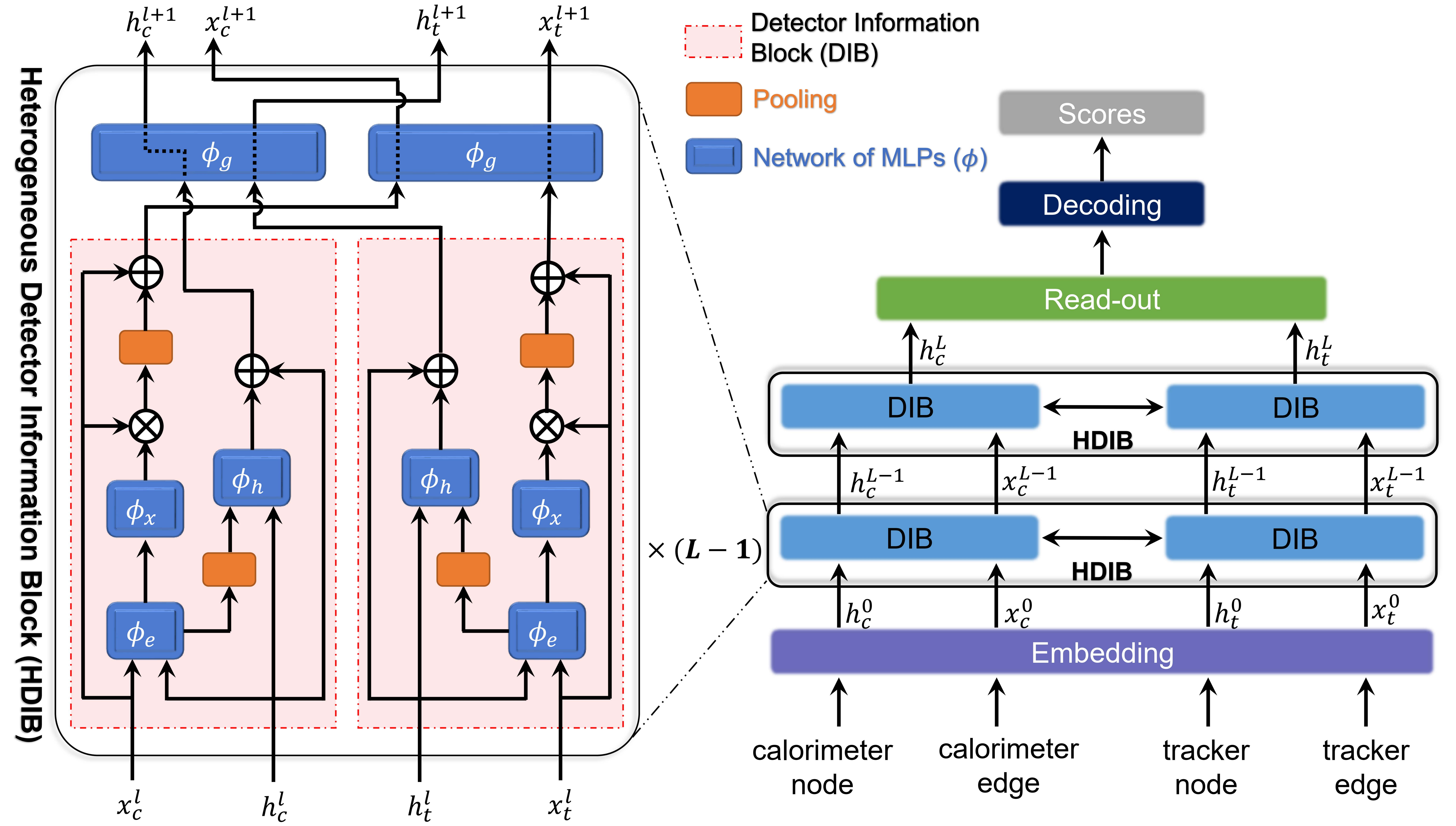}
    \caption{
    Architecture of the heterogeneous GNN.
    $h_{t}$ and $h_{c}$ denote the node embedding, $x_{t}$ and $x_{c}$ denote the edge embedding. The subscript letter $l$ represents the $l$-th layer. The subscript letter $t$ represents the tracker and the letter $c$ represents the calorimeter. $\phi_{e}$, $\phi_{x}$, $\phi_{h}$ and $\phi_{g}$ are neural networks of MLPs.
    }
    \label{fig:gnn_structure}
\end{figure*}

Event graphs are then input into a GNN-based heterogeneous architecture, as illustrated in Figure~\ref{fig:gnn_structure}. 
Features of different types of nodes and edges are first embedded into a high-dimension latent space and then forwarded to a Heterogeneous Detector Information Block (HDIB). The HDIB consists of two Detector Information Blocks (DIBs)~\cite{lorentznet, satorras2021n}and two multilayer perceptrons (MLPs).
The HDIB adopts a parameter-sharing design between the tracker and the calorimeter types, specifically tailored for integrating information from different types of detectors.
After the total $L$ layers of HDIBs, the final node embeddings of the tracker and the calorimeter are aggregated and forwarded to the decoding layer to form the classification scores.

The GNN training is performed with 8 GPUs and the batch size is 256 per GPU. The initial learning rate is $10^{-4}$, and a dropout rate of 0.1 is used to avoid overfitting. 
The cross-entropy loss is minimized using the Adam optimizer without weight decay.
The GNN model is trained with 30 epochs, and the network is validated at the end of each epoch.
The model demonstrating the minimum validation loss value on the validation dataset is then applied to the final test dataset.
The distributions of the training and testing losses for GNN are shown in Appendix~\ref{app:training_loss}. 

\begin{table*}[!htb]
\centering
\caption{Node and edge features defined in the heterogeneous graph.}
\label{tab:gnn-fea}
\resizebox{1.0\textwidth}{!}{
\begin{tabular}{cccc}
    \toprule
     Features & Variable & Definition  \\ 
     \midrule
    \multirow{6}*{calorimeter type node $i$} & $|x_{i}^{\mu}|$ & the space-time interval \\
    ~ & $|p_{i}^{\mu}|$ & the invariant mass\\ 
    ~ & $N_{i}$ & the number of hits \\
    ~ & $\eta_{i}$ & $\frac{1}{2}\ln{\frac{1+\frac{p_{z}}{p}}{1-\frac{p_{z}}{p}}}$  \\
    ~ & $\phi_{i}$ & arctan$\frac{p_{y}}{p_{x}}$ \\
    ~ & $\mathcal{R}_{i}$ & $\sqrt{\eta^{2}+\phi^{2}}$ \\ 
    \midrule
    \multirow{2}*{calorimeter type edge between node $i$ and $j$} & \multicolumn{2}{c}{$x_{i}^{\mu}x_{j\mu}$, $p_{i}^{\mu}p_{j\mu}$, $x_{i}^{\mu}p_{j\mu}$, $p_{i}^{\mu}x_{j\mu}$}  \\
    ~ & \multicolumn{2}{c}{$|x_{i}^{\mu}-x_{j}^{\mu}|$, $|p_{i}^{\mu}-p_{j}^{\mu}|$, $\eta_{i}-\eta_{j}$, $\phi_{i}-\phi_{j}$, $\mathcal{R}_{i}-\mathcal{R}_{j}$} \\ 
    \midrule
    \multirow{5}*{tracker type node $i$} & $|r|$ & euclidean distance  \\
    ~ & $N_{i}$ & the number of hits \\
    ~ & $\eta_{i}$ & $\frac{1}{2}\ln{\frac{1+\frac{z}{r}}{1-\frac{z}{r}}}$ \\
    ~ & $\phi_{i}$ & arctan$\frac{y}{x}$ \\
    ~ & $\mathcal{R}_{i}$ & $\sqrt{\eta^{2}+\phi^{2}}$ \\ 
    \midrule
    \multirow{1}*{tracker type edge between node $i$ and $j$} & \multicolumn{2}{c}{$|r_{i}-r_{j}|$, $r_{i}r_{j}$, $\eta_{i}-\eta_{j}$, $\phi_{i}-\phi_{j}$, $\mathcal{R}_{i}-\mathcal{R}_{j}$}  \\
    \bottomrule
\end{tabular}
}
\end{table*}

\section{Systematic Uncertainties}
\numberedfootnotes
The dominant uncertainty source comes from the training variability of neural networks. 
To assess the robustness of our machine-learning models, 50 CNNs are trained with different initial seeds. The training uncertainty for the neural networks is estimated as half of the difference between the maximum and minimum efficiencies observed, which amounts to approximately \(1.7\%\)~\myfootnote{A similar procedure was applied to the GNN models, yielding a training uncertainty of approximately 2.3\%, although this is not included in the final results since the CNN model serves as the baseline model.}.The theoretical uncertainty of the ZH cross section is estimated to be 3\%~\cite{Freitas:2022hyp}.
The uncertainty originating from the integrated luminosity for CEPC at $\sqrt{s} = 240$~GeV is estimated to be \( 0.13\% \)~\cite{lum_sys}, which is negligible in this study.
\symbolfootnotes

\section{Results}
\begin{figure}[!htb]
    \centering
    \includegraphics[width=0.65\columnwidth]{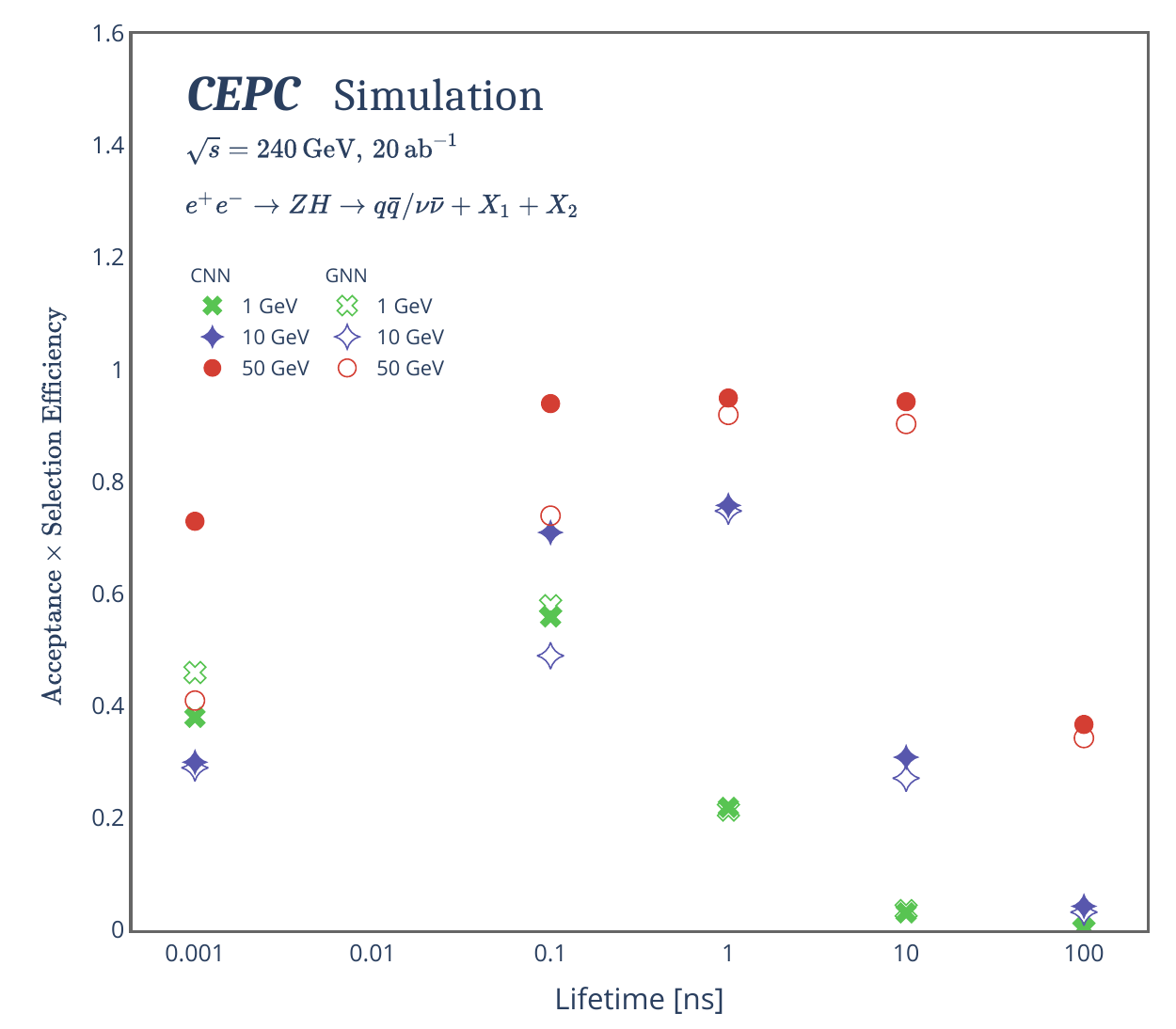}
    \caption{
    Efficiency and acceptance of LLP Detection with CEPC detector.
    The product of geometry acceptance and selection efficiency for LLPs with varying masses and lifetimes. 
    Solid symbols represent data from a CNN-based approach, while hollow symbols are from a GNN-based approach. 
    The varied shapes correspond to different assumed LLP masses, with circles for 50 GeV, diamonds for 10 GeV, and crosses for 1 GeV. 
    The error bars representing statistical uncertainties have been omitted from the figure as their values are too small for effective visual representation.
    }
    \label{fig:combined_eff}
\end{figure}

Both CNN and GNN methods have been evaluated for the efficiency of LLP signals (type I and II signals combined) from the XGboost output score while keeping background free, 
the efficiency results are summarized in Table~\ref{tab:combined_eff}. The signal efficiencies of the CNN method range from 30\% to 95\%, while the GNN method efficiencies range from 29\% to 92\% for various LLP masses and lifetimes. 
The signal efficiency is multiplied by the signal acceptance to obtain the signal yield, as shown in Figure~\ref{fig:combined_eff}.

\begin{table*}[htbp] 
  \centering
  \caption{
  Signal efficiencies for different LLPs masses and lifetimes obtained with CNN-based and GNN-based methods.
  All efficiency values are presented with their corresponding statistical uncertainties.
  }
  \resizebox{1.0\textwidth}{!}{
\begin{tabular}{ccccccc}
  \toprule
  \multirow{2}{*}{Approach} & Efficiency (\%) & \multicolumn{5}{c}{Lifetime [ns]} \\
  \cmidrule{2-7}
  & Mass [GeV] & 0.001 & 0.1 & 1 & 10 & 100 \\ 
  \midrule
  \multirow{3}{*}{CNN} & 1 & $38 \pm 0.1$ & $56 \pm 0.1$ & $45 \pm 0.2$ & $48 \pm 0.6$ & $52 \pm 1.9$ \\
  & 10 & $30 \pm 0.1$ & $71 \pm 0.1$ & $76 \pm 0.1$ & $66 \pm 0.1$ & $69 \pm 0.5$ \\
  & 50 & $73 \pm 0.1$ & $ 94 \pm 0.1$ & $95 \pm 0.0$ & $95 \pm 0.0$ & $91 \pm 0.1$ \\
  \midrule
  \multirow{3}{*}{GNN}
  & 1 & $46 \pm 0.1$ & $58 \pm 0.1$ & $44 \pm 0.2$ & $54 \pm 0.6$ & $43 \pm 1.8$ \\
   & 10 & $29 \pm 0.1$ & $49 \pm 0.1$ & $75 \pm 0.1$ & $58 \pm 0.2$ & $52 \pm 0.5$ \\
  & 50 & $41 \pm 0.1$ & $74 \pm 0.1$ & $92 \pm 0.1$ & $91 \pm 0.1$ & $85 \pm 0.1$ \\
  \bottomrule
\end{tabular}
}

  \label{tab:combined_eff}
\end{table*}

The signal efficiencies for both the CNN and GNN approaches were evaluated. Their performance is comparable for low-mass LLP scenarios; however, at intermediate and higher LLP masses, the CNN consistently outperforms the GNN. Consequently, CNN is chosen as the baseline model for the subsequent sensitivity analysis and exclusion limit calculations, as reflected in Table~\ref{tab:combined_eff}.

To estimate the sensitivity, we generate 10,000 pseudo-experiments under the null hypothesis (no LLP signal), using the Asimov dataset. In each pseudo-experiment, event yields are randomly sampled from Poisson distributions with means equal to the expected Standard Model background yields, thereby realistically modeling statistical fluctuations. The profile likelihood ratio is employed as the test statistic, defined as:
\(
q_\mu = -2\ln\frac{\mathcal{L}(\mu, \hat{\hat{\theta}})}{\mathcal{L}(\hat{\mu},\hat{\theta})},
\)
where \(\mu\) is the signal strength under test, \(\hat{\mu}\) and \(\hat{\theta}\) denote the best-fit signal strength and nuisance parameters, and \(\hat{\hat{\theta}}\) represents the nuisance parameters' conditional maximum likelihood estimates for a given \(\mu\).

\begin{figure}[!htbp]
    \footnotesize
    \centering
        \subfloat{\includegraphics[width=0.45\textwidth]{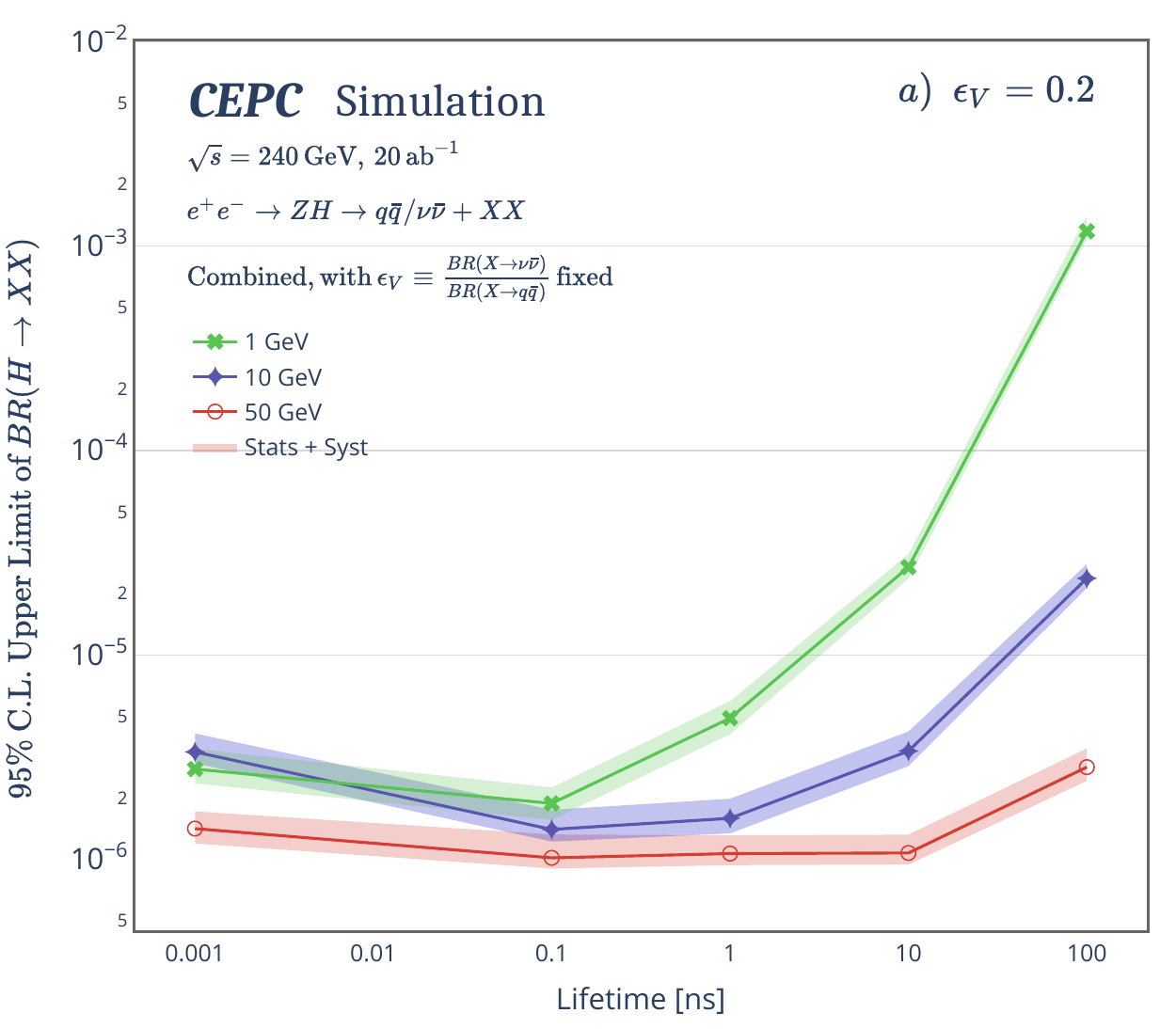}}
        \subfloat{\includegraphics[width=0.45\textwidth]{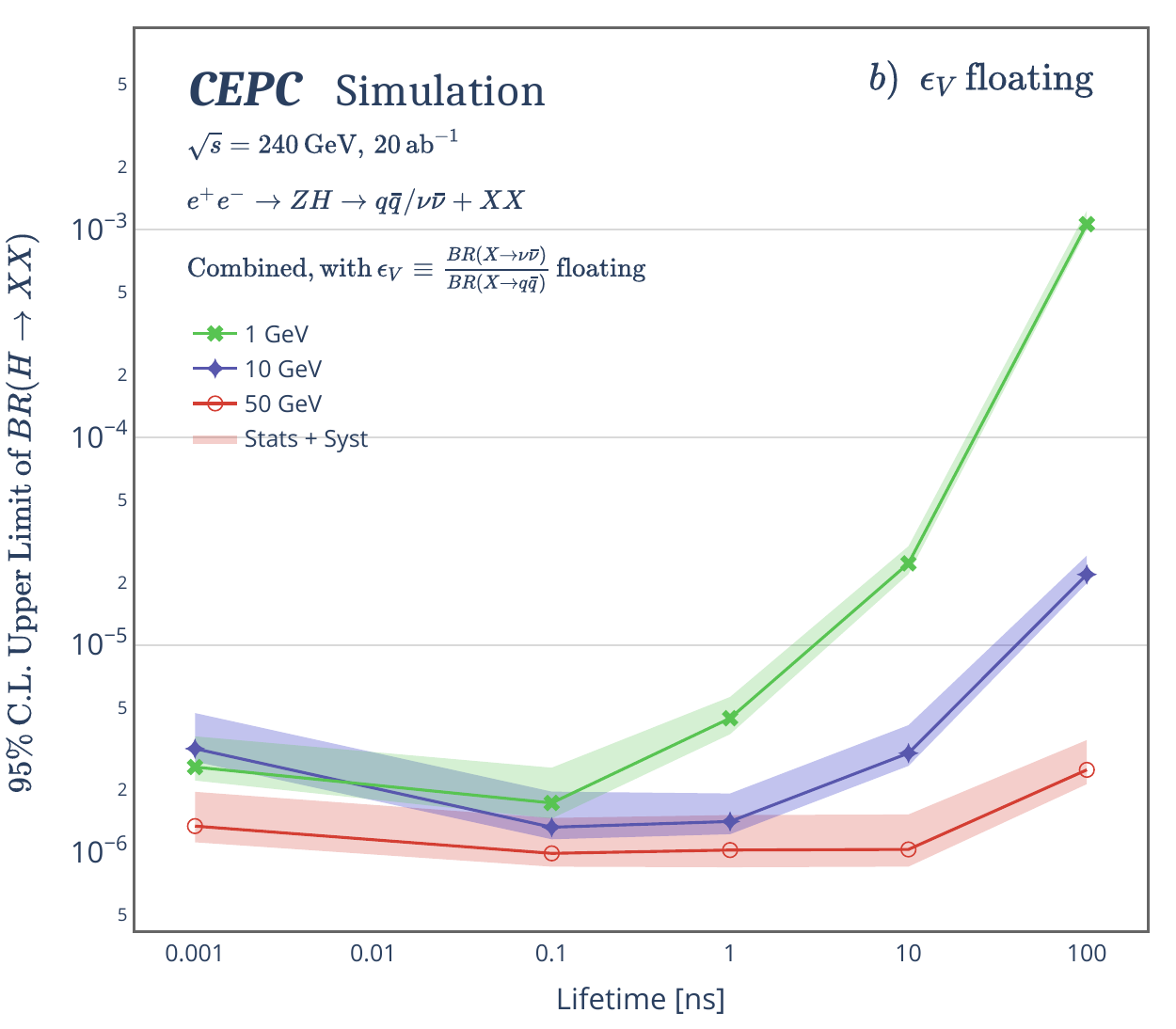}}
    \caption{
    One-dimensional constraints on Higgs boson decays to LLPs.
    95\% C.L. upper limit on the branching ratio (BR) for the Higgs boson (\( H \)) decay into pairs of LLPs (\( XX \)), where \( \epsilon_{V} \) is the ratio \( \frac{BR(X \to \nu\bar{\nu})}{BR(X \to q\bar{q})} \).
    \textbf{a)}: a fixed ratio \( \epsilon_{V} = 0.2 \), 
    \textbf{b)}: a floating \( \epsilon_{V} \). 
    The shaded areas indicate statistical and systematic uncertainties combined.
    }
    \label{fig:limit_total}
\end{figure}

The 95\% confidence level upper limits on the branching ratio \(\mathcal{B}(H \rightarrow XX)\) are derived using the CLs method~\cite{CLs}, comparing the observed test statistics against the expected distributions under the null and alternative (with LLP signal) hypotheses. The analyzed samples correspond to a luminosity of \(20~\text{ab}^{-1}\), corresponding to approximately \(4 \times 10^6\) Higgs bosons~\cite{CEPCStudyGroup:TDR1}. We consider two LLP signal scenarios:

\begin{itemize}
    \item Type I and Type II signal yields have a fixed ratio defined by the parameter \(\epsilon_{V}:= \frac{BR(X \to \nu\bar{\nu})}{BR(X \to q\bar{q})}\), which is set to 0.2. A one-dimensional 95\% confidence level upper limit on \(\mathcal{B}(H \rightarrow XX)\) is derived and shown in Figure~\ref{fig:limit_total}(a). Further detailed results from the pseudo-experiments under the null hypothesis are provided in Appendix~\ref{app:sensitivity}.
     
    \item Type I and Type II signal yields have a floating ratio  \( \epsilon_V \) with an allowed range from $10^{-7}$ to $1$. A one-dimensional 95\% Confidence Level upper limit on \( \mathcal{B}(H \rightarrow XX) \) is derived and shown in Figure~\ref{fig:limit_total}b).
 
 \item The ratio of Type I to Type II signal yields is floating, and two-dimensional 95\% Confidence Level upper limits on $\mathcal{B}_{\textrm{2-jet}}$ and  $\mathcal{B}_{\textrm{4-jet}} $ are derived. A bivariate statistical fit is performed to derive the upper limits and results are shown in Figure~\ref{fig:limit_2d}. 
More detailed results on 2-D upper limits can be found in Appendix~\ref{app:2d_limit}.
\end{itemize}

Both 1-D and 2-D exclusion limits are summarized in Table~\ref{tab:limit_combined}. As the results indicate, statistical uncertainties dominate in this analysis.

In addition, we relaxed the background-free assumption to evaluate the lower bound of the experimental sensitivity. The results under the assumption of two background events are presented in Appendix~\ref{sec:sensitivity_2bkg}.

\begin{figure*}[!htb]
   \includegraphics[width=0.32\linewidth]{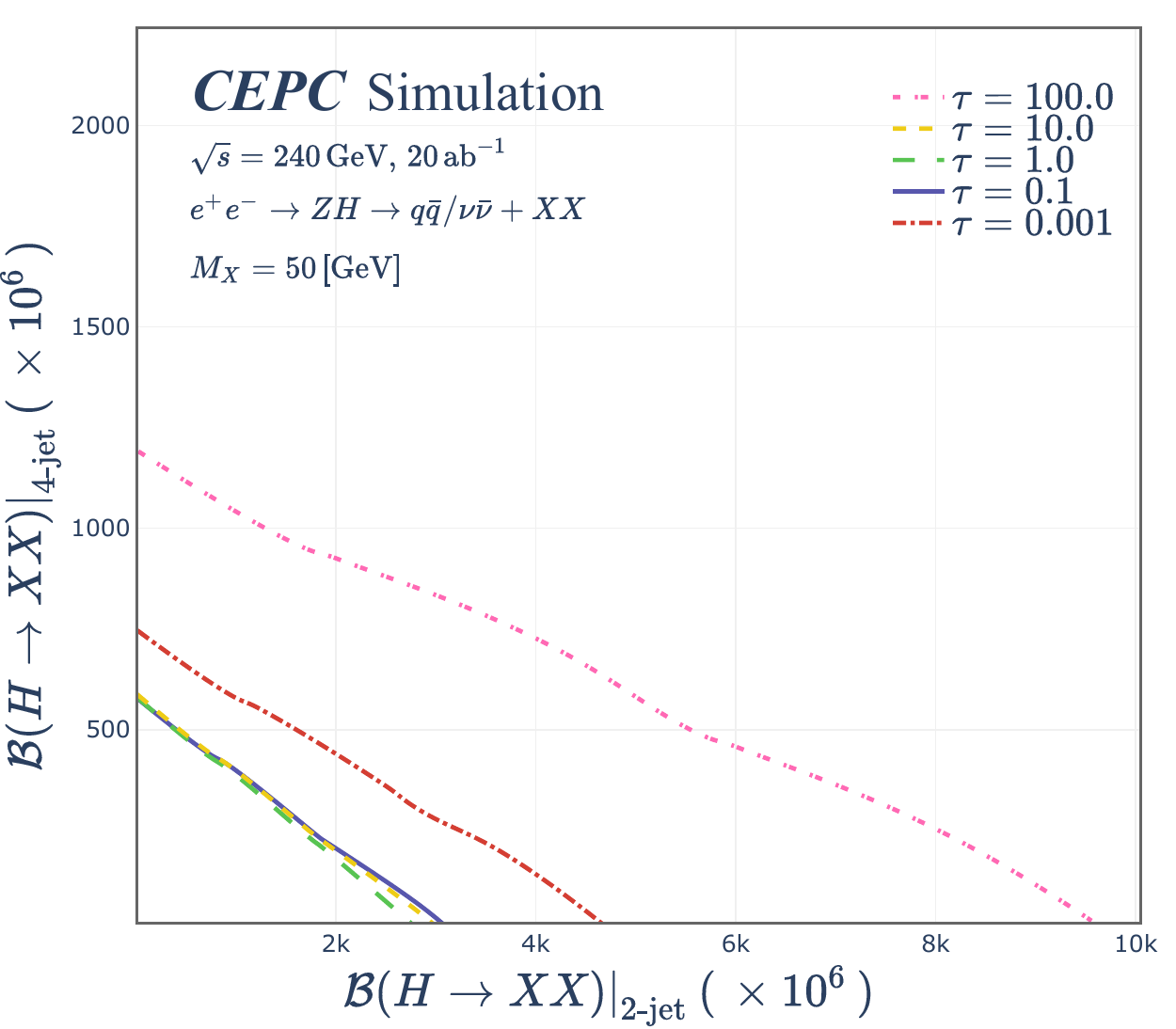}
   \includegraphics[width=0.32\linewidth]{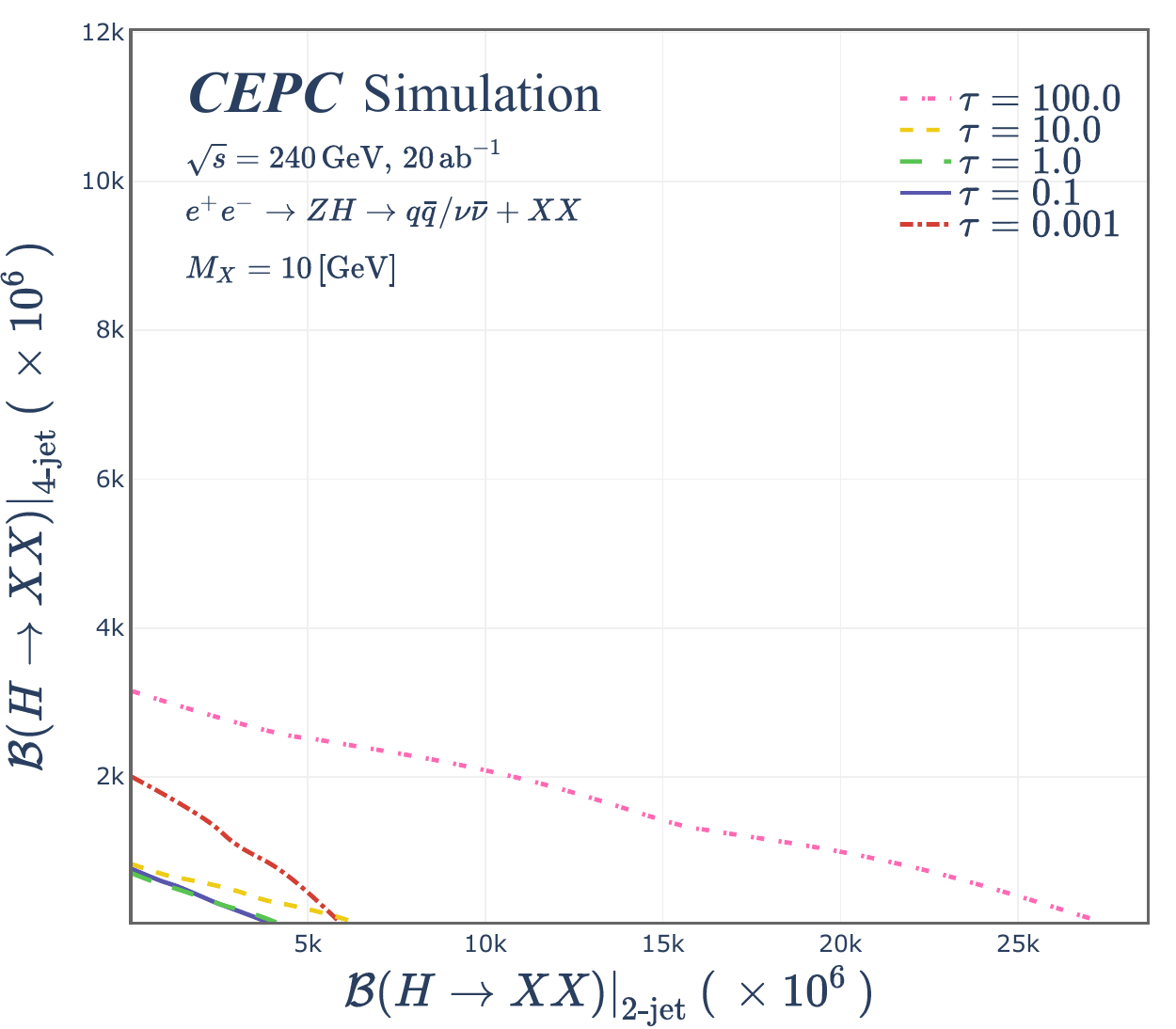}
   \includegraphics[width=0.32\linewidth]{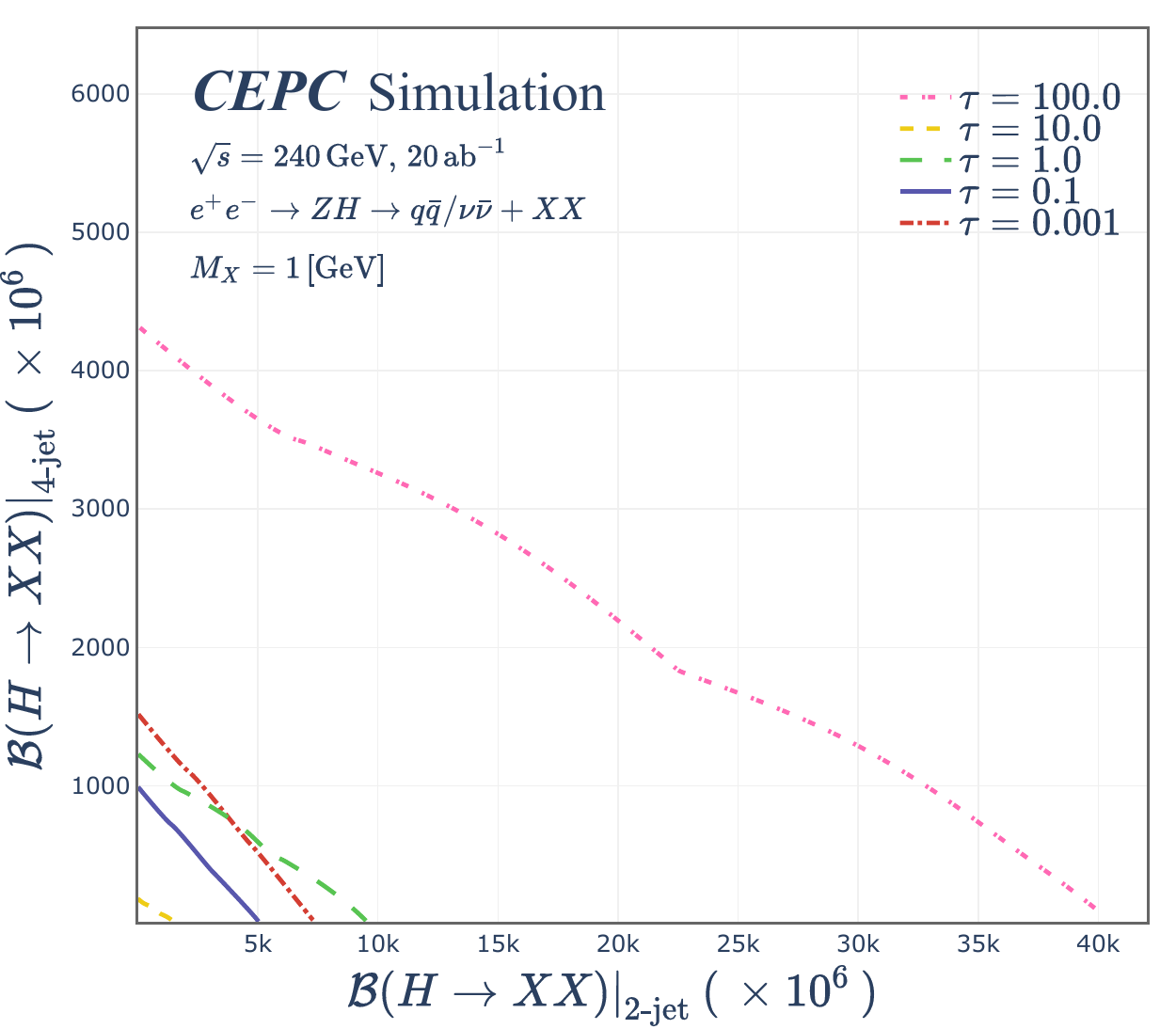}
   \caption{ 
   Two-dimensional constraints on Higgs Boson decays to LLPs.
   The 95\% C.L. 2-D upper limit on ($\mathcal{B}_{\textrm{2-jet}}$, $\mathcal{B}_{\textrm{4-jet}}$) for three LLPs masses 50~GeV (left), 10~GeV (middle), 1~GeV (right). Different colored lines indicate different LLP lifetimes. The uncertainties on the limits are omitted and a few limits are scaled by a factor for better visibility: 1) 4 for $M_X=10$, $\tau = 100$; 2) 0.2 for $M_X=1$, $\tau=10$; 3) 4 for $M_X=50$, $\tau=100$.
   }
   \label{fig:limit_2d}
\end{figure*}

\begin{table}[!htb] 
  \centering
  \caption{
  The 95\% C.L. exclusion limit on BR($h \to XX$) for all signal channels with both fixed and floating $\epsilon_{V}$.  
  The limits include $\pm 1 \sigma$ uncertainties after taking into account both statistical and systematic contributions.}
  
  \resizebox{0.8\textwidth}{!}{
  \begin{tabular}{ccccccc}
  \toprule
   \multirow{2}{*}{Scenario} & $\mathcal{B}$  ($\times 10^{-6}$) & \multicolumn{5}{c}{Lifetime [ns]} \\
  \cmidrule{2-7}
   & Mass [GeV] & 0.001 & 0.1 & 1 & 10 & 100 \\ 
   \midrule
   \multirow{3}{*}{Fixed} & 1 &$2.77^{+0.70}_{-0.41}$ & $1.88^{+0.37}_{-0.32}$ & $4.92^{+1.03}_{-0.81}$ & $26.94^{+4.20}_{-3.23}$ & $1181.45^{+217.49}_{-100.45}$ \\
   & 10 & $3.36^{+0.78}_{-0.40}$  & $1.40^{+0.35}_{-0.18}$  & $1.59^{+0.39}_{-0.25}$   & $3.39^{+0.84}_{-0.53}$ & $23.73^{+4.12}_{-2.42}$ \\
   & 50 & $1.42^{+0.30}_{-0.22}$ & $1.02^{+0.31}_{-0.12}$ & $1.07^{+0.24}_{-0.13}$ & $1.08^{+0.25}_{-0.13}$ & $2.83^{+0.66}_{-0.41}$ \\
  \midrule
   \multirow{3}{*}{Floating} & 1 & $2.59^{+1.05}_{-0.36}$ & $1.74^{+0.83}_{-0.28}$ & $4.46^{+1.18}_{-0.72}$ & $24.80^{+5.17}_{-2.78}$ & $1063.64^{+172.59}_{-68.69}$ \\
   & 10 & $3.18^{+1.54}_{-0.42}$ & $1.33^{+0.64}_{-0.17}$ & $1.42^{+0.52}_{-0.19}$  & $3.02^{+1.10}_{-0.40}$ & $21.93^{+5.09}_{-2.03}$ \\
   & 50 & $1.35^{+0.62}_{-0.22}$ & $1.00^{+0.48}_{-0.14}$ & $1.03^{+0.49}_{-0.18}$ & $1.04^{+0.49}_{-0.18}$ & $2.51^{+0.98}_{-0.37}$ \\
  \bottomrule
  \end{tabular}
  }
  \label{tab:limit_combined}
\end{table}

\section{Discussion}

\subsection{Comparison between CNN-based and GNN-based Methods and Selection-based Method}

From the signal efficiencies listed in Table~\ref{tab:combined_eff}, both the CNN and GNN approaches deliver excellent performance across the parameter space. Notably, their efficiencies are comparable at low LLP masses; however, as the mass increases, the CNN consistently outperforms the GNN. This mass-dependent difference likely stems from CNNs' ability to effectively capture spatially localized features in the two-dimensional image representation, which become more pronounced at higher masses. Consequently, we adopt CNN as the baseline for our subsequent sensitivity evaluation and exclusion limit calculations, while acknowledging that GNN remains a competitive alternative—particularly in low-mass regimes. 
Both methods demonstrate superior capabilities in deciphering the complex data distributions present in the detector space, capturing intricate event structures and topologies more effectively when comparing with the traditional selection-based approach. Detailed comparisons with the selection-based method is available in Appendix~\ref{app:sec_cut_based_ana}.

CNN is adept at recognizing local features in two-dimensional space, whereas GNN delves into the relationships and structural nuances of nodes within a higher-dimensional framework, effectively identifying displaced vertices, a key characteristic of LLPs. In practical application, CNNs process events as two-dimensional images, while GNNs require an initial transformation of event information into graph structures. This preprocessing step can affect GNN's performance, contingent on the algorithm's ability to retain crucial event information. In terms of training speed, GNNs generally achieve faster speeds due to the reduced number of nodes compared to the number of pixels in CNN images. Nevertheless, CNNs tend to achieve quicker convergence, usually within five epochs, as opposed to GNN's requirement of 15-20 epochs.

To further enhance network performance, GNN could benefit from refined algorithms that improve graph conversion and granularity, all the while preserving data integrity and minimizing graph complexity. For CNN, the accuracy of the network is significantly influenced by image resolution, highlighting computational power as a primary challenge.

\subsection{Comparisons with other searches}

We evaluate the effectiveness of our ML-based approach for LLPs detection at future lepton colliders by comparing our results with results at hadron colliders with the ATLAS experiment~\cite{ATLAS2024LLP,ATLAS_llp_muon}
and CMS experiment~\cite{CMS:llp2023,CMS_llp_2024} as well as future HL-LHC experiments~\cite{HL-LHC}. The comparison is based on four primary metrics: signal acceptance, selection efficiency, analysis strategy and signal yields:
\begin{itemize}
    \item \textbf{Signal Acceptance:}
   Both ATLAS and CMS results have limited signal acceptance, typically a few percent, as they focus on LLPs decaying in the muon detector. In contrast, our ML-based approach covers the entire detector, resulting in 100\% signal acceptance except for LLPs with long lifetime ($> 10$~ns) and low mass ($< 10$~GeV).
    
    \item \textbf{Selection Efficiency:}
    For hadron colliders, LLPs events typically trigger on displaced decays and/or large missing transverse energy. The LLPs trigger efficiency at the ATLAS experiment is estimated to be between \(10^{-3}\) and  \(0.3\)~\cite{ATLAS2024LLP}. Besides trigger efficiency, there are additional efficiencies involved such as displaced vertex/object reconstruction efficiencies which are typically in the order of a few percent. In contrast, LLPs event selection at lepton colliders can adapt to a triggerless-equivalent approach~\cite{trigger1,trigger2} owing to the clean environment. The ML-based approach can be applied directly with low-level detector information without any event-level reconstruction. As a result, our ML-based approach with lepton colliders can achieve an overall selection efficiency as high as \(95\%\), an improvement of several orders in magnitude when compared with LHC or HL-LHC efficiencies.

    \item \textbf{Analysis Strategy:}
    Traditionally, analyses of LLPs conducted elsewhere have employed a selection-based method, which involves categorizing events into multiple subsets with different decay modes and orthogonal signal types. These analyses necessitate manual re-tuning and re-optimization for each subset and different LLPs mass and lifetime configurations. In contrast, deep neural networks can be retrained in automation with a similar setup for each LLPs mass and lifetime, resulting in a simplified analysis framework and higher efficiencies compared to the selection-based method. Additional comparisons can also be found in Appendix~\ref{app:sec_cut_based_ana}.
    
    \item \textbf{Signal Yields and Upper Limits Comparison:}
    LHC and HL-LHC can produce a significantly larger number of Higgs bosons compared to lepton colliders. Despite this, higher signal acceptance and selection efficiencies in our ML-based approach compensate for the relatively low number of Higgs bosons. We achieve upper limits as low as \(1.0 \times 10^{-6}\) on \( \mathcal{B}(H \rightarrow XX) \) with \(4 \times 10^6\) Higgs bosons. This upper limit is approximately three orders of magnitude better than the \(10^{-3}\) limit observed at the ATLAS and CMS experiments with \(10^7\) Higgs bosons and it is comparable to the projected HL-LHC limit with about \(10^8\) expected Higgs bosons.
     
\end{itemize}

Besides comparing with LLPs searches at hadron colliders, we have also compared our result with a preliminary LLPs study~\cite{jeanty2022sensitivity} on the ILC~\cite{ILC:2019gyn} sensitivity with a traditional selection-based method. The ILC sensitivity study searches for long-lived dark photons produced in Higgstrahlung events via the Higgs portal. We have compared our result with the hadronic decay dark photo result since the event signature is similar. We have seen that the signal acceptance factors are similar between ILC and CEPC detectors but the signal efficiencies differ significantly. The signal efficiencies in the ILC study range from \(0.1\%\) to \(10\%\), which is at least an order of magnitude lower than ours. The upper limits on \( \mathcal{B}(H \rightarrow XX) \) in the ILC study are derived under the assumption of \(100\%\) truth-level signal efficiency. Under this assumption, two results show similar sensitivities only in the low lifetime region (\(< 1~ns\)) of LLPs. In the long lifetime region, for example, for a 1 GeV LLP with a lifetime of a few nanoseconds, our result yields an upper limit of about \(5 \times 10^{-6}\) which is an order of magnitude better than the ILC’s upper limit of \(10^{-4}\). 

\subsection{Comparison with an External Detector}
Finally, we quantify the sensitivity gain from deploying an additional external detector alongside the primary detector.
As shown in Appendix~\ref{app:external_detector}, this configuration can improve the LLP detection reach by up to a factor of~13.7 in certain mass–lifetime regimes, reinforcing the potential value of such extensions.

\section{Conclusion}
In summary, we have investigated the search for LLPs at future lepton colliders using machine learning techniques with full simulation data samples. We have demonstrated that the efficiency of LLP signal detection can reach up to 95\% for LLPs with a mass around 50 GeV and a lifetime of approximately 1 nanosecond, surpassing traditional selection-based methods on almost all fronts. 
Our methodology's effectiveness is not confined to theoretical constructs but is evidenced by detailed simulations. With a luminosity of \(20~\text{ab}^{-1}\) and around \(4 \times 10^6\) Higgs bosons analyzed, this study sets new benchmarks for sensitivity in LLP searches with lepton colliders, the branching ratio of Higgs decaying into LLPs reaches \(1.0 \times 10^{-6}\).

This work not only demonstrates the substantial potential of deep learning techniques in particle physics research but also sets a solid foundation for future explorations at lepton colliders such as CEPC, ILC, FCC, and CLIC. The adaptability of our machine learning approach across different collider environments augurs well for its application in broader physics analyses, promising a new era of discoveries.

\section{Acknowledgement}
The authors thank theorists Zhen Liu, Tao Liu, Jia Liu, Ziping Wang, as well as experimentalists Manqi Ruan and Gang Li for their useful suggestions and inputs about the paper. This work was supported by 
 National Key R\&D Program of China (Grant No.~2024YFA1610603,~2023YFA1606003);
National Natural Science Foundation of China (Grants No.~12305217, 12375072 and 12475108).

%\section*{References}
%\nocite{*}
\bibliography{ref}
\bibliographystyle{unsrt}

\clearpage

\appendix
\setcounter{section}{1}
\appendix

\section{Pile-up and cosmic ray backgrounds}
\label{pile_up_and_cosmic_ray}

Unlike hadron colliders, lepton colliders typically exhibit a significantly lower level of pile-up due to the cleaner nature of lepton interactions and the generally lower total cross-sections involved. The pile-up (\(\mu\)) at a lepton collider can be quantitatively estimated using the formula \(\mu = L \cdot \sigma \cdot \tau\), where \(L\) is the luminosity, \(\sigma\) is the total cross-section, and \(\tau\) is the bunch spacing.

Given the parameters for the CEPC~\cite{CEPCStudyGroup:CDR2}, with \(L = 5 \times 10^{34} \, \textrm{cm}^{-2} \textrm{s}^{-1}\), \(\sigma = 1000 \times 10^{-36} \, \textrm{cm}^2\), and \(\tau = 636 \times 10^{-9} \, \textrm{s}\), the computed pile-up rate is approximately \(3 \times 10^{-5}\). This low pile-up rate underscores the minimal impact of pile-up on the CEPC's operation, making the pile-up background negligible.

The cosmic ray background, particularly from muons, is effectively mitigated in lepton colliders through several ways, making it a negligible concern in most analyses. 
Firstly, the timing method uses the fact that cosmic muons do not coincide with the precise timing of collision events, allowing for a clear differentiation based on the synchronization of particle detection with beam crossings. 
Secondly, the distinct trajectories of cosmic muons, which are typically vertical, are different from the varied directions of muons originating from collision points. 
Moreover, the underground placement of detectors significantly shields them from cosmic interference, complemented by additional physical barriers that absorb cosmic radiation. The combination of these strategies effectively minimizes the cosmic ray background noise. This background is also ignored in other LLPs search experiments such as the MATHUSLA experiment~\cite{MATHUSLA_2019}.

\section{Training losses}
\label{app:training_loss}
\begin{figure*}[htbp]
  \includegraphics[width=0.30\linewidth]{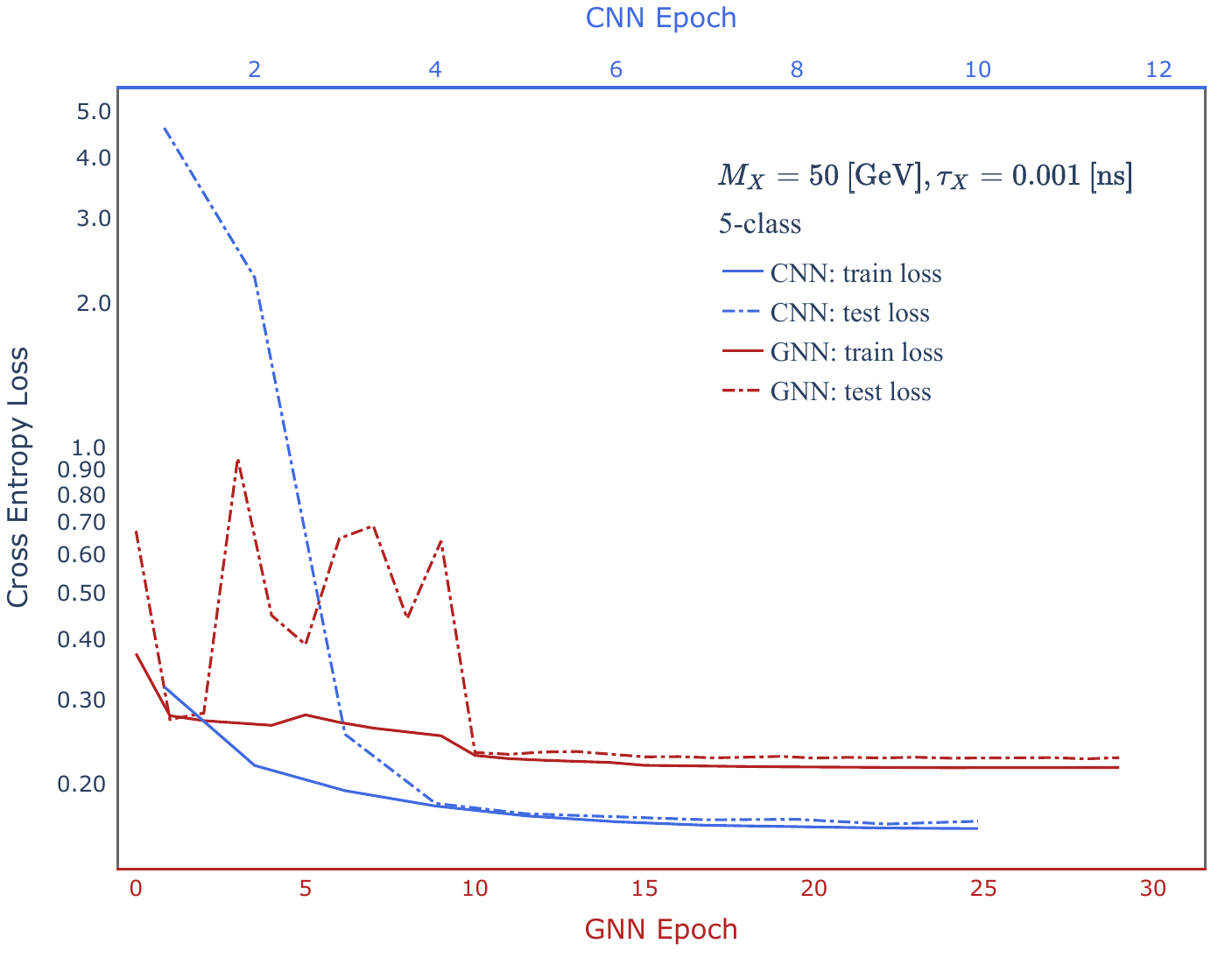}
  \includegraphics[width=0.30\linewidth]{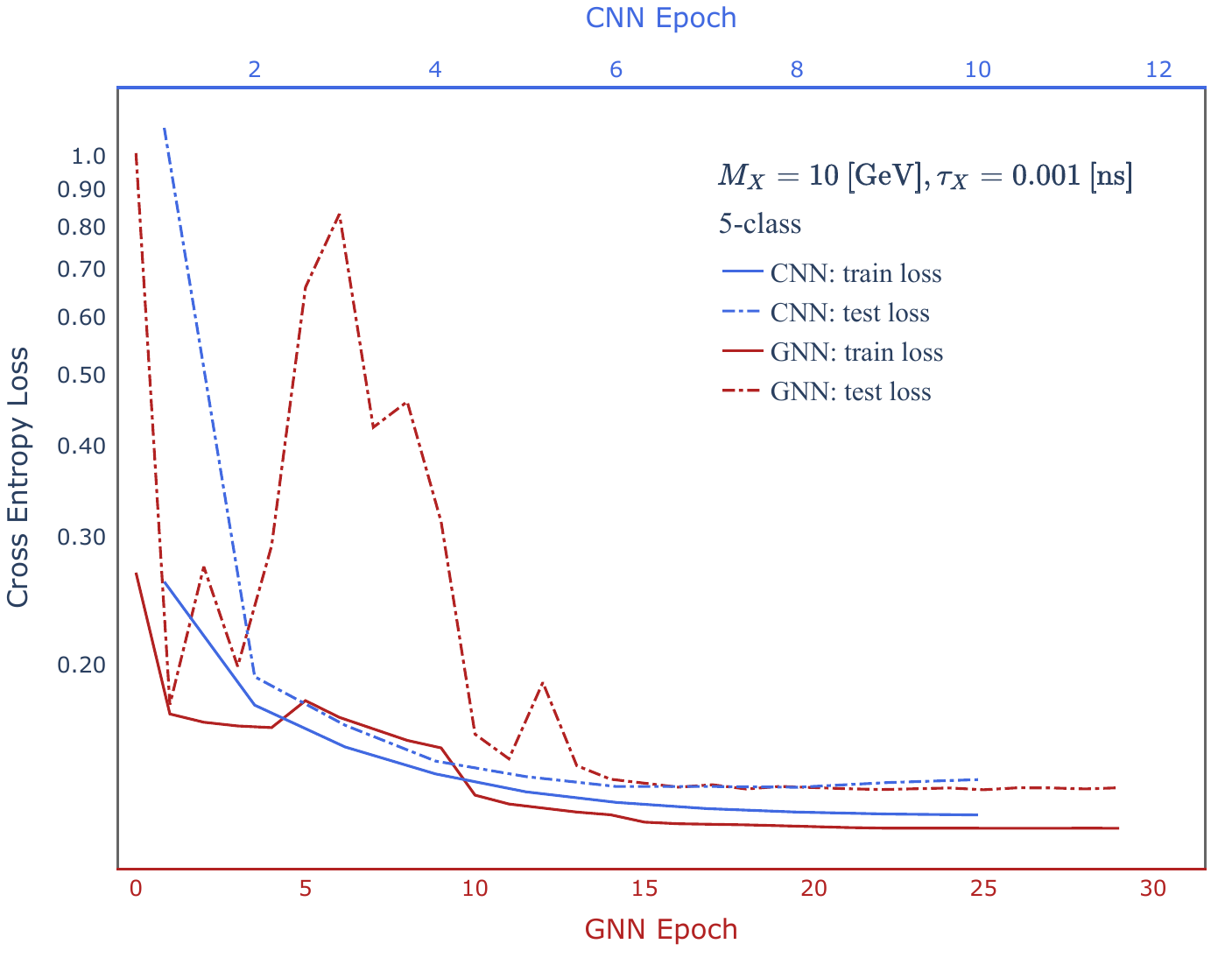}
  \includegraphics[width=0.30\linewidth]{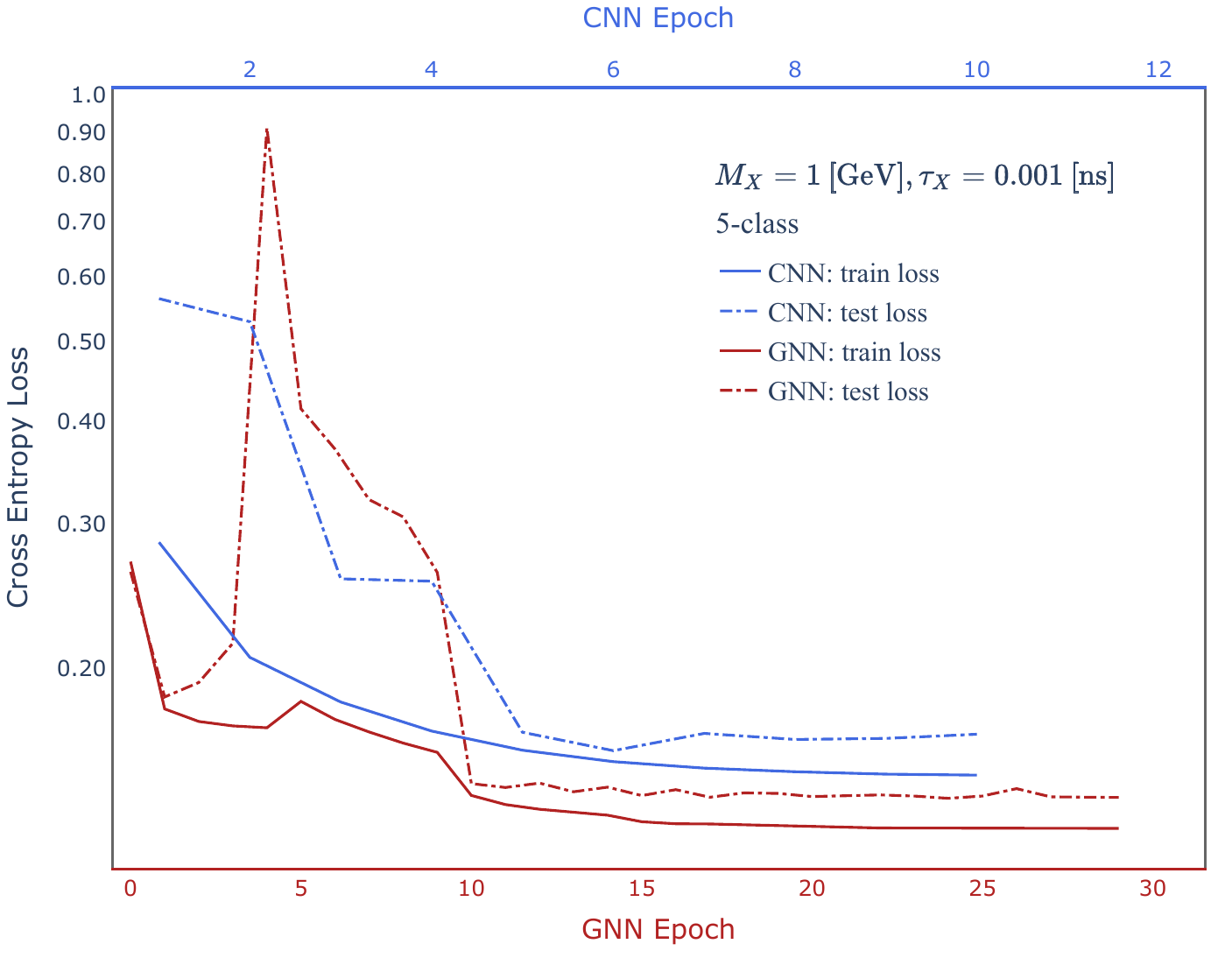}\\
  
  \includegraphics[width=0.30\linewidth]{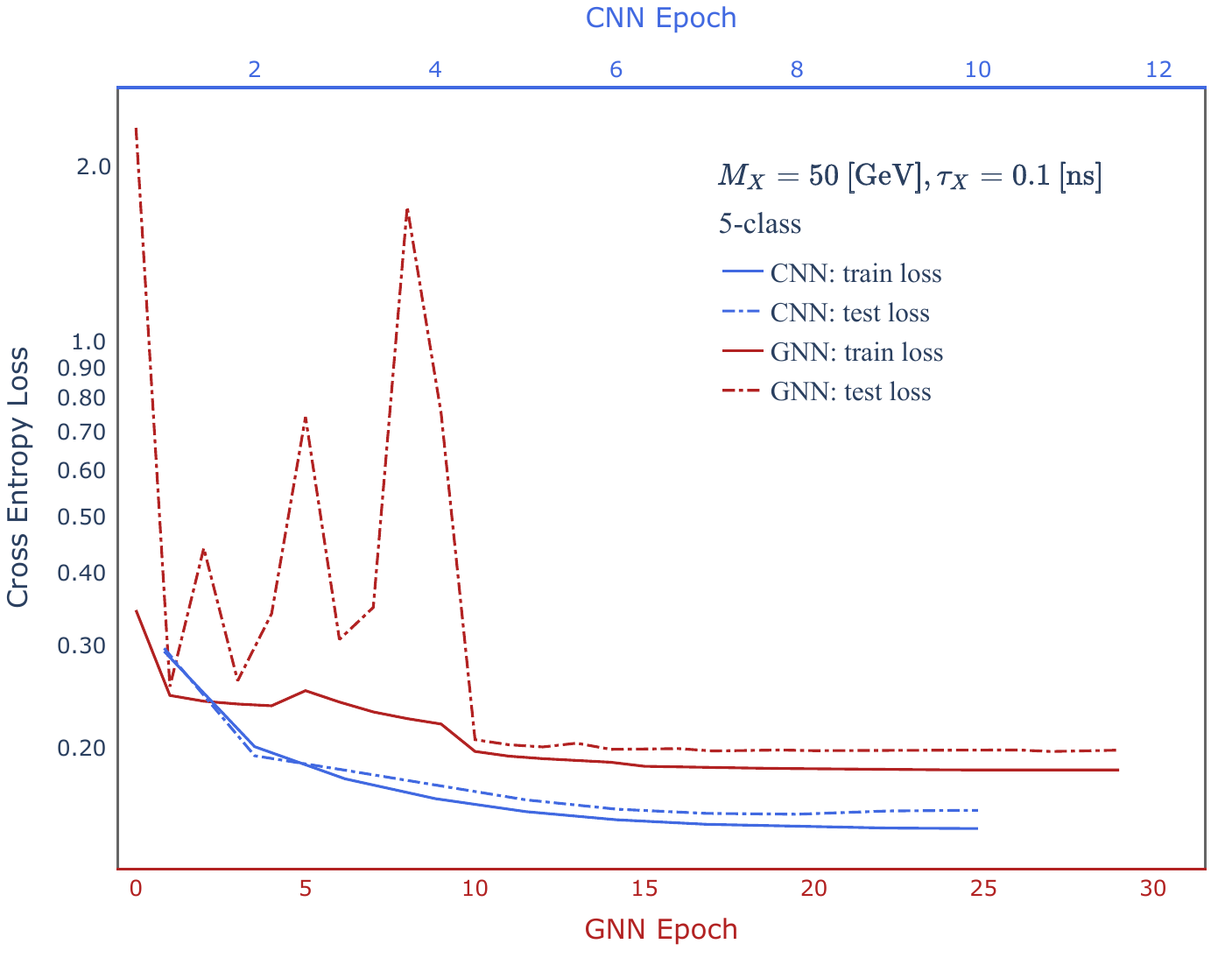}
  \includegraphics[width=0.30\linewidth]{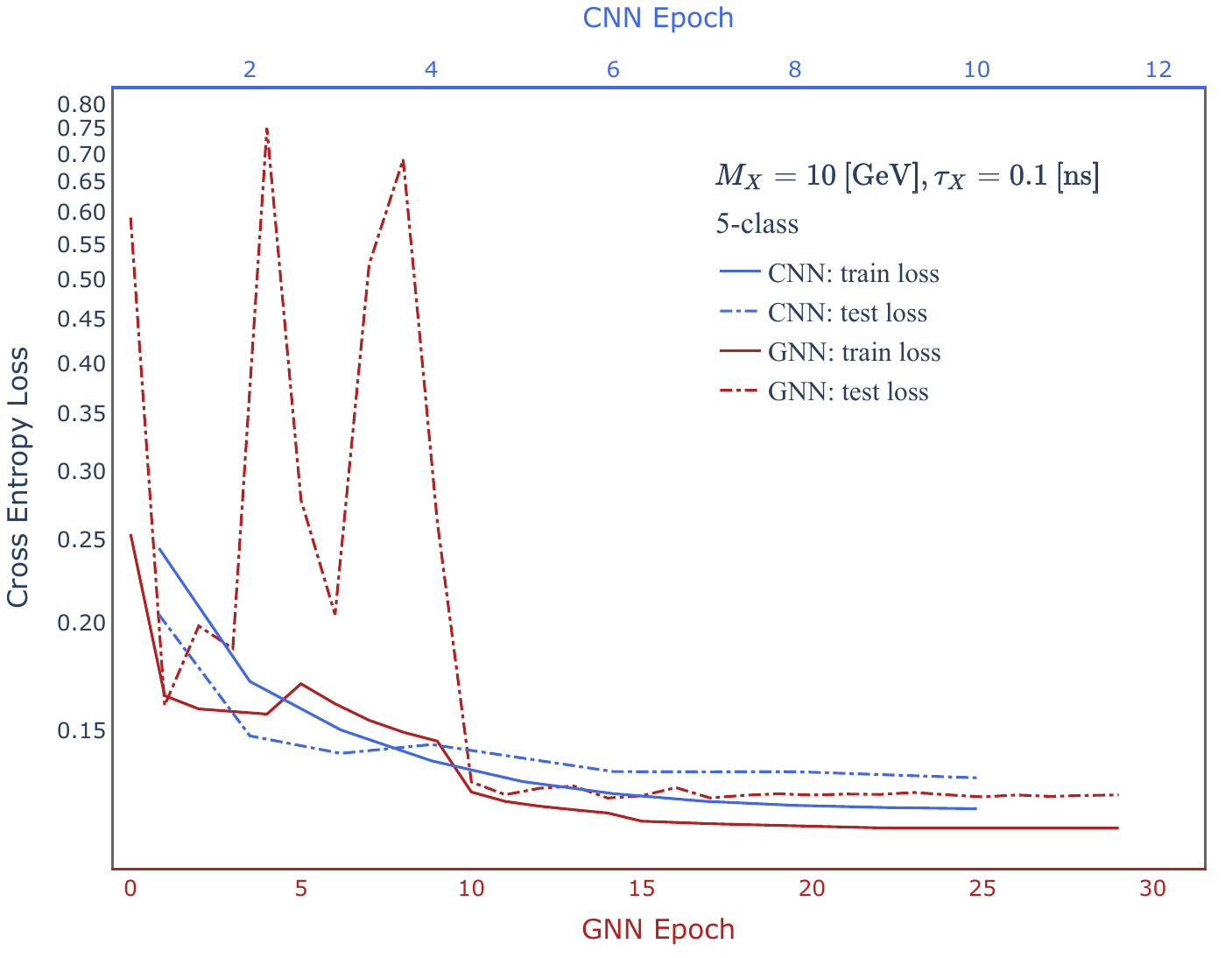}
  \includegraphics[width=0.30\linewidth]{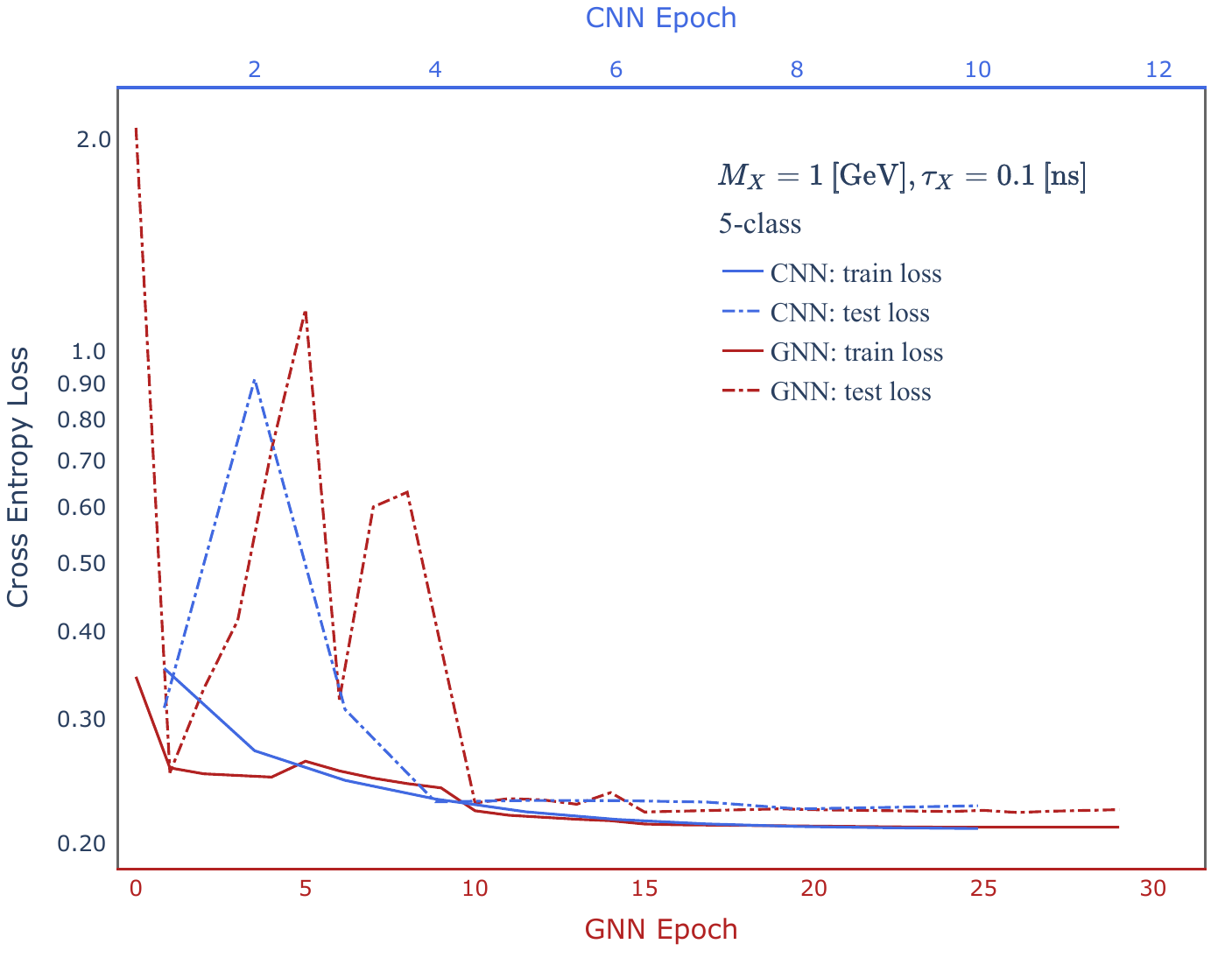}\\

  \includegraphics[width=0.30\linewidth]{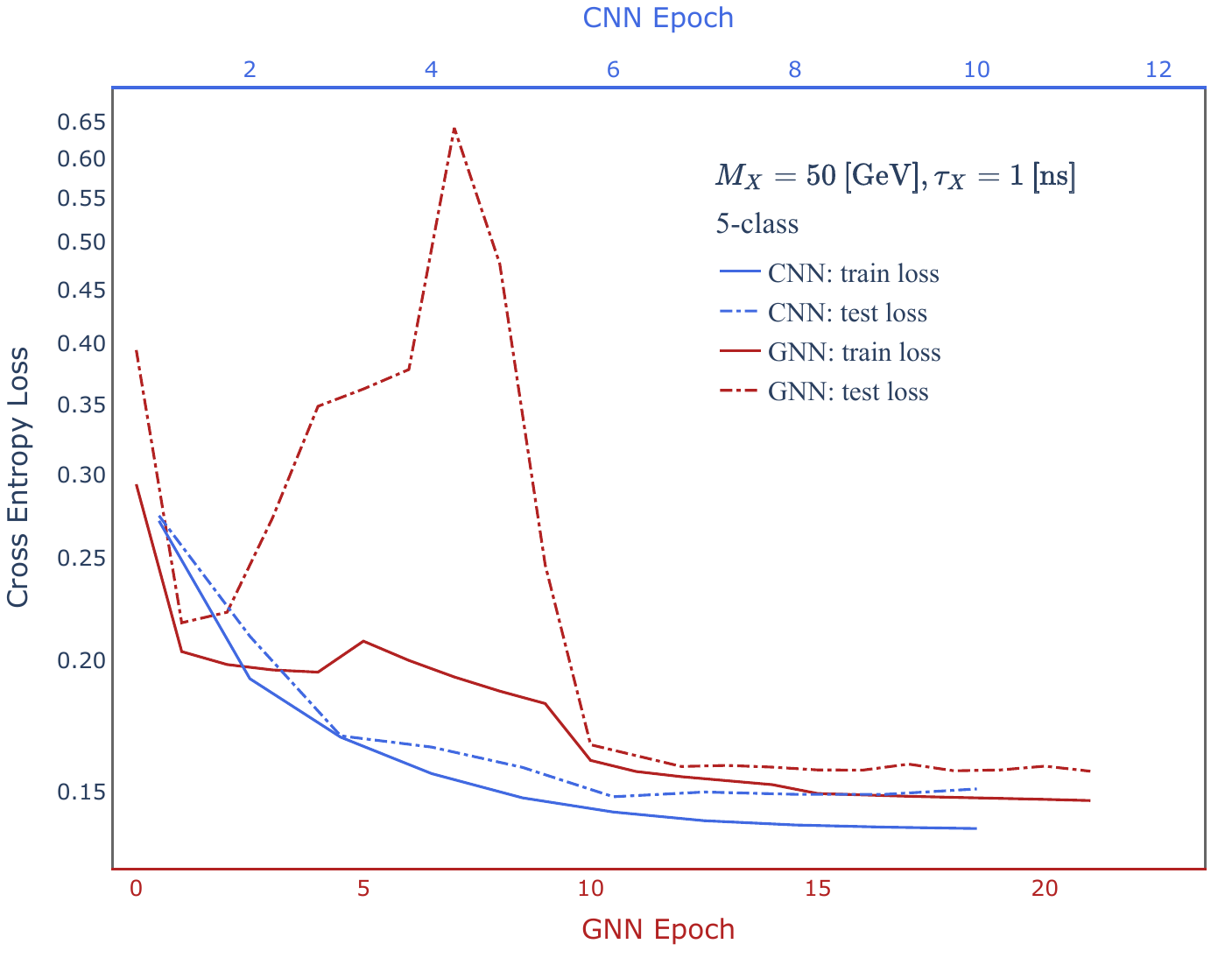}
  \includegraphics[width=0.30\linewidth]{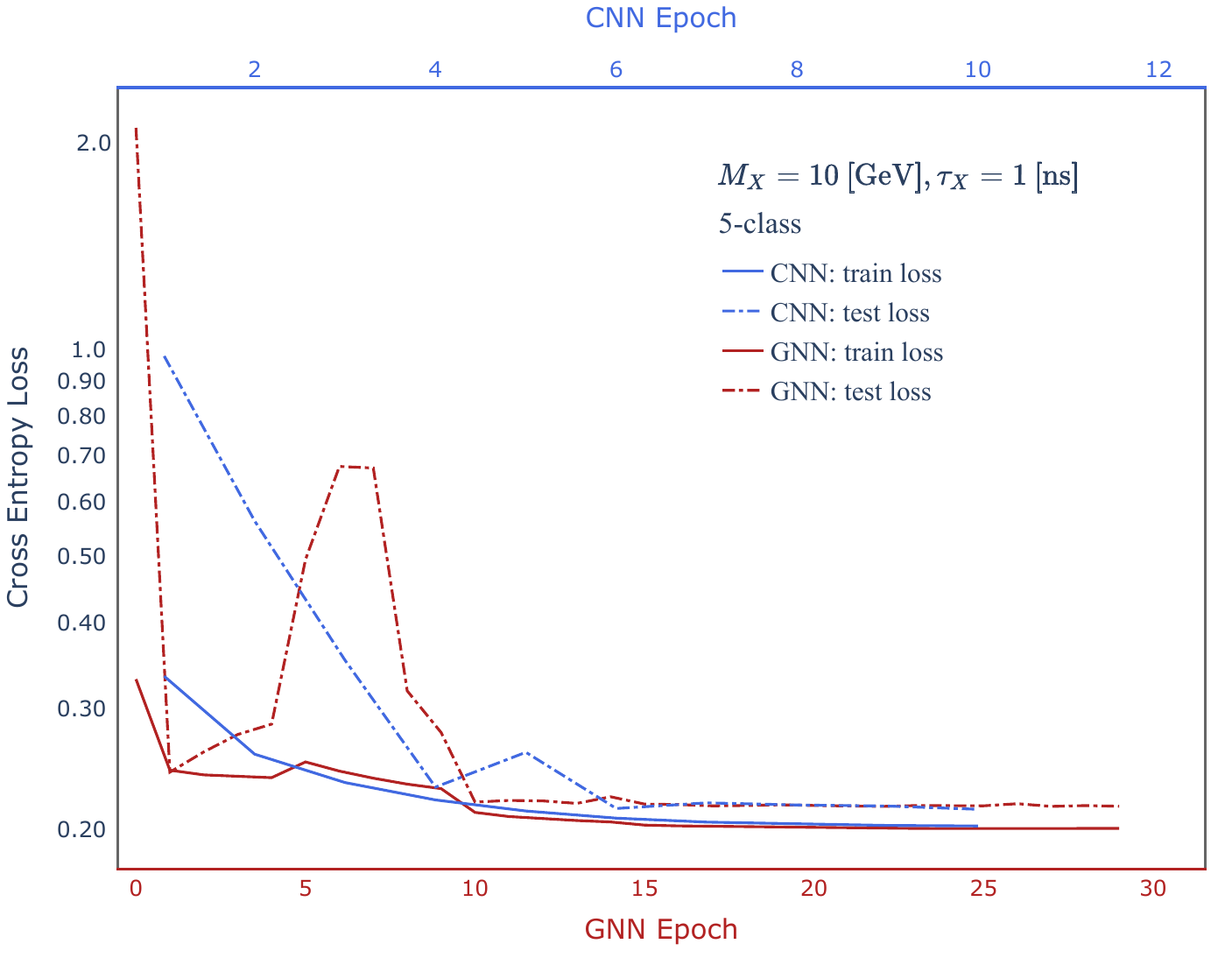}
  \includegraphics[width=0.30\linewidth]{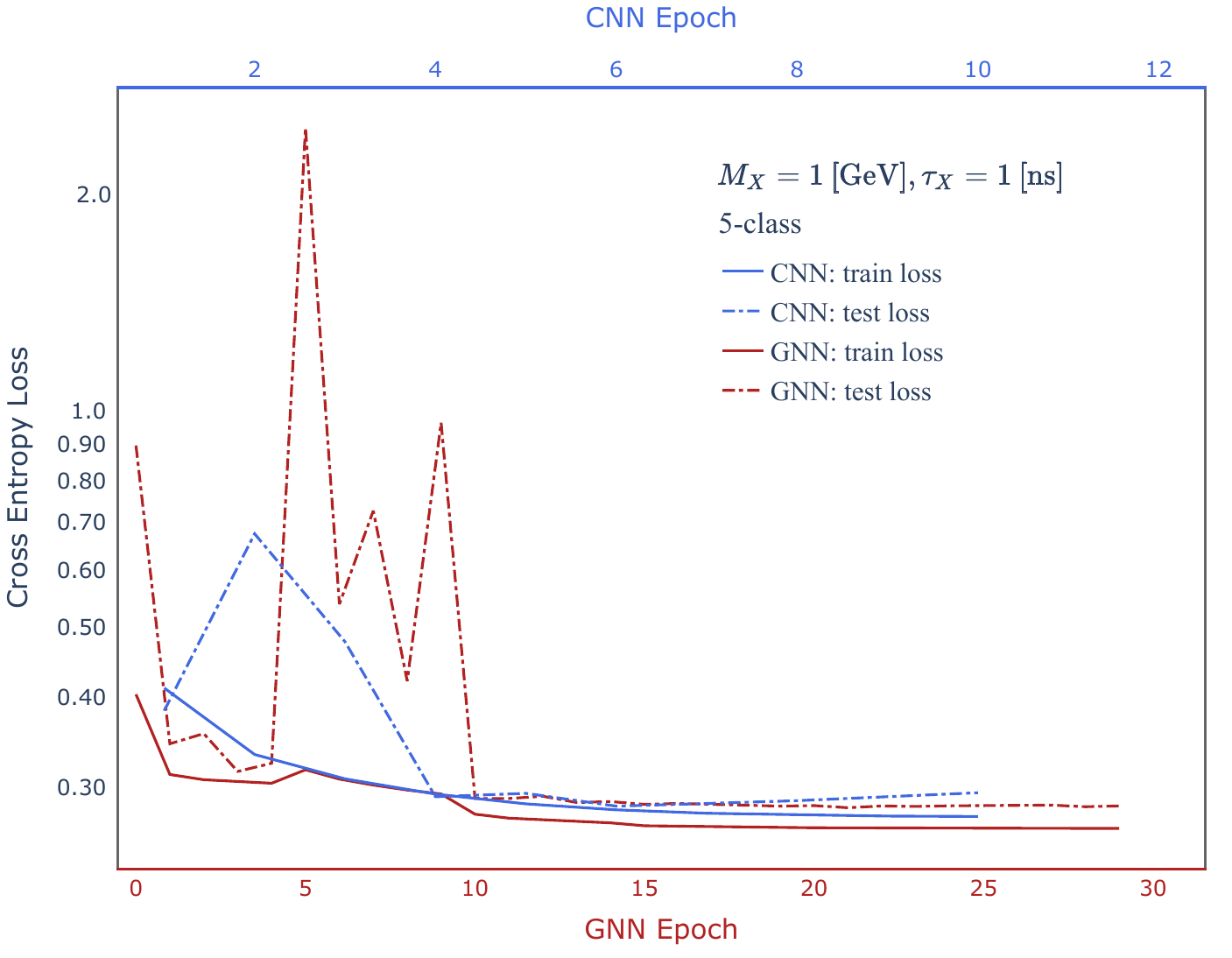}\\
  
  \includegraphics[width=0.30\linewidth]{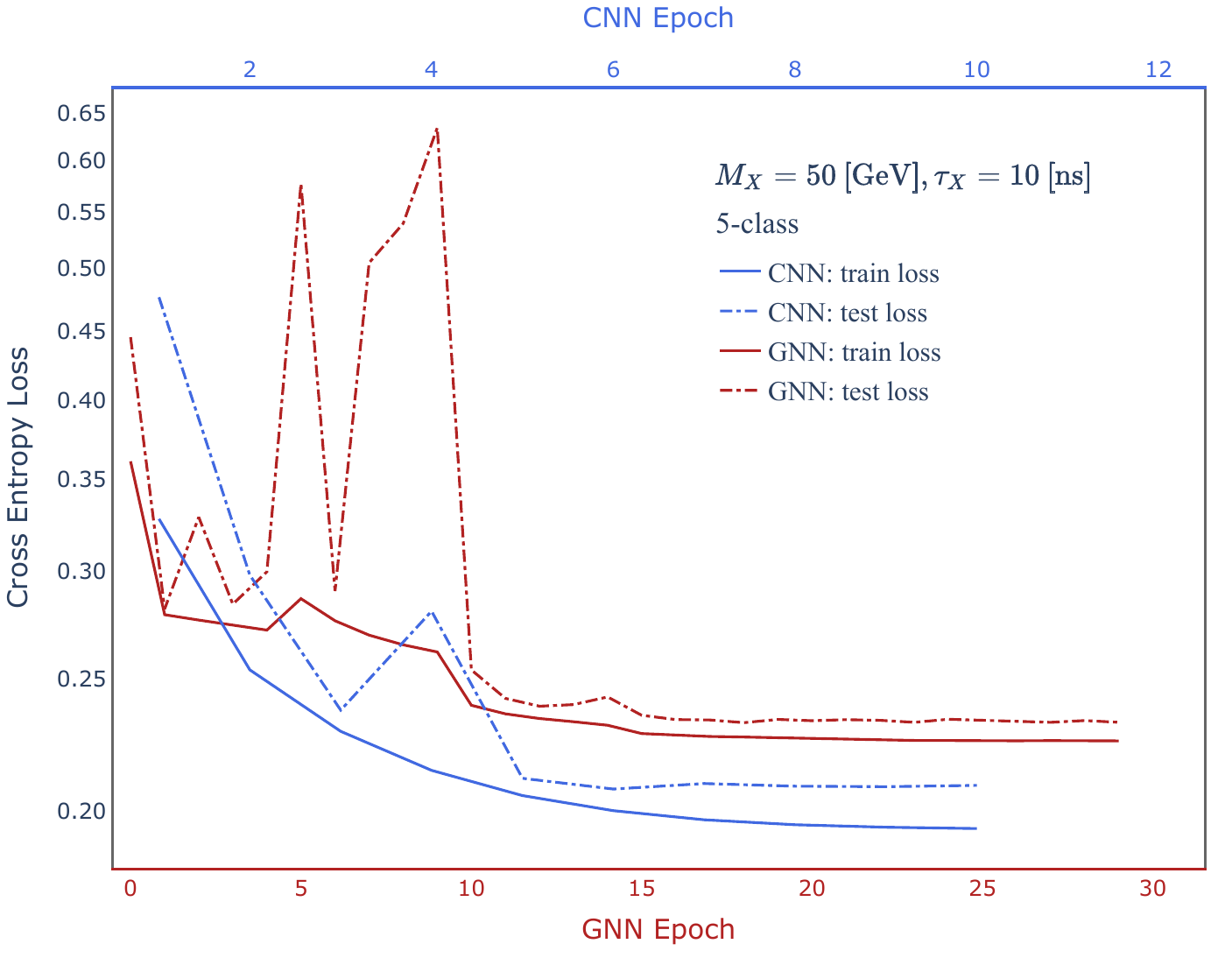}
  \includegraphics[width=0.30\linewidth]{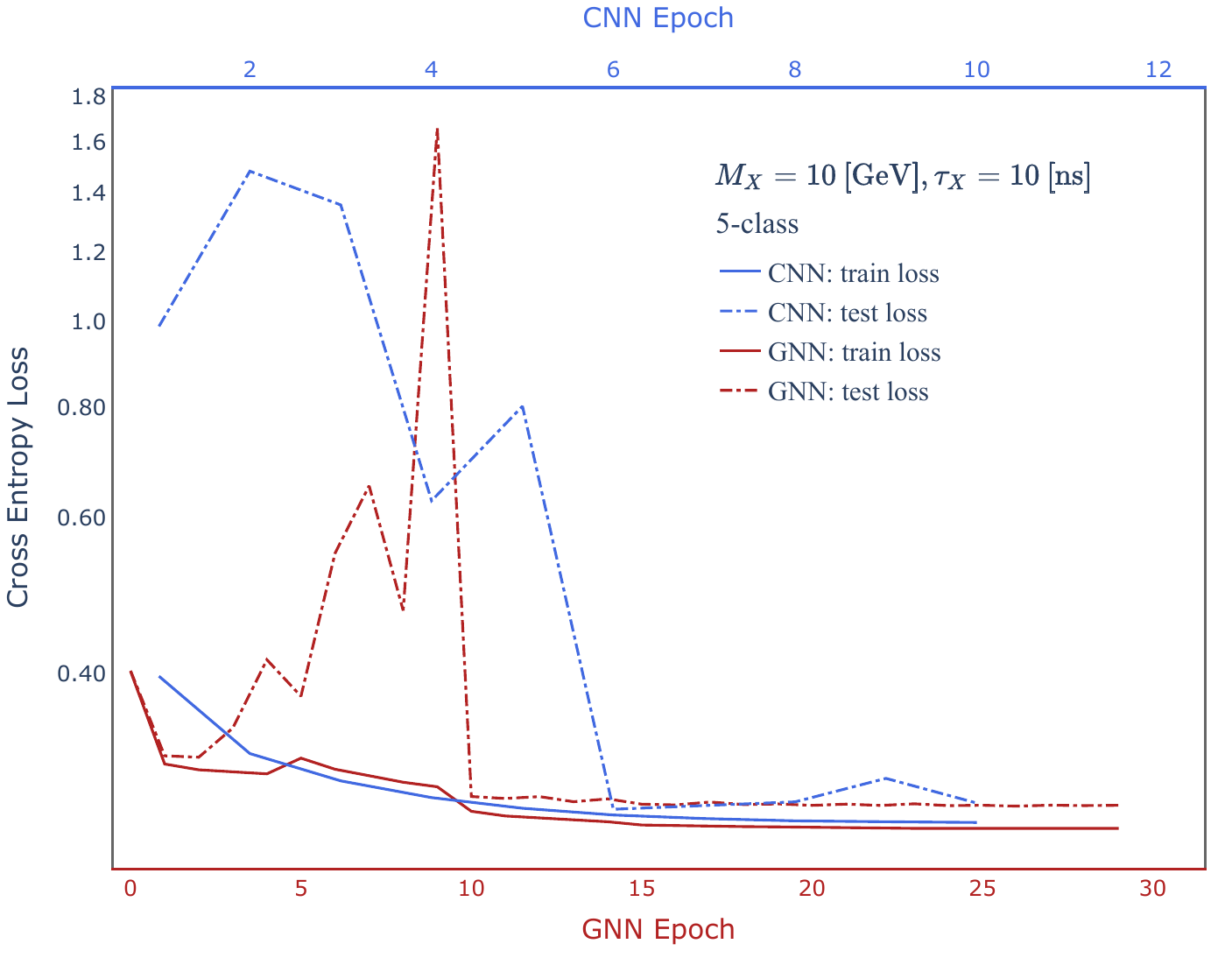}
  \includegraphics[width=0.30\linewidth]{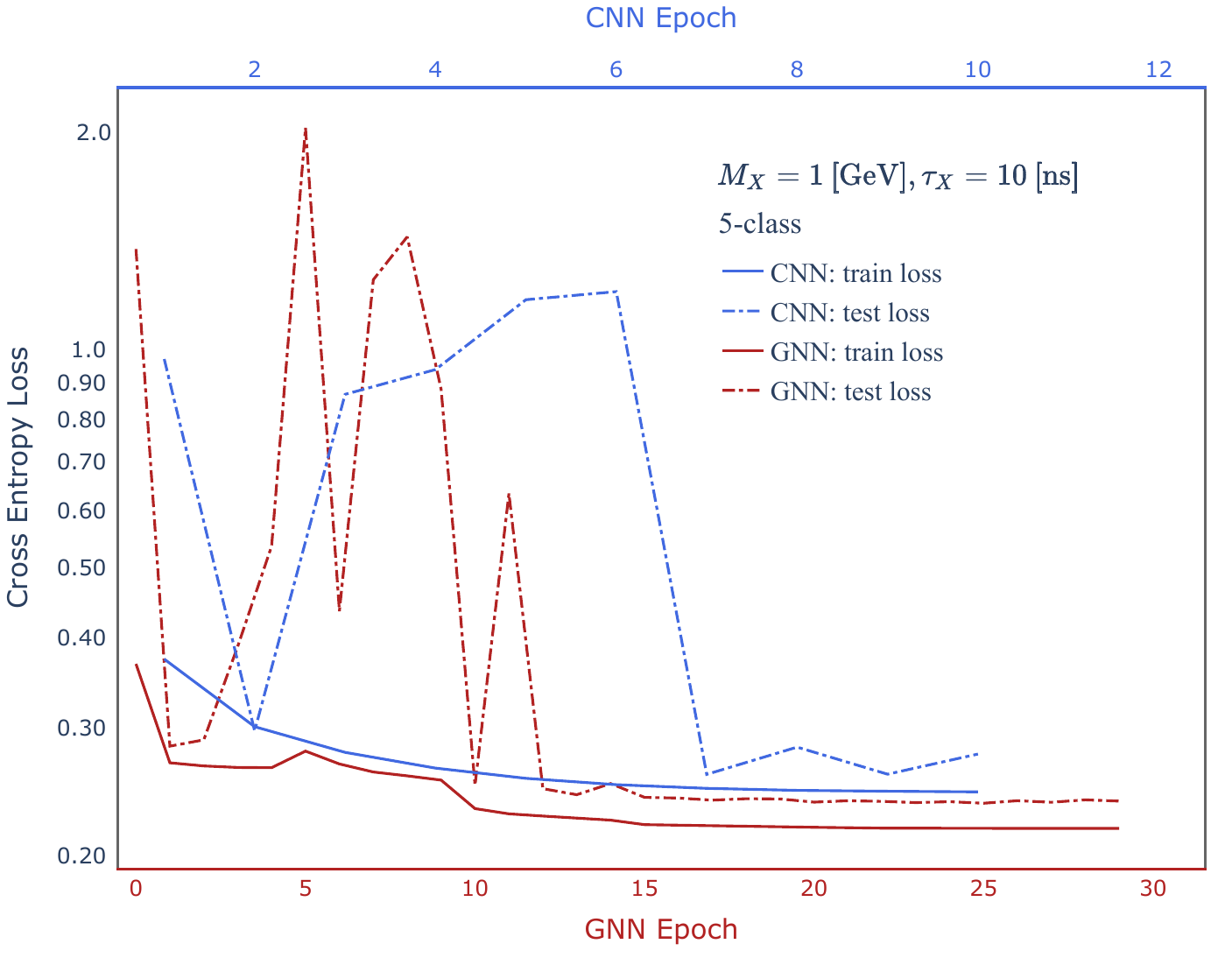}\\
  
  \includegraphics[width=0.30\linewidth]{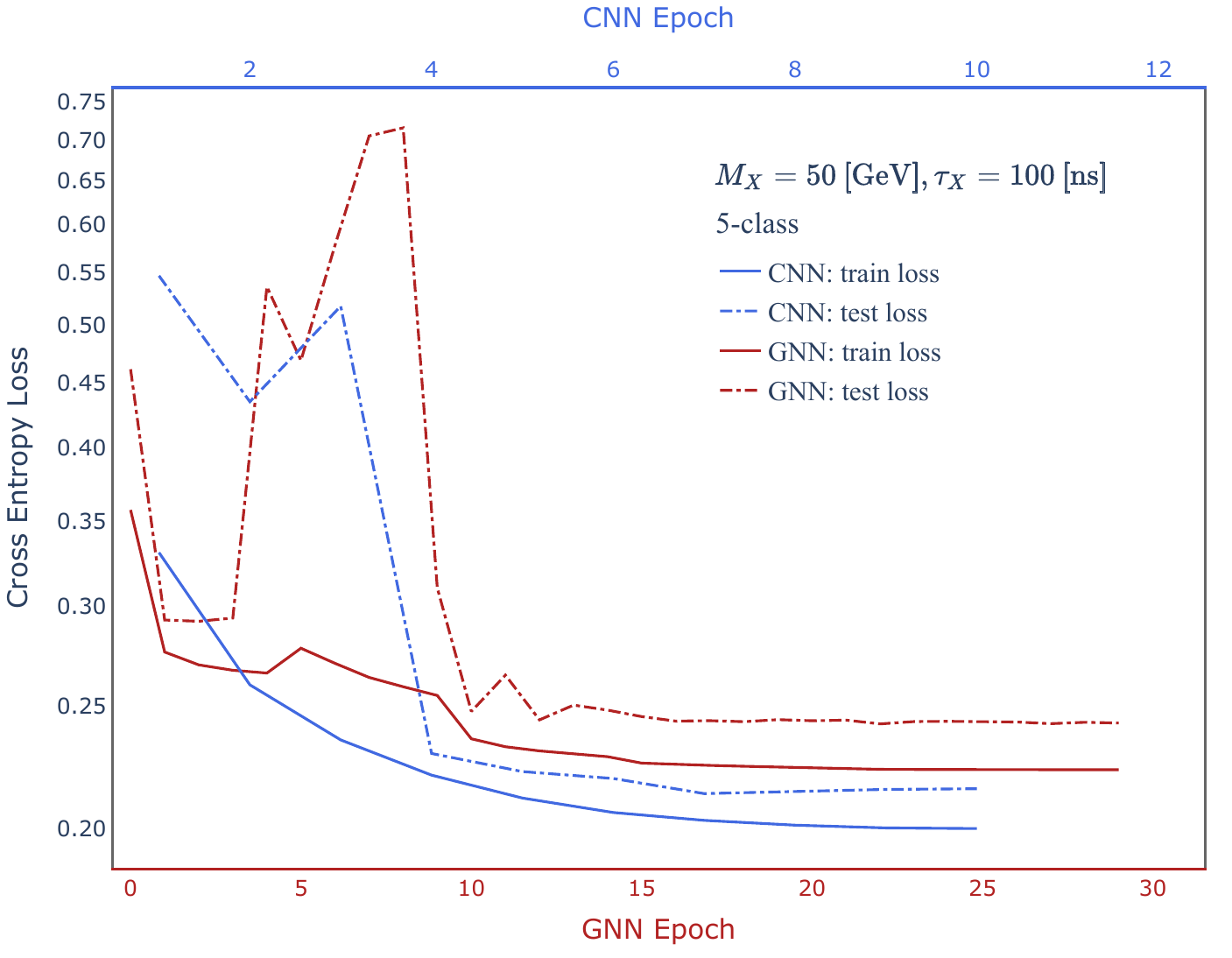}
  \includegraphics[width=0.30\linewidth]{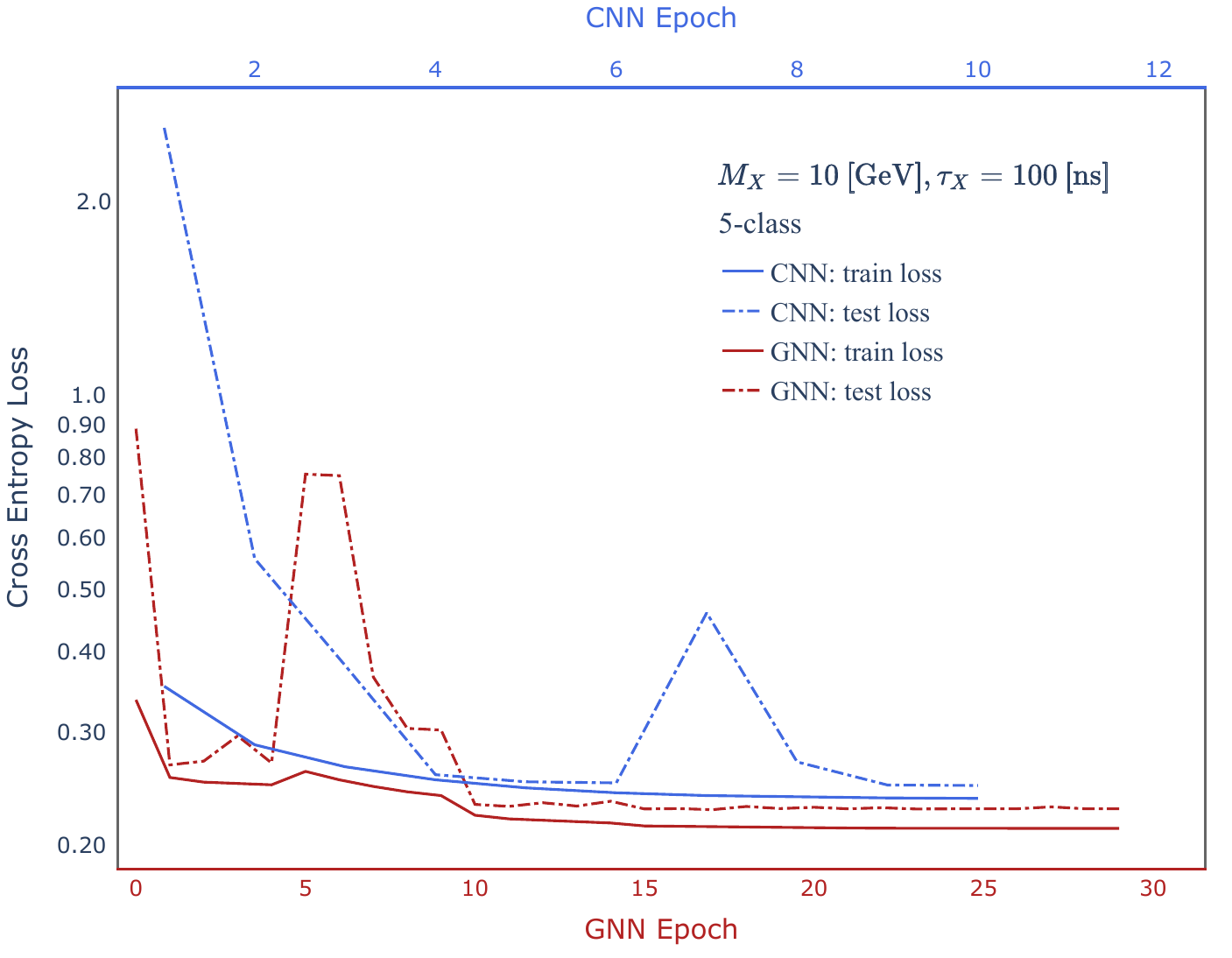}
  \includegraphics[width=0.30\linewidth]{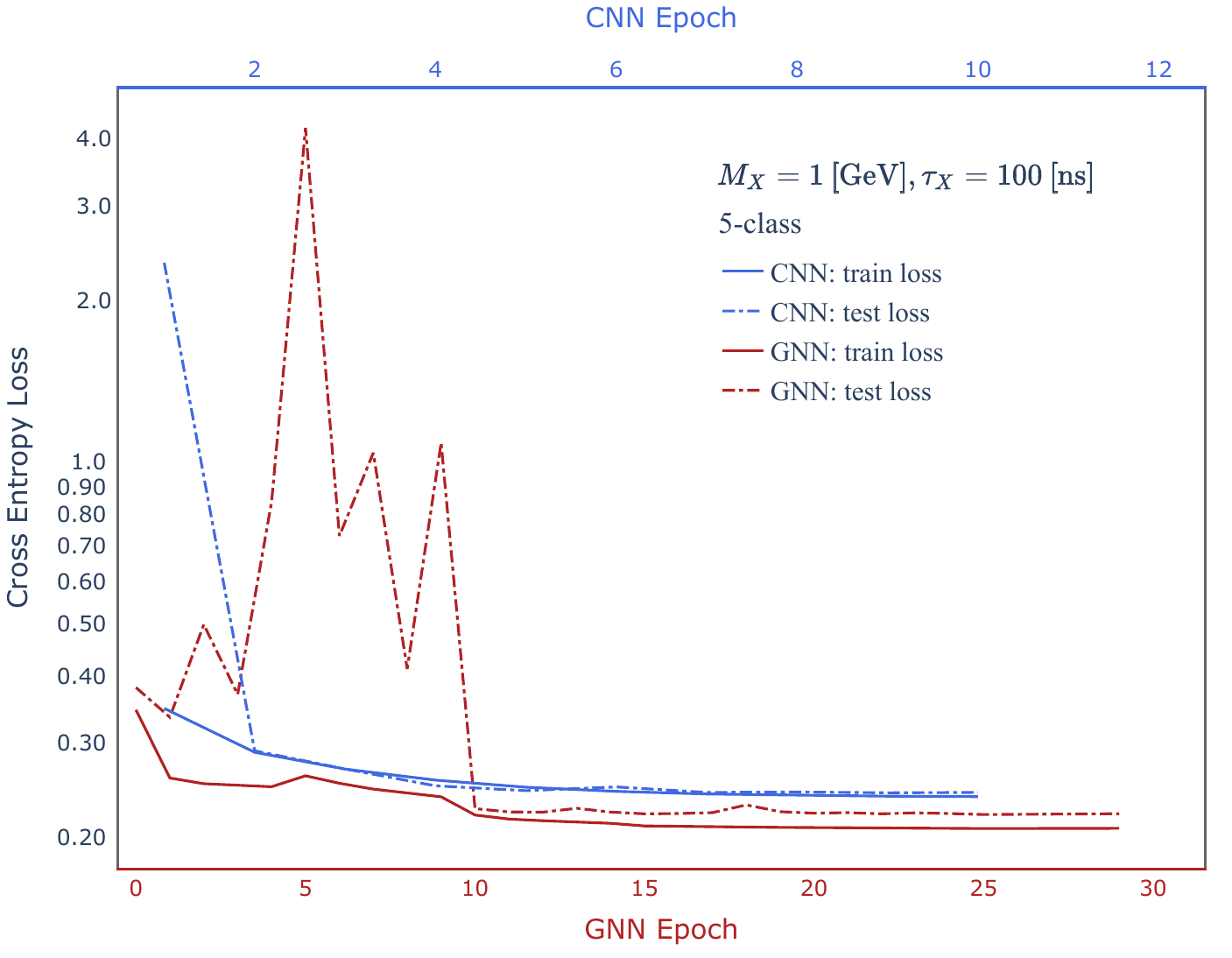}\\
  \caption{The loss function converges as training epochs increase when training the CNN and GNN. From top to bottom, the LLPs' lifetime in corresponding samples are 0.001, 0.1, 1, 10, 100 nanoseconds; From left to right, the LLPs' mass in corresponding samples are 50, 10, 1 GeV.}
  \label{fig:loss_all}
\end{figure*}

Figure~\ref{fig:loss_all} illustrates values of the loss function  variation with the training epochs as training the CNN and GNN. A clear convergence can be seen during training various simulation samples corresponding to LLPs of different  masses and lifetimes.

\label{sec_loss}

 \section{Sensitivity study on $\mathcal{B}(H \rightarrow XX)$}\label{app:sensitivity}
\begin{figure*}[!htbp]
 \includegraphics[width=0.30\linewidth ]{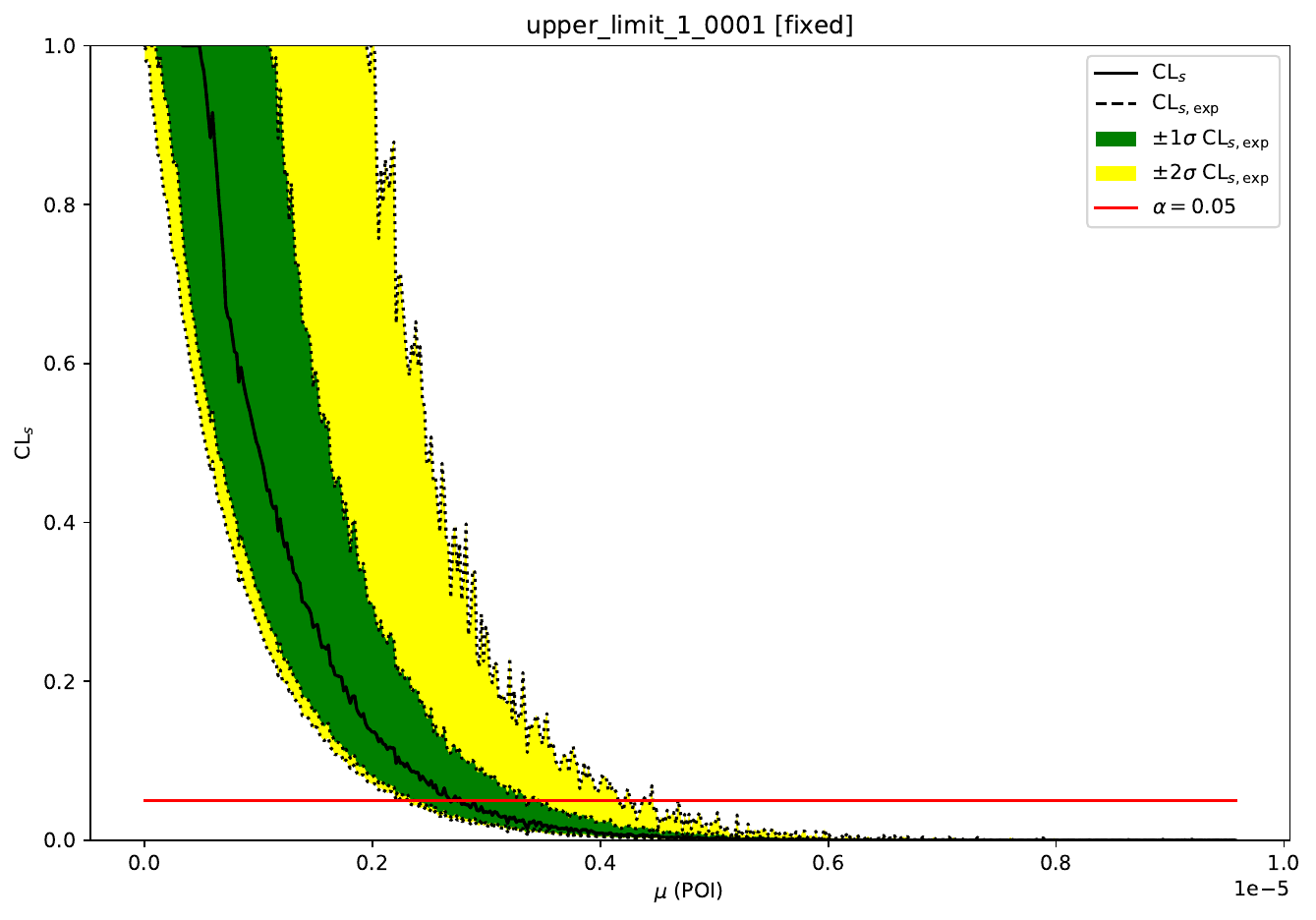}
 \includegraphics[width=0.30\linewidth ]{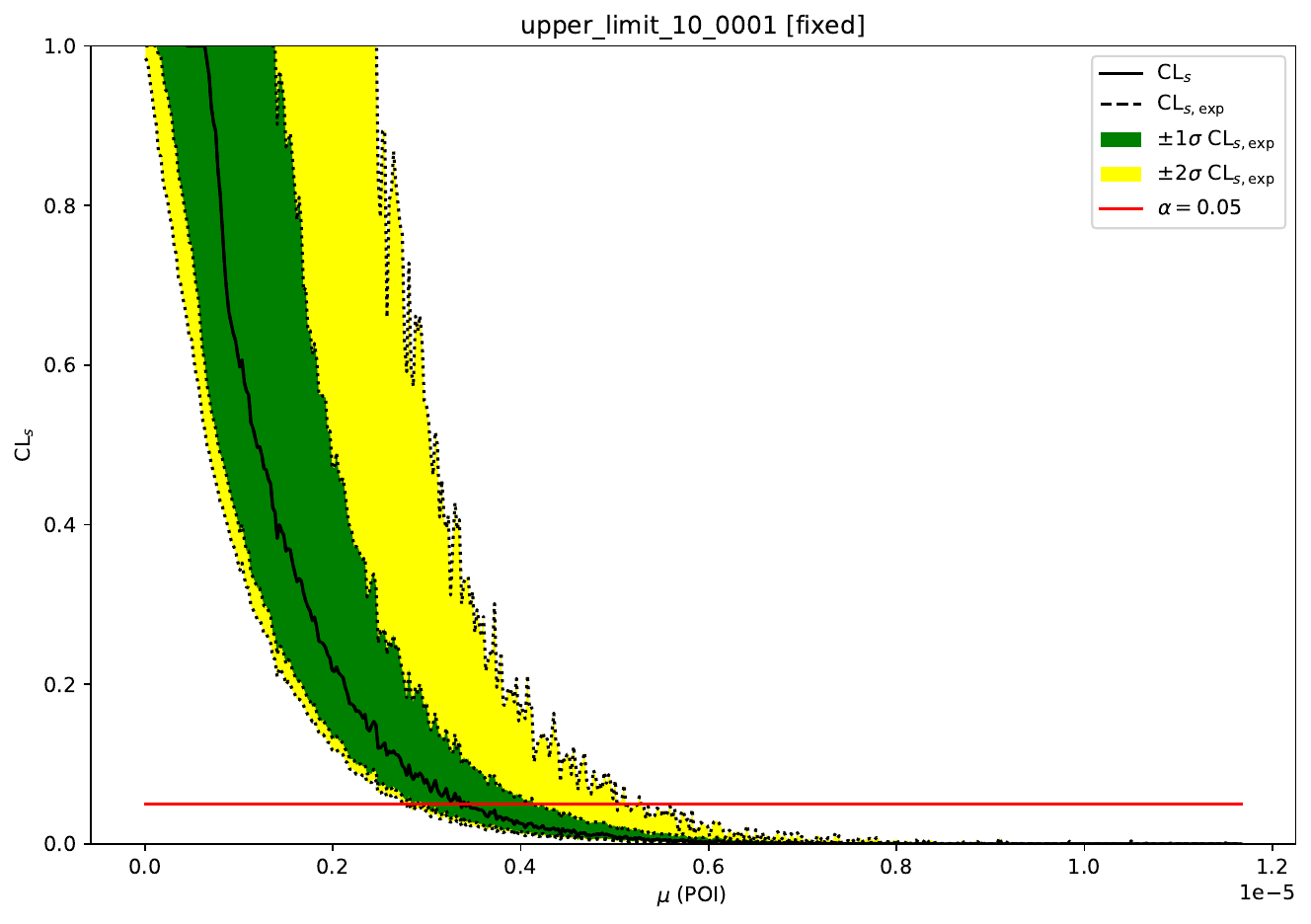}
 \includegraphics[width=0.30\linewidth ]{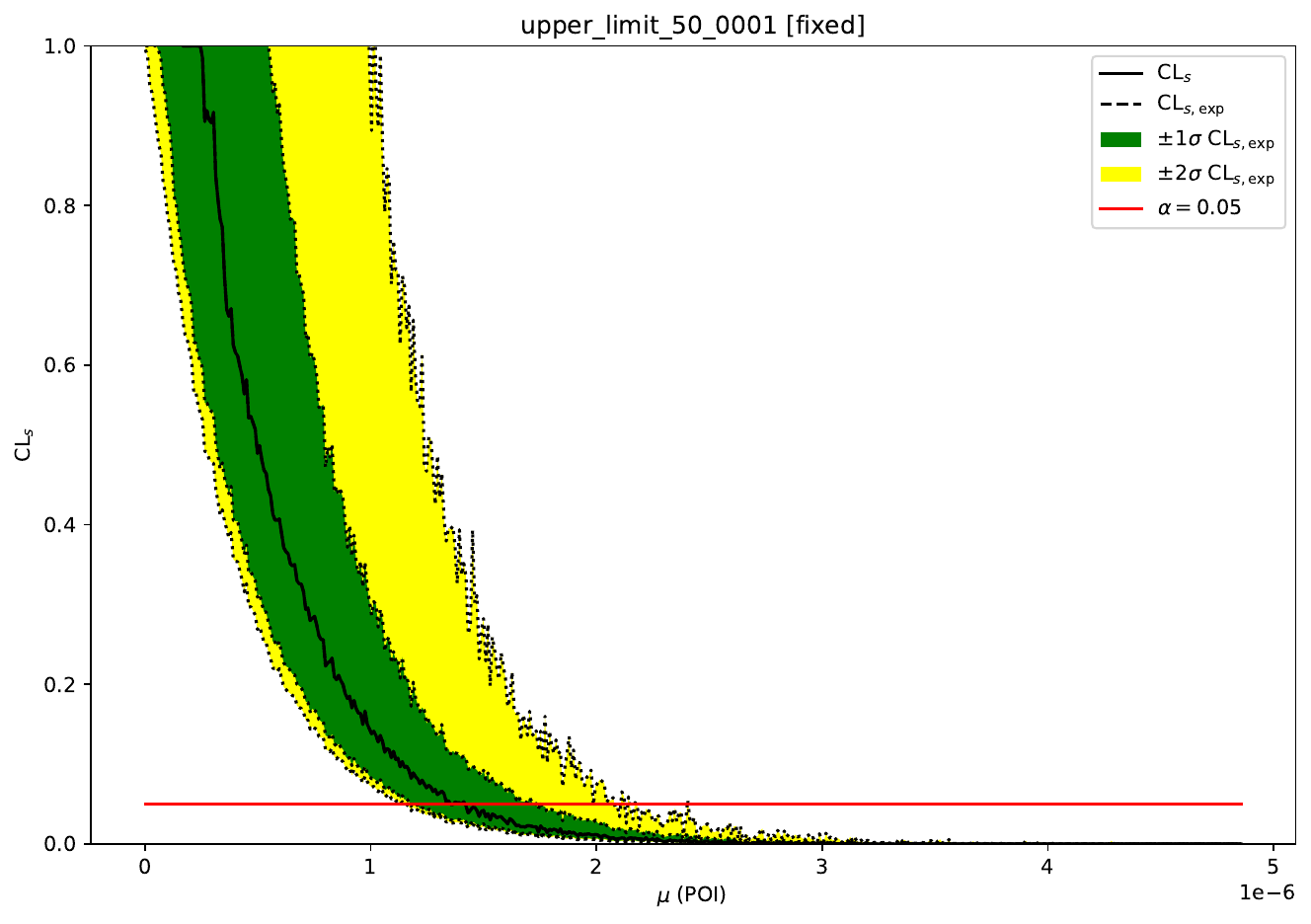}
 \\
 \includegraphics[width=0.30\linewidth ]{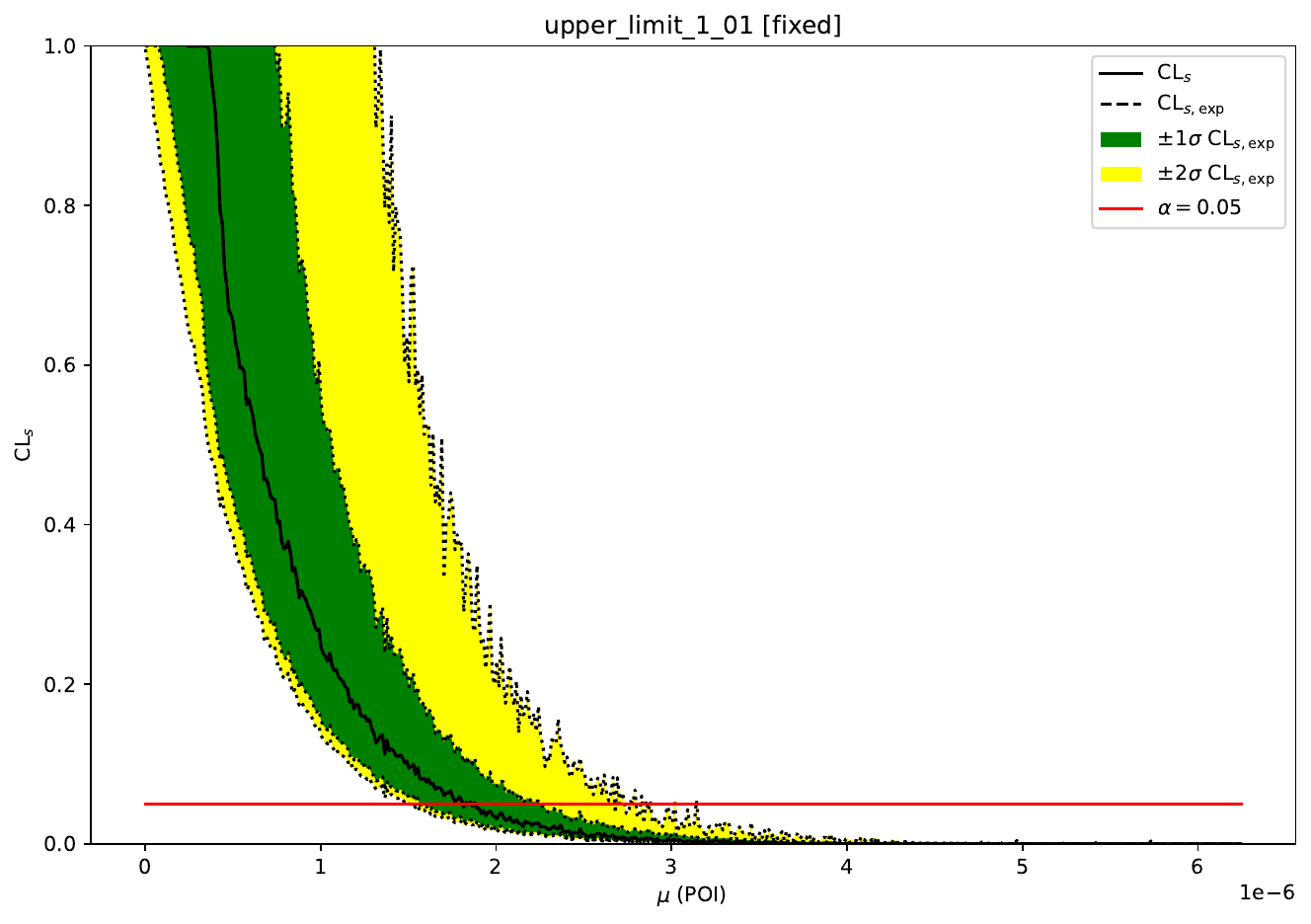}
 \includegraphics[width=0.30\linewidth ]{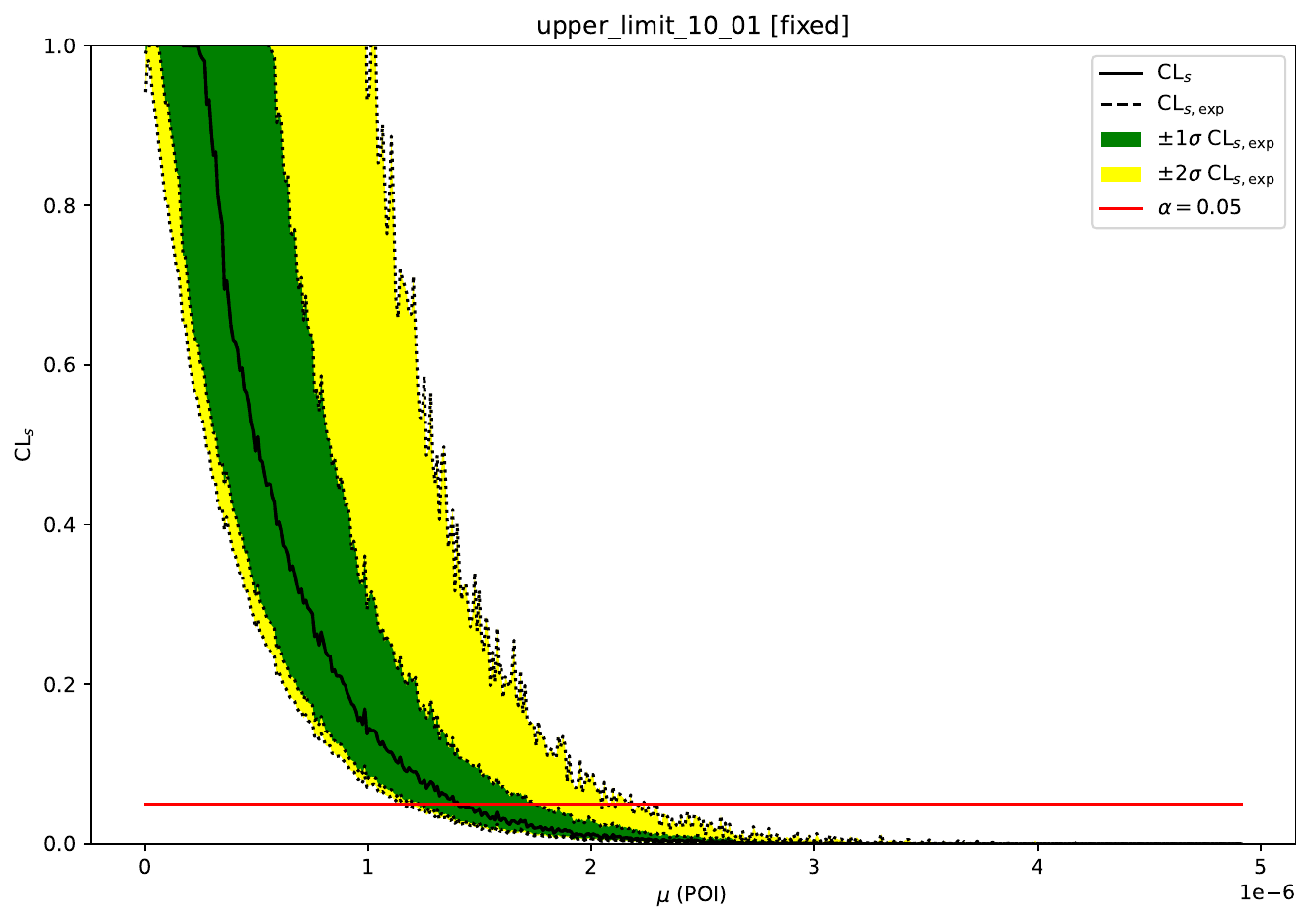}
 \includegraphics[width=0.30\linewidth ]{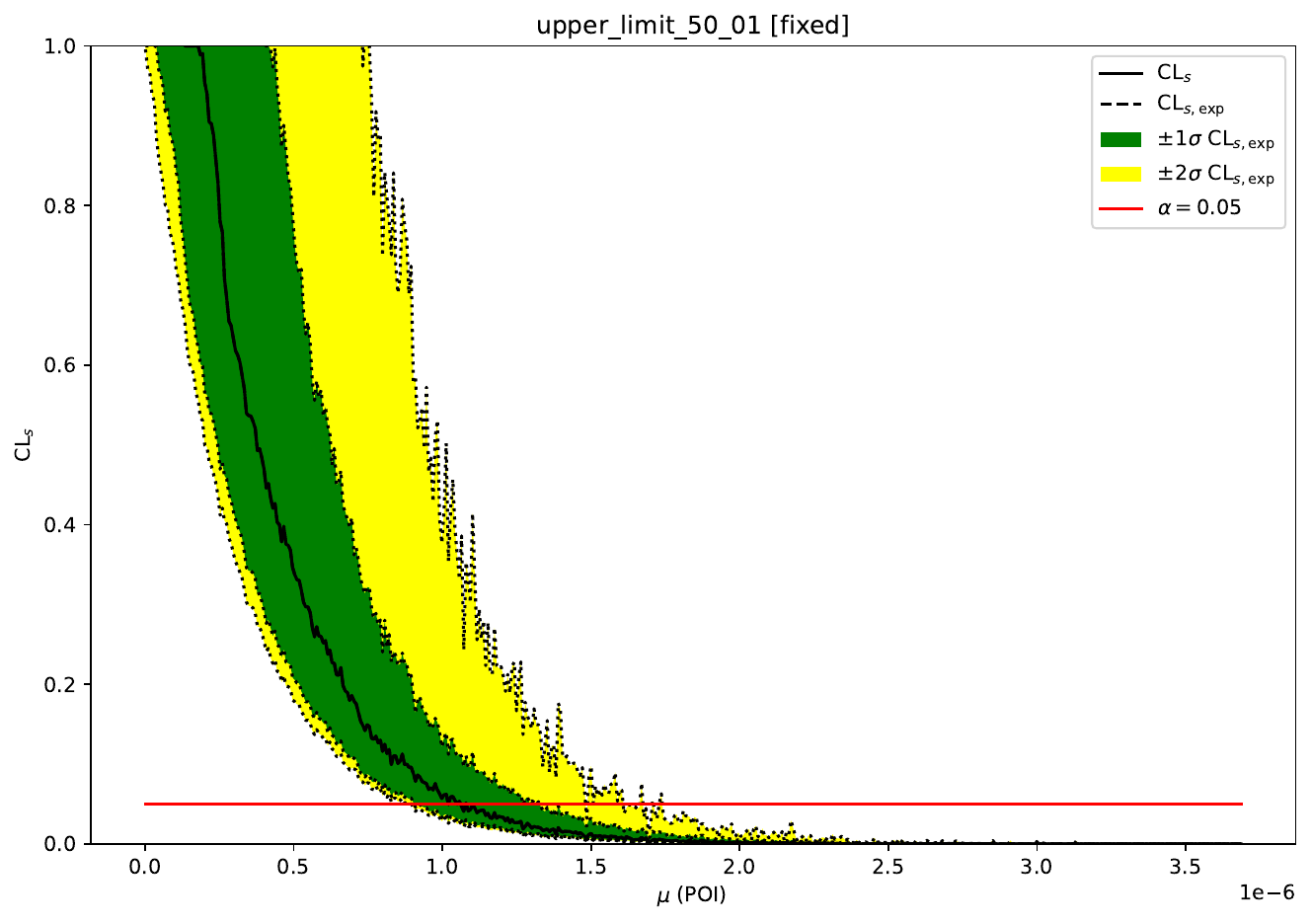}
 \\
  \includegraphics[width=0.30\linewidth ]{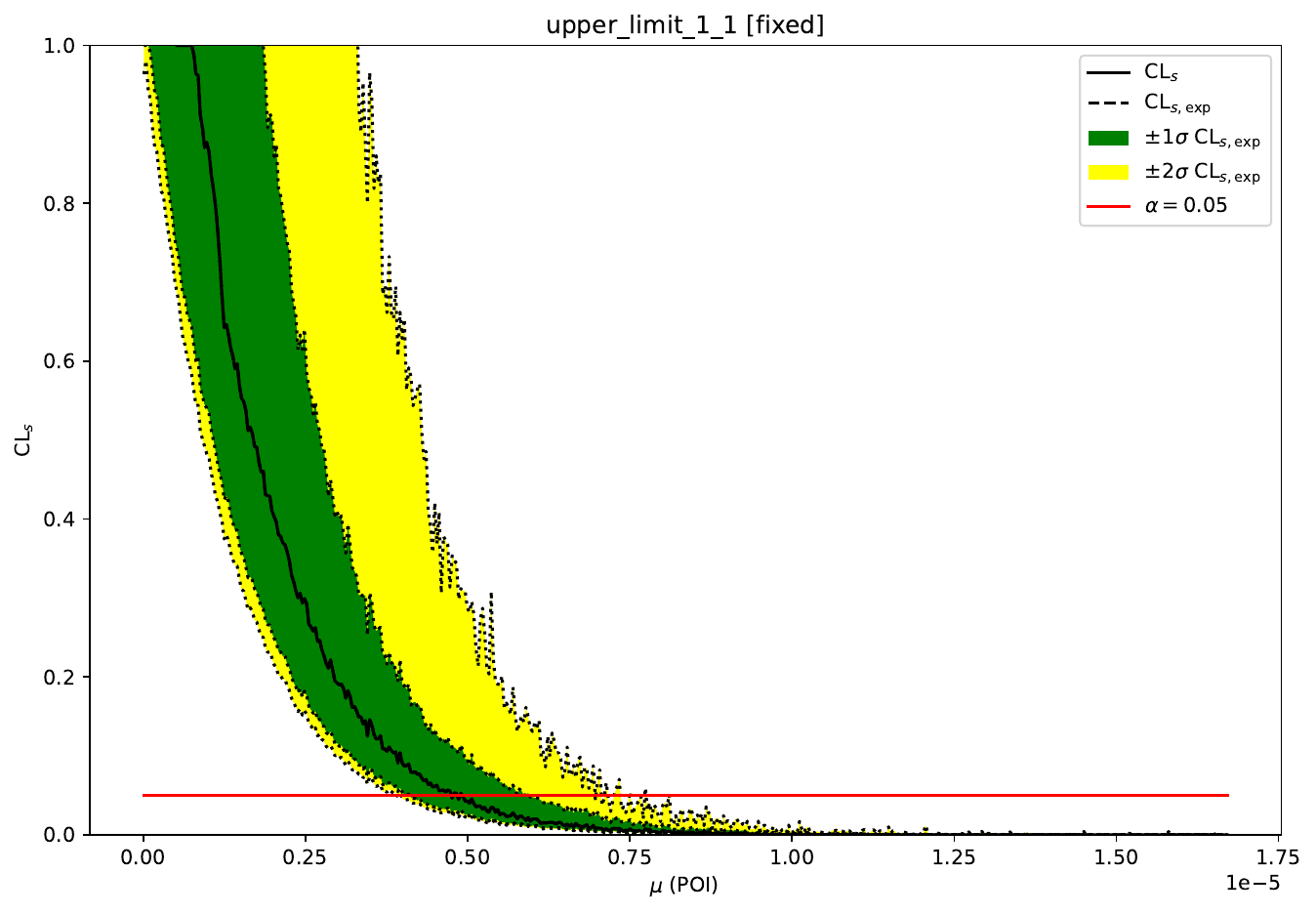}
 \includegraphics[width=0.30\linewidth ]{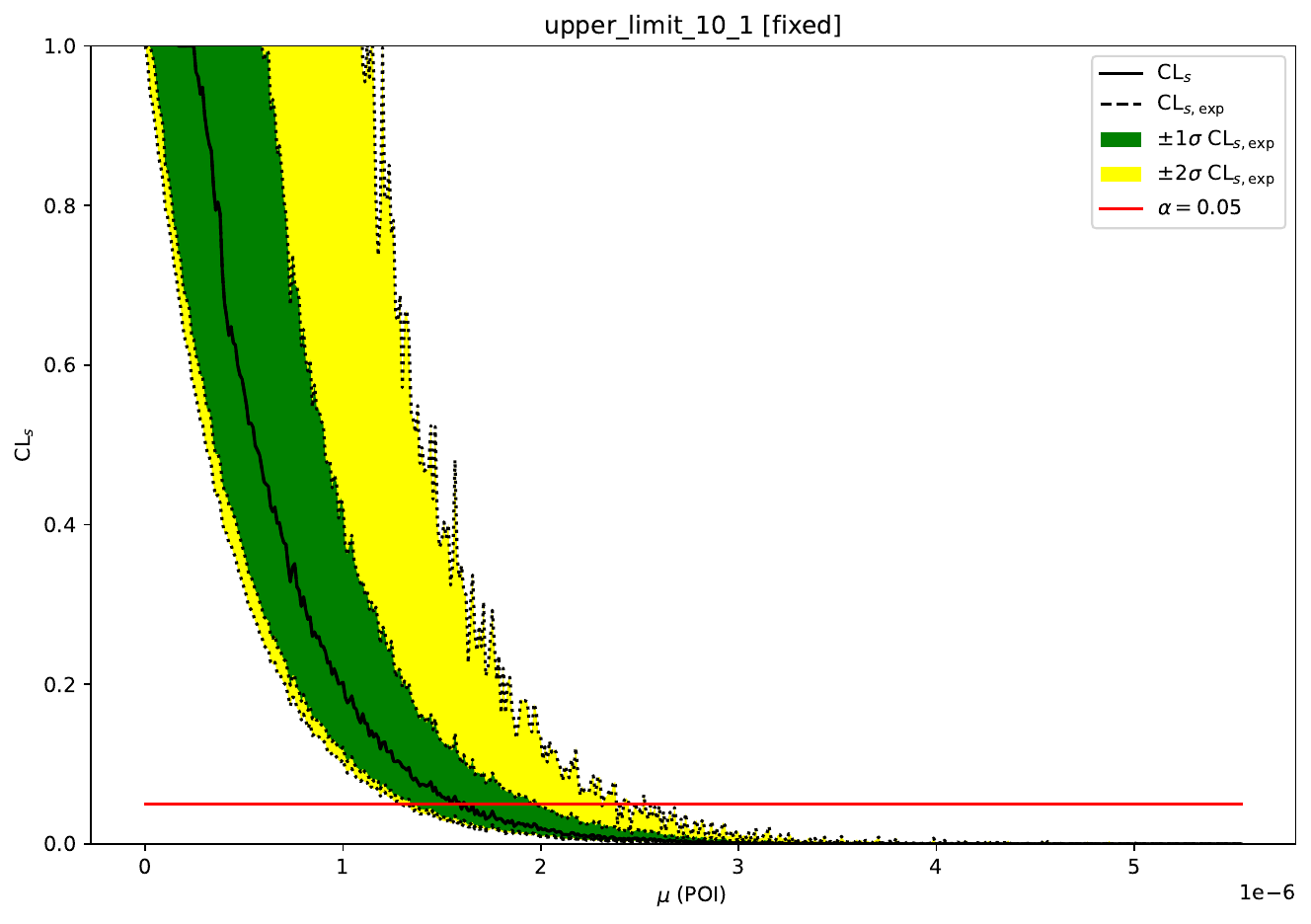}
 \includegraphics[width=0.30\linewidth ]{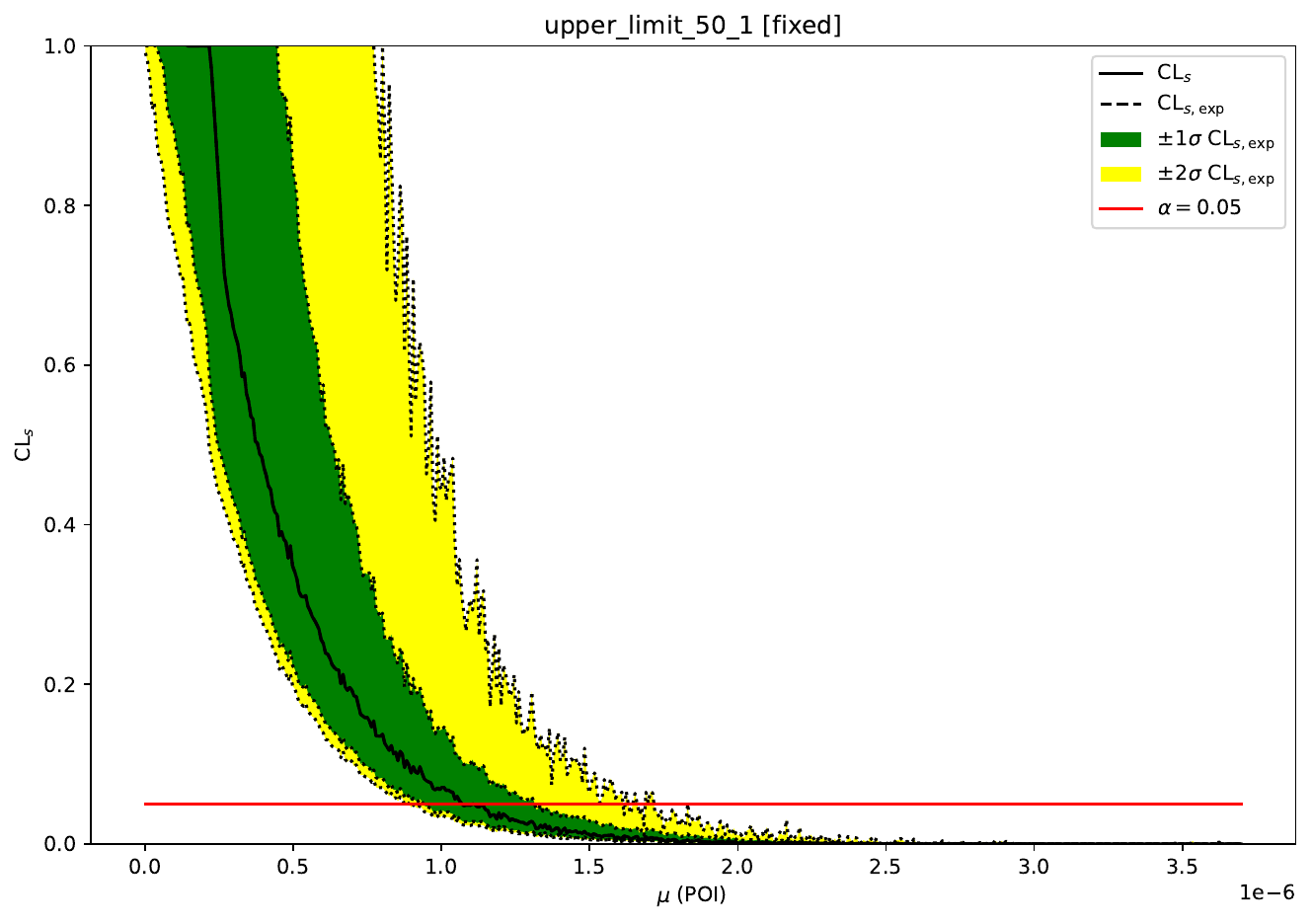}
 \\
  \includegraphics[width=0.30\linewidth ]{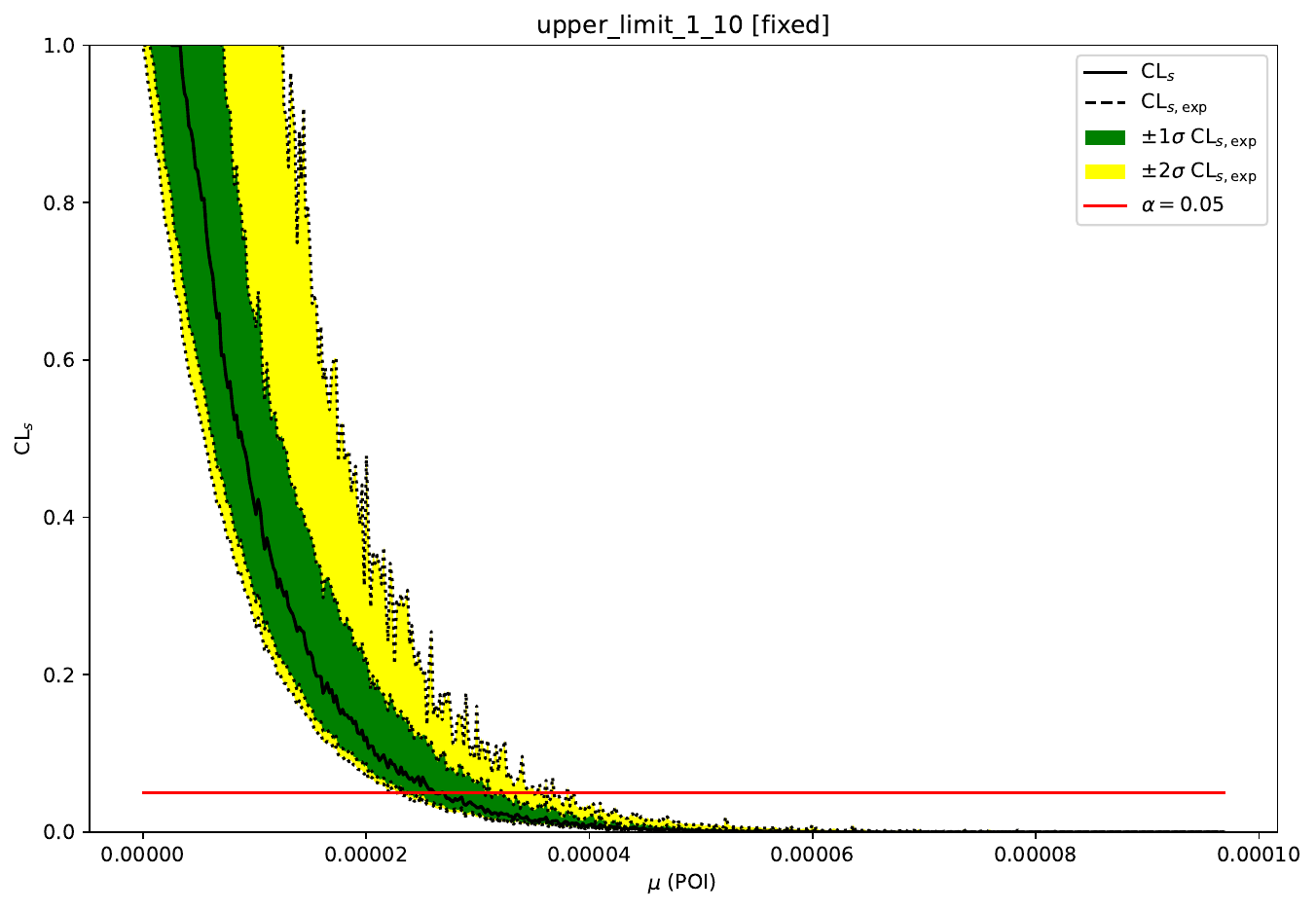}
 \includegraphics[width=0.30\linewidth ]{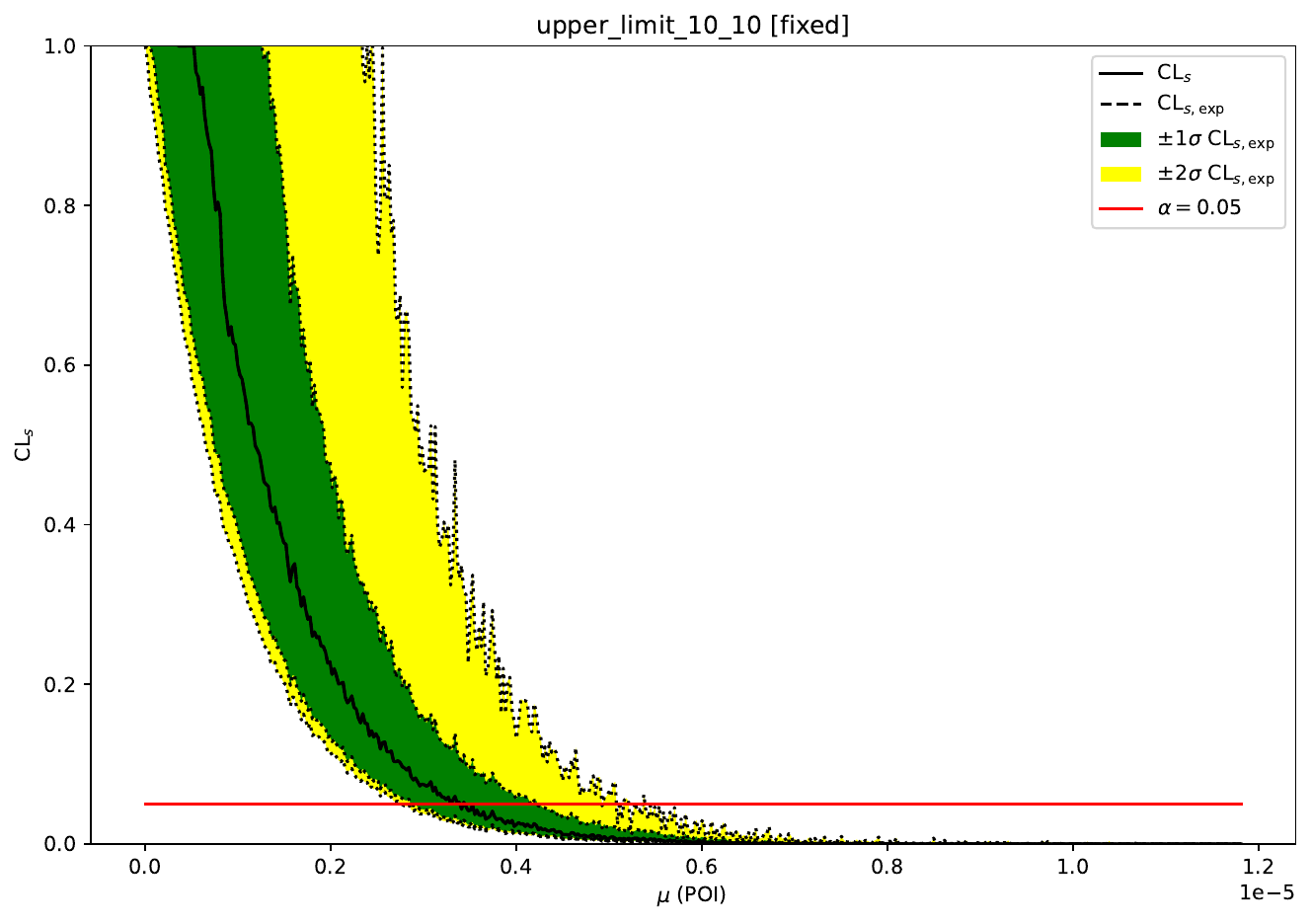}
 \includegraphics[width=0.30\linewidth ]{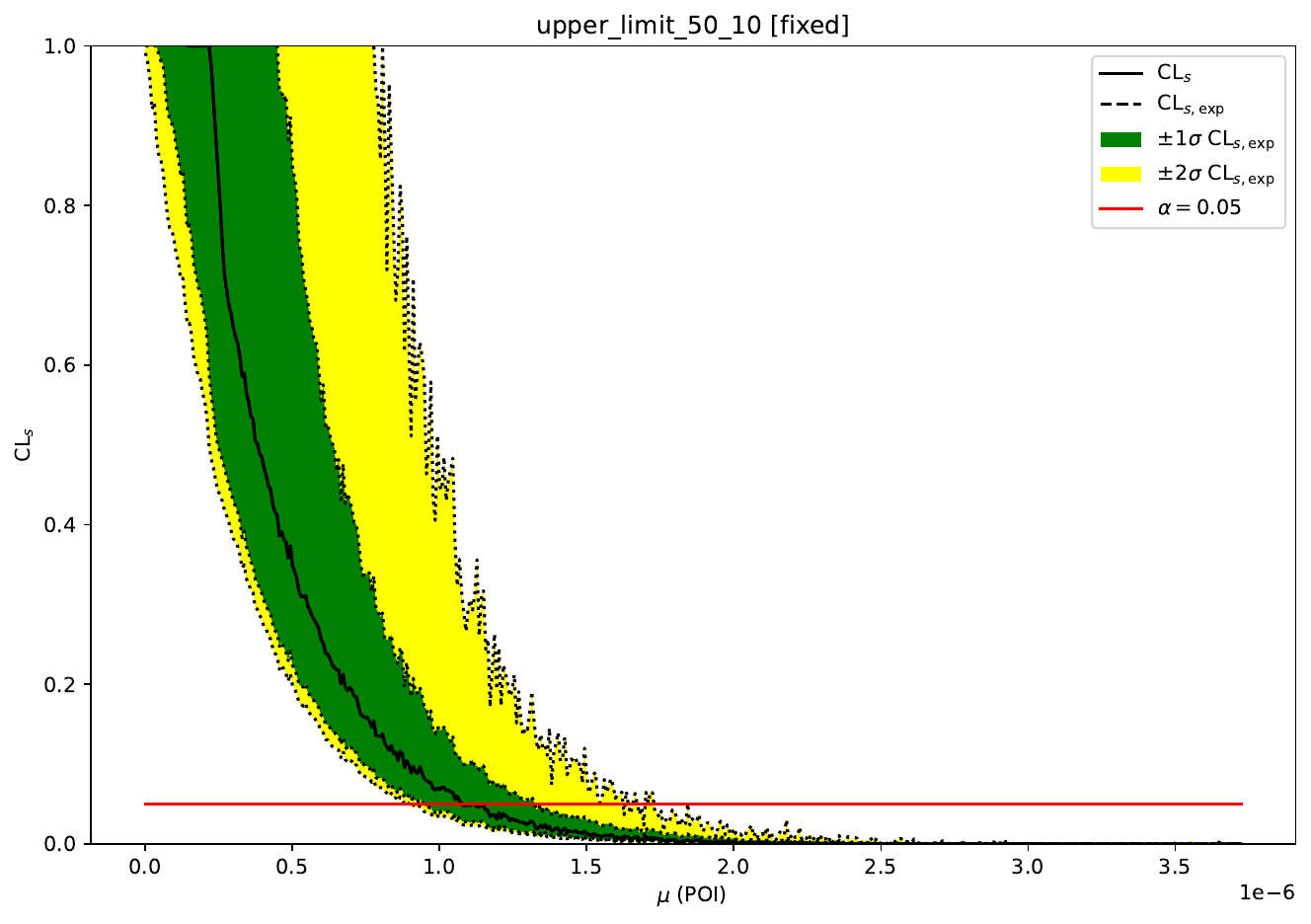}
 \\
   \includegraphics[width=0.30\linewidth ]{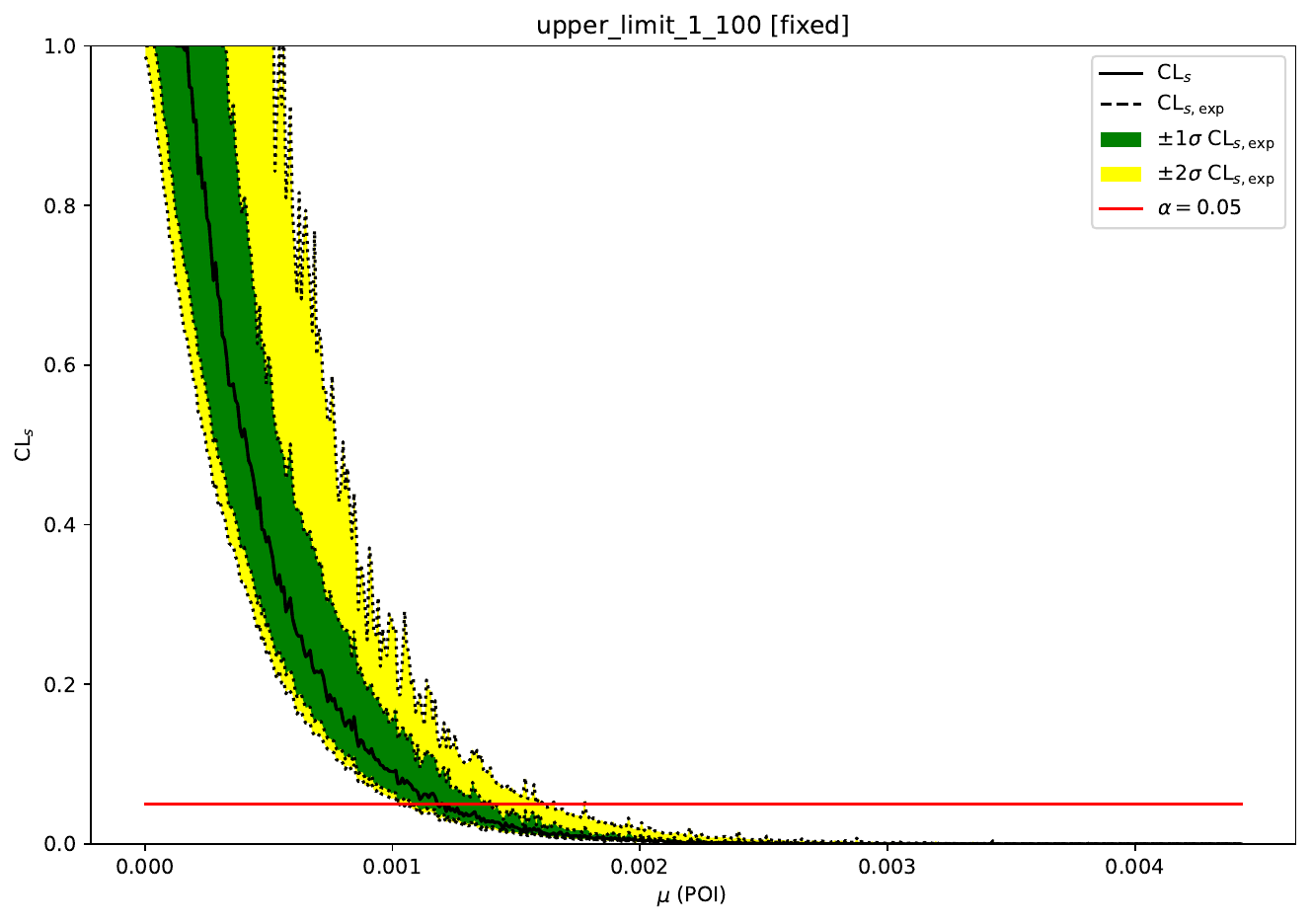}
 \includegraphics[width=0.30\linewidth ]{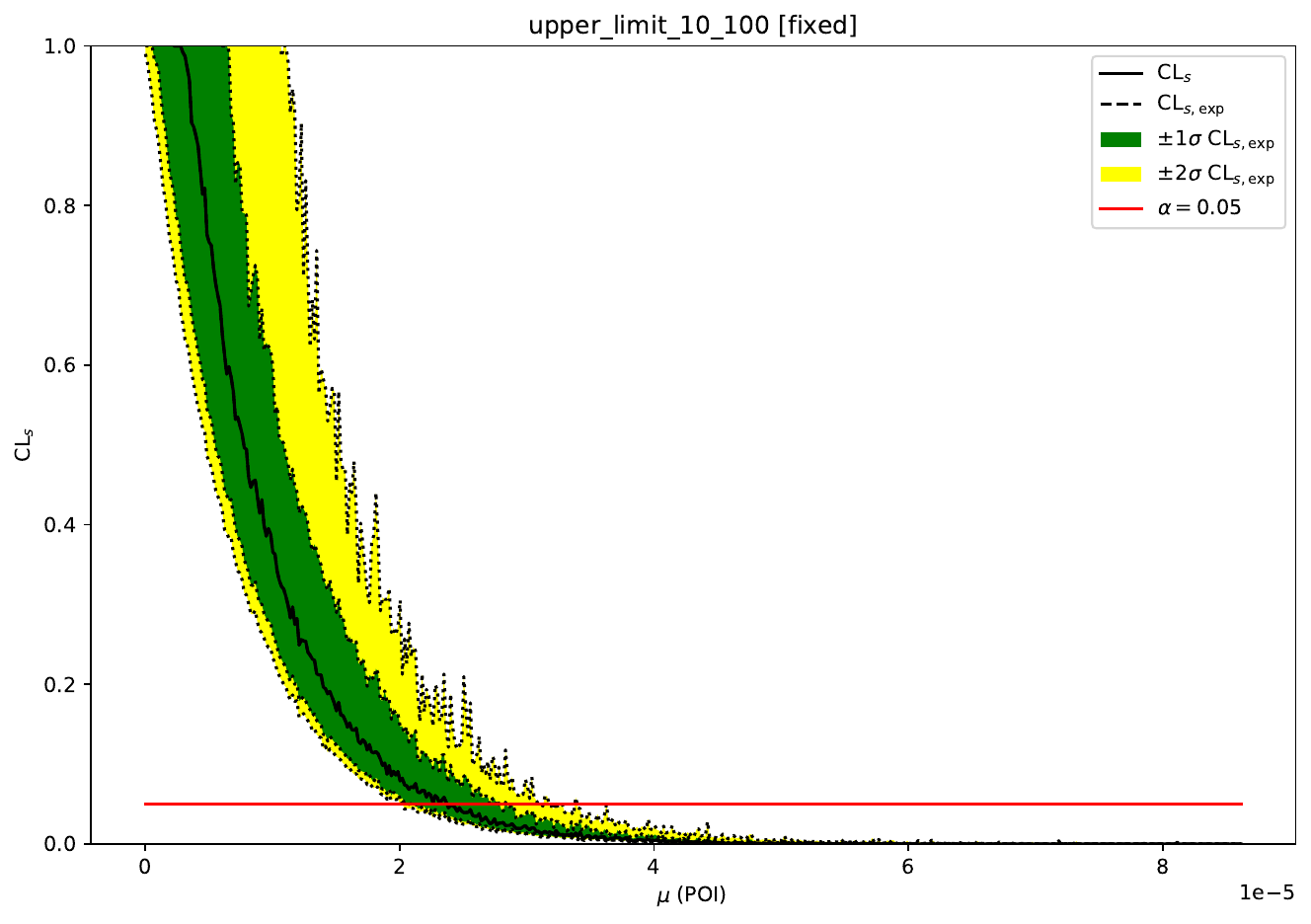}
 \includegraphics[width=0.30\linewidth ]{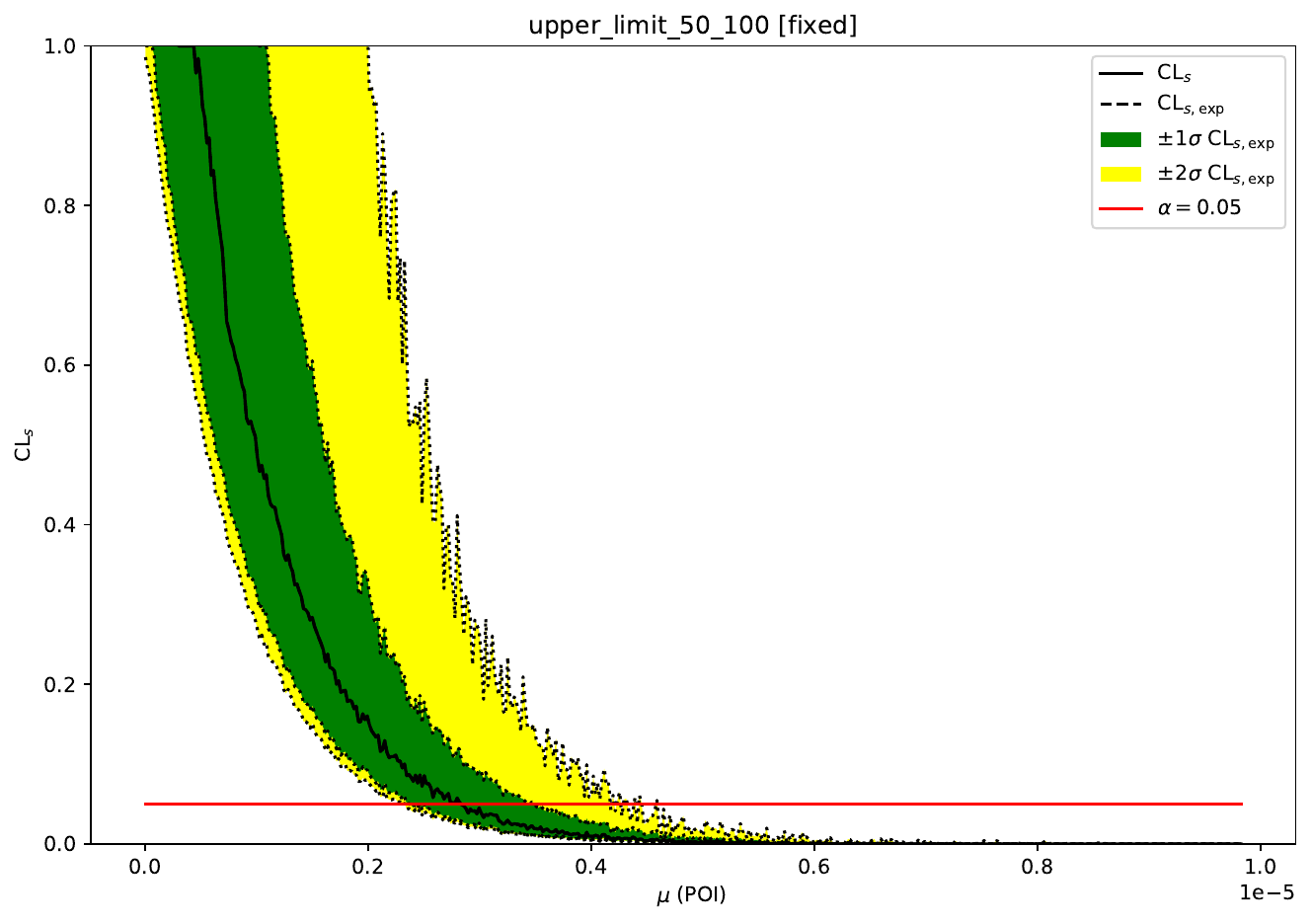}
 \caption{Confidence limits obtained from pseudo-experiments under the null hypothesis, as a function of the signal strength ($\mu$), with a fixed ratio $\epsilon_{V}= \frac{BR(X \to \nu\bar{\nu})}{BR(X \to q\bar{q})}=0.2$. Columns correspond to LLP masses of 50, 10, and 1 GeV, respectively (left to right), and rows correspond to LLP lifetimes of 0.0001, 0.1, 1.0, 10, and 100 nanoseconds (top to bottom).} \label{fig:cl_limit_fix}
 \end{figure*}

\begin{figure*}[htbp]
 \includegraphics[width=0.30\linewidth ]
 {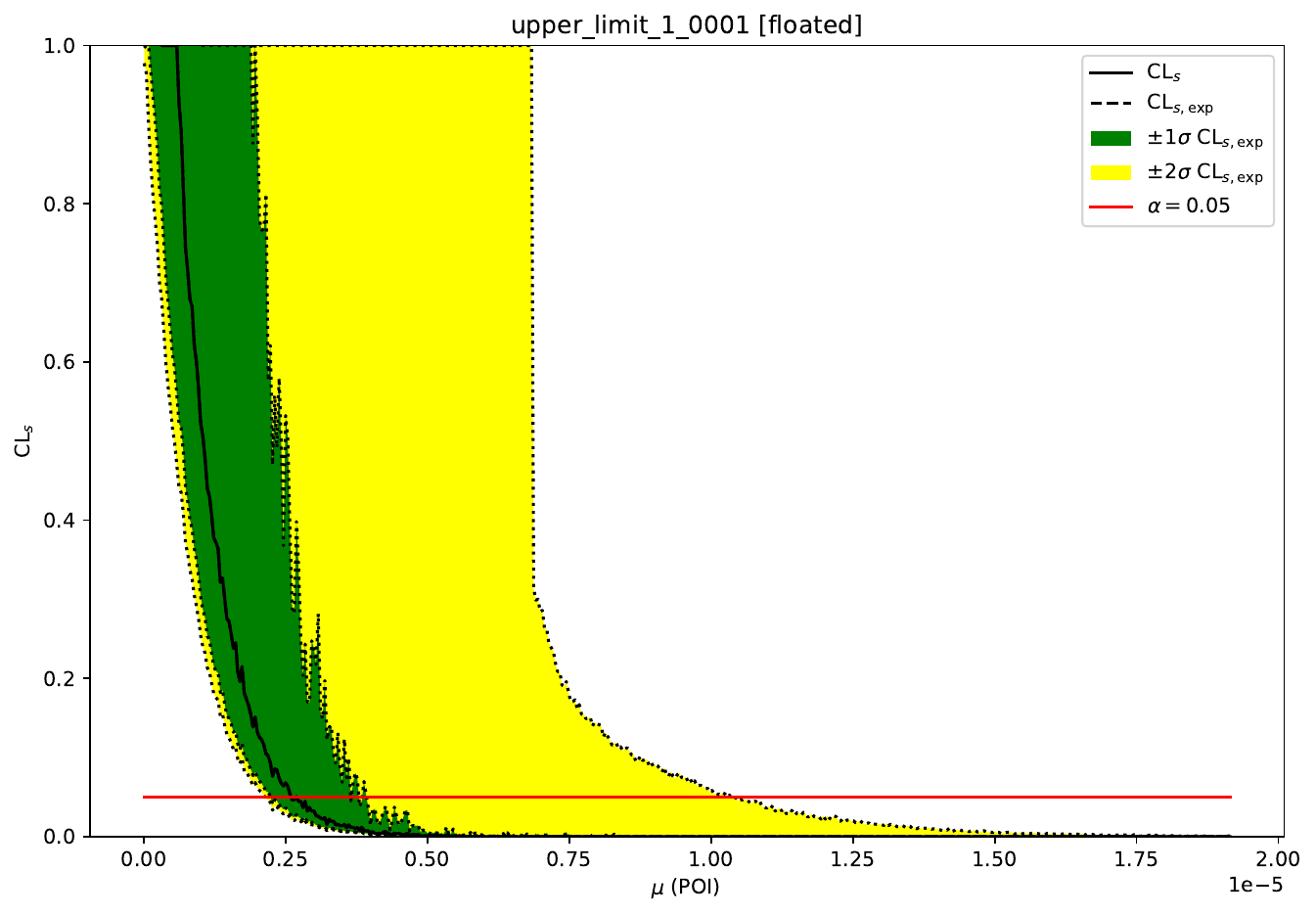}
 \includegraphics[width=0.30\linewidth ]{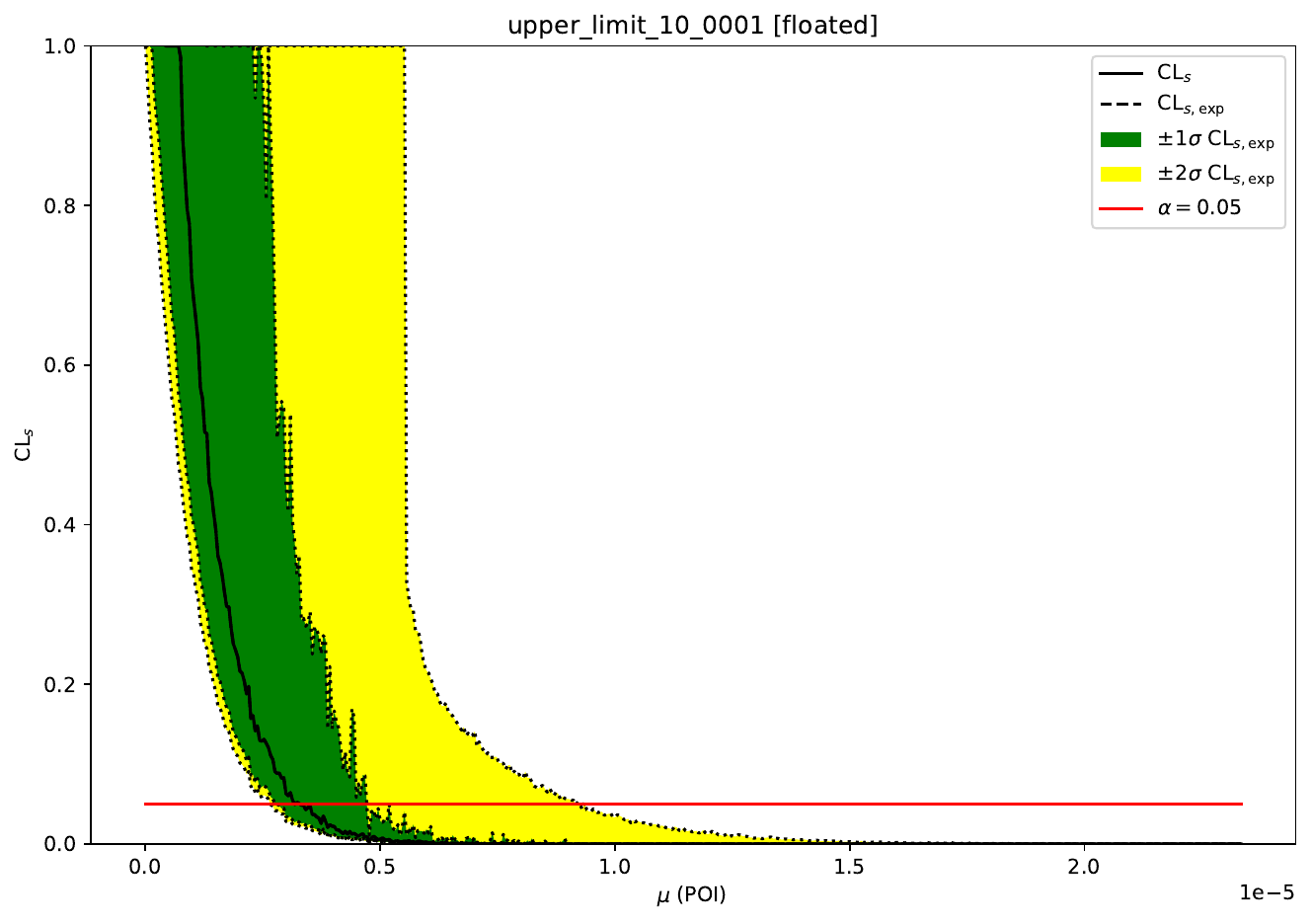}
 \includegraphics[width=0.30\linewidth ]{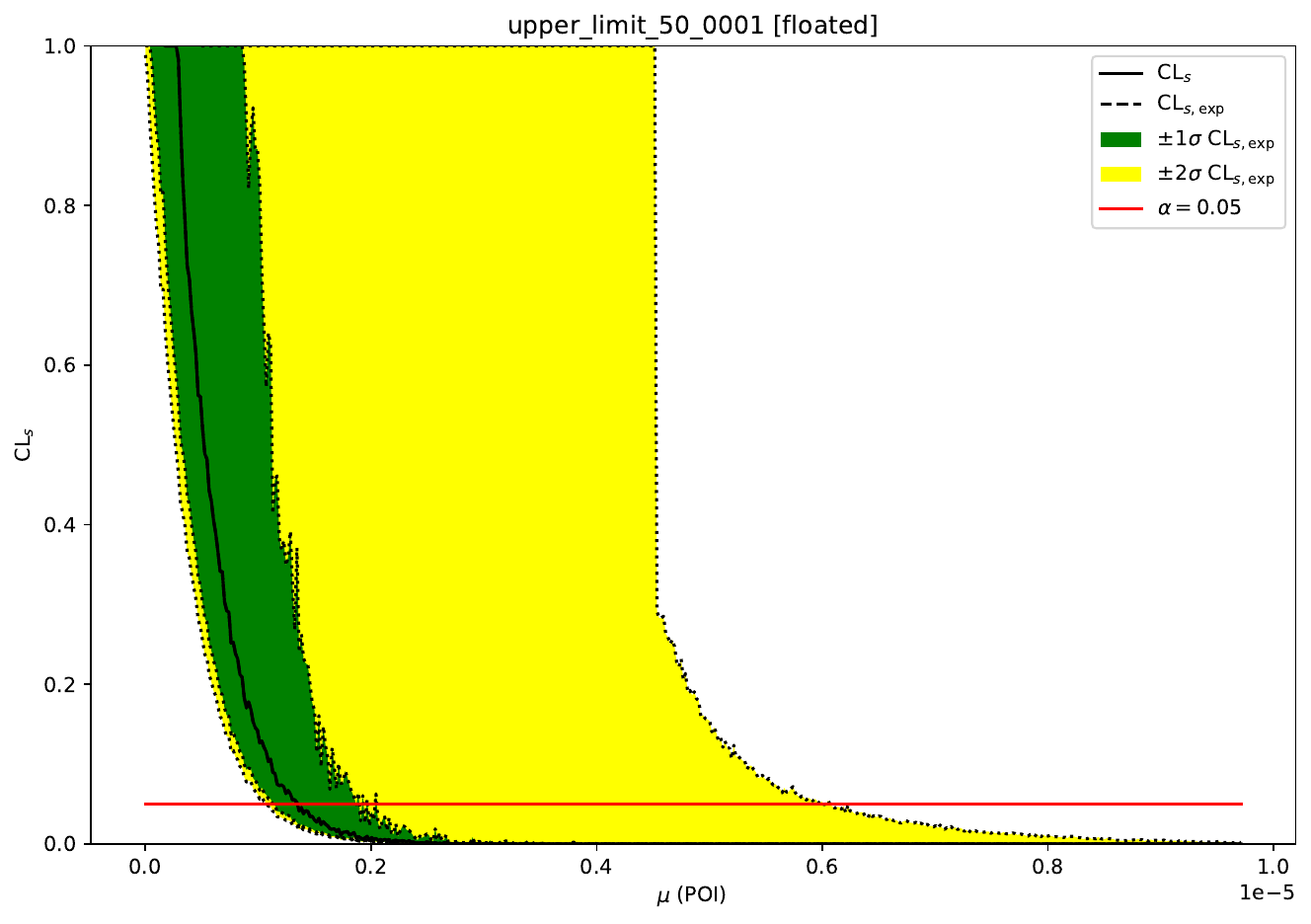}
 \\
 \includegraphics[width=0.30\linewidth ]{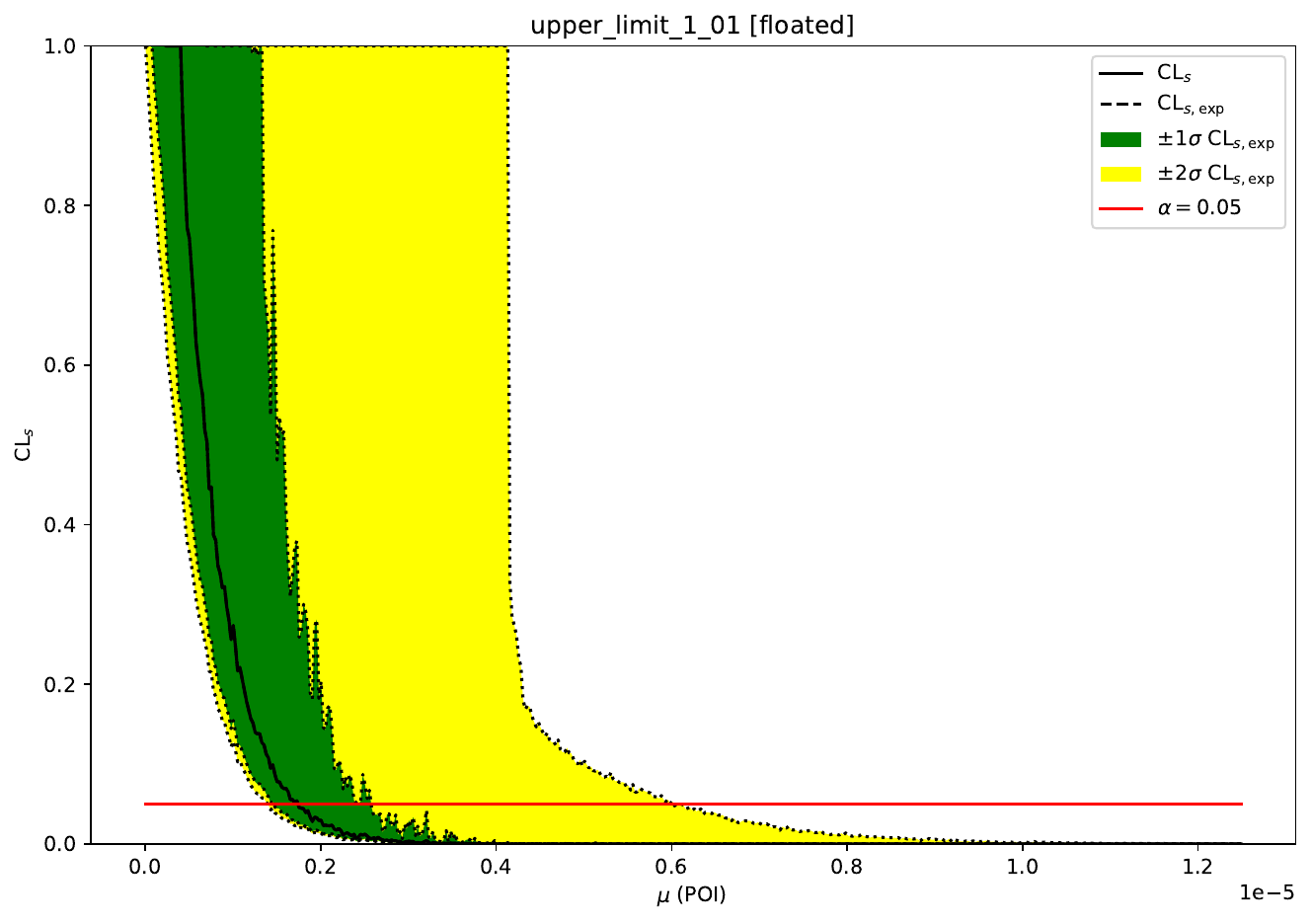}
 \includegraphics[width=0.30\linewidth ]{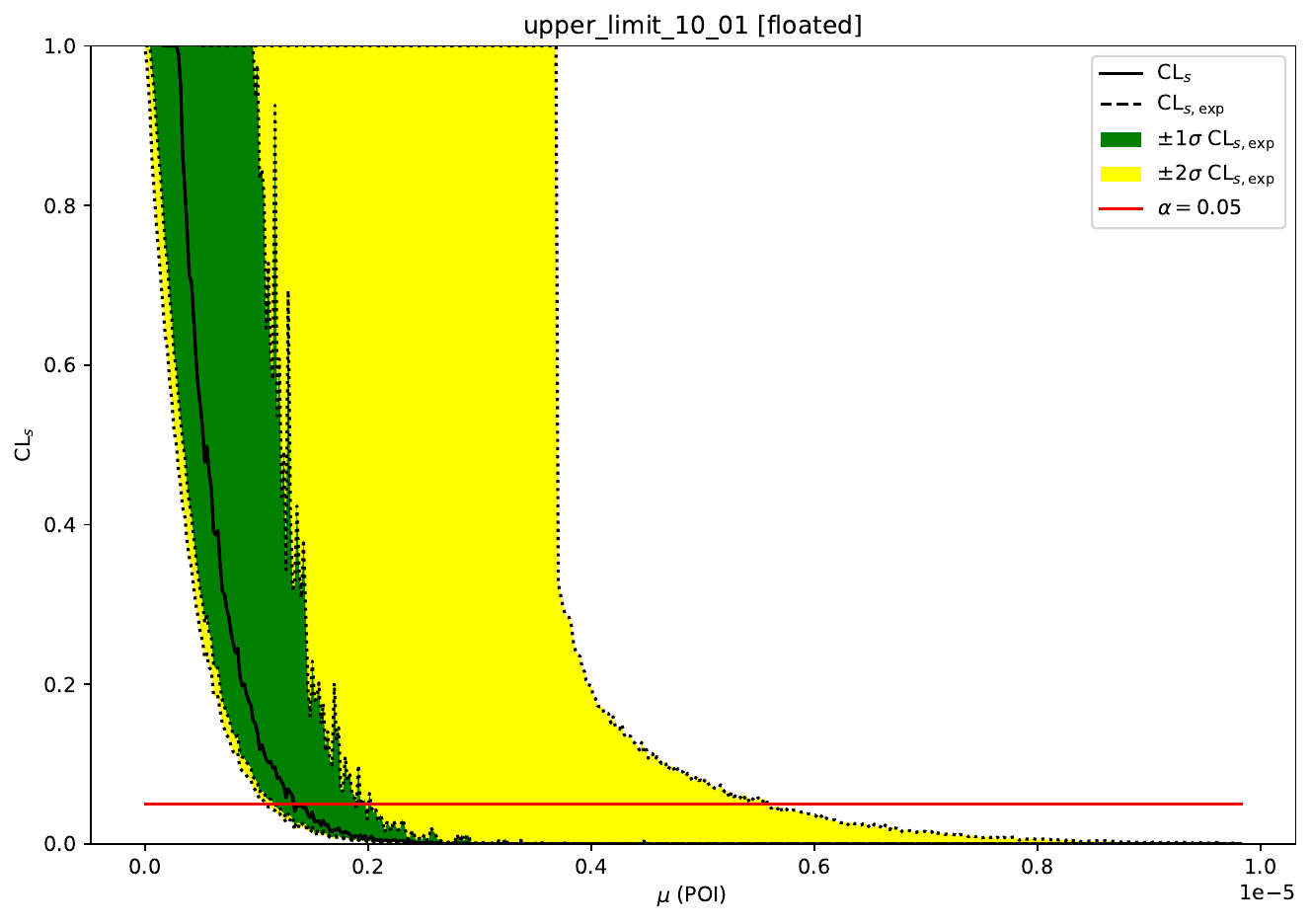}
 \includegraphics[width=0.30\linewidth ]{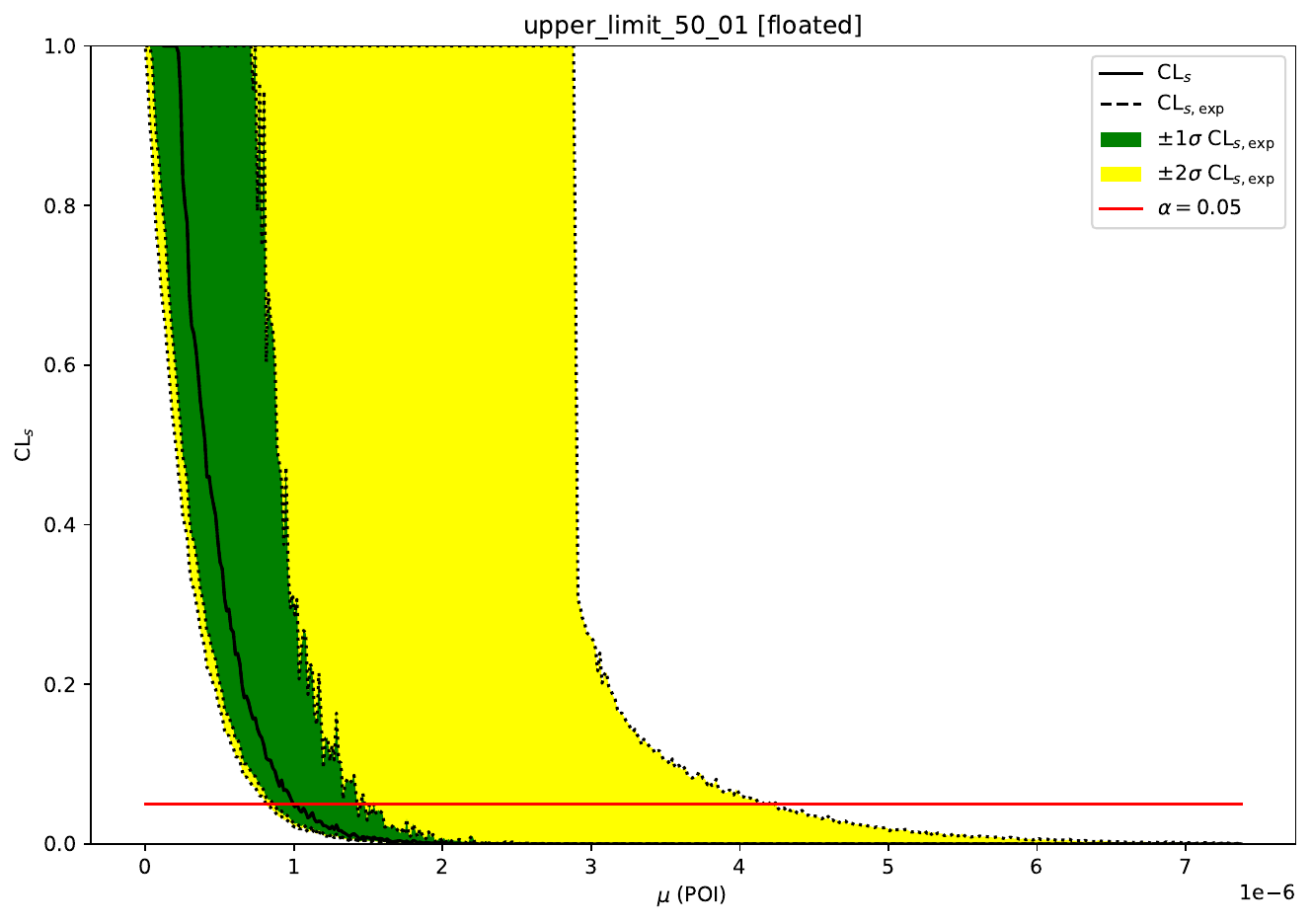}
 \\
 \includegraphics[width=0.30\linewidth ]{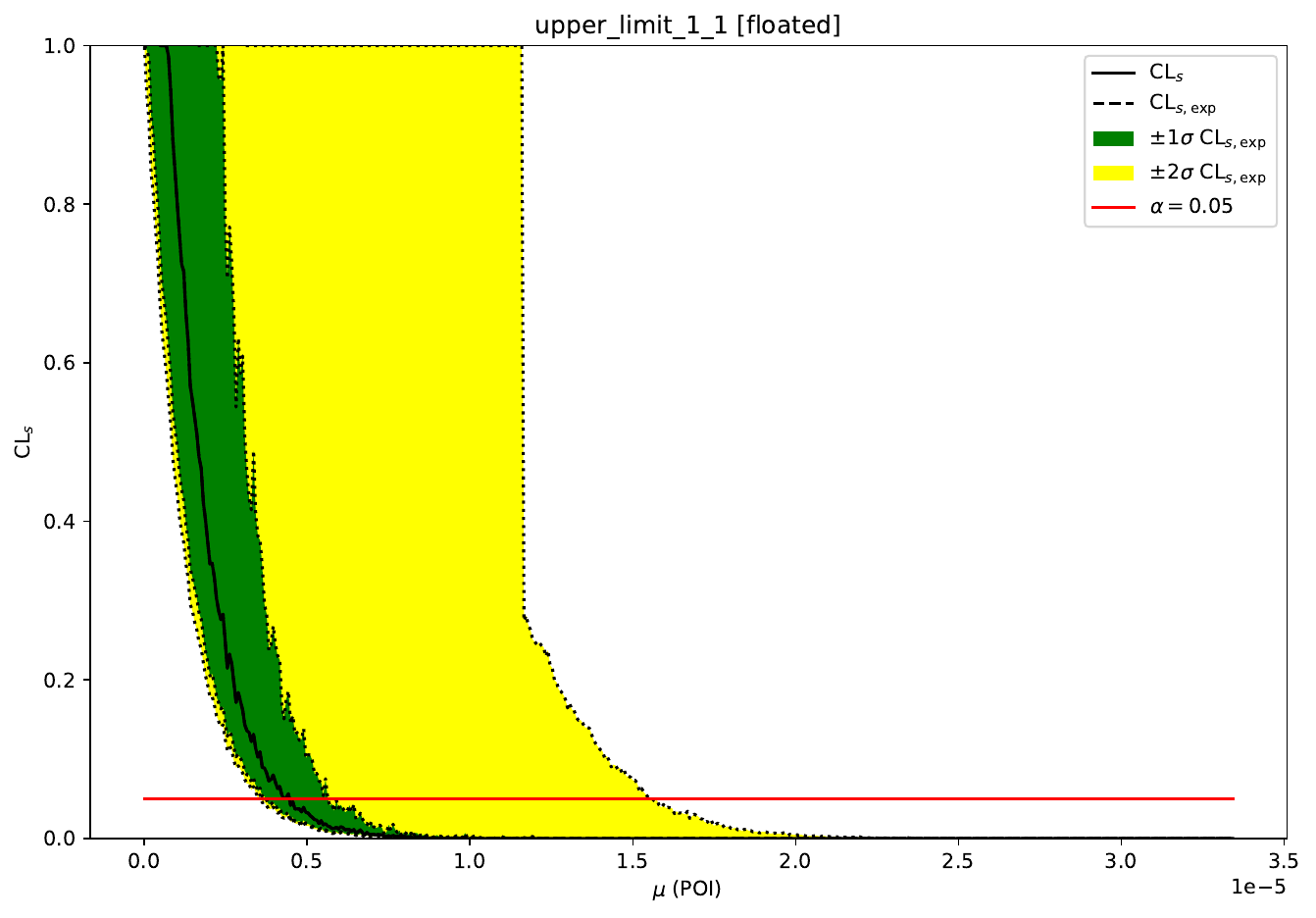}
 \includegraphics[width=0.30\linewidth ]{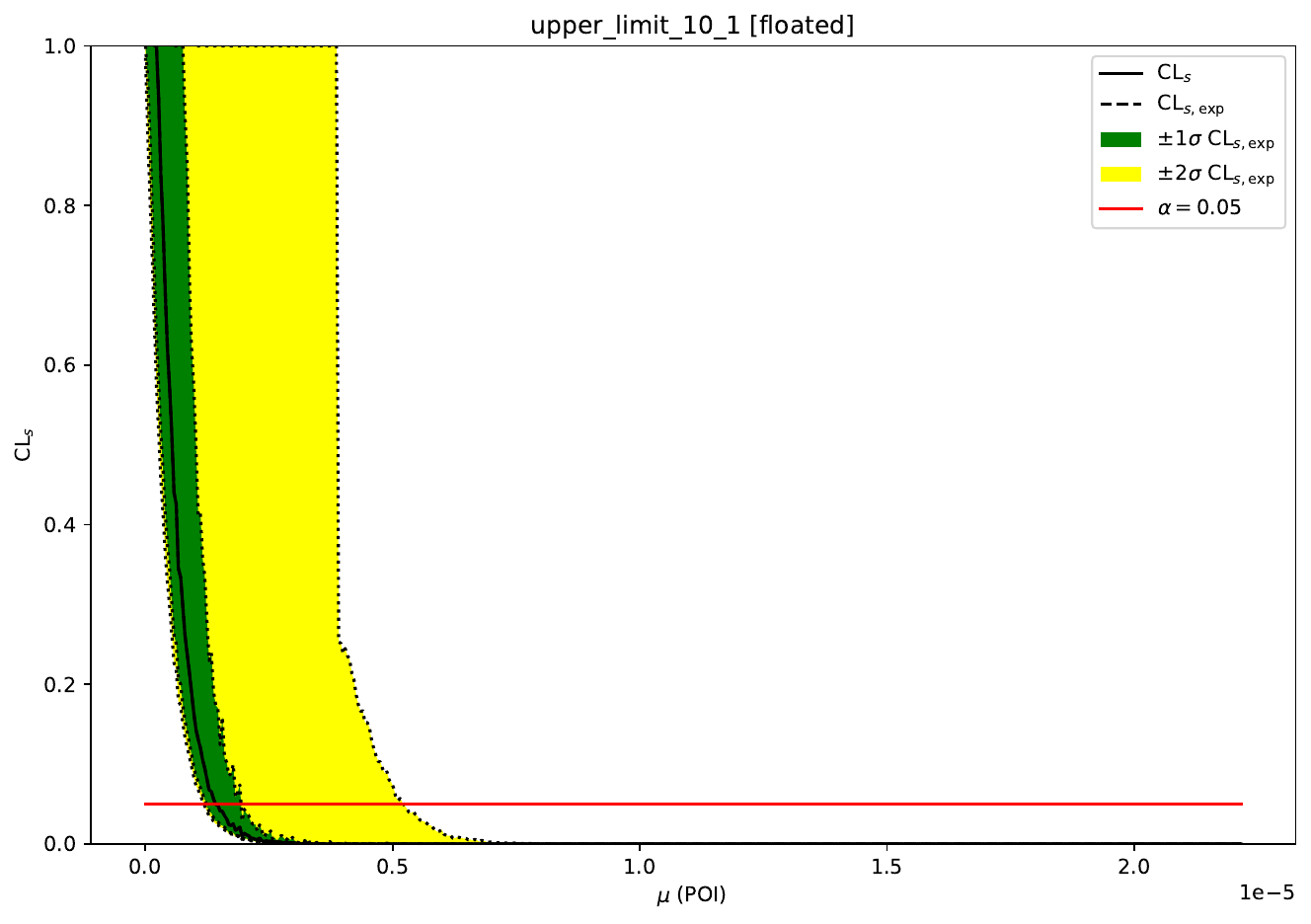}
 \includegraphics[width=0.30\linewidth ]{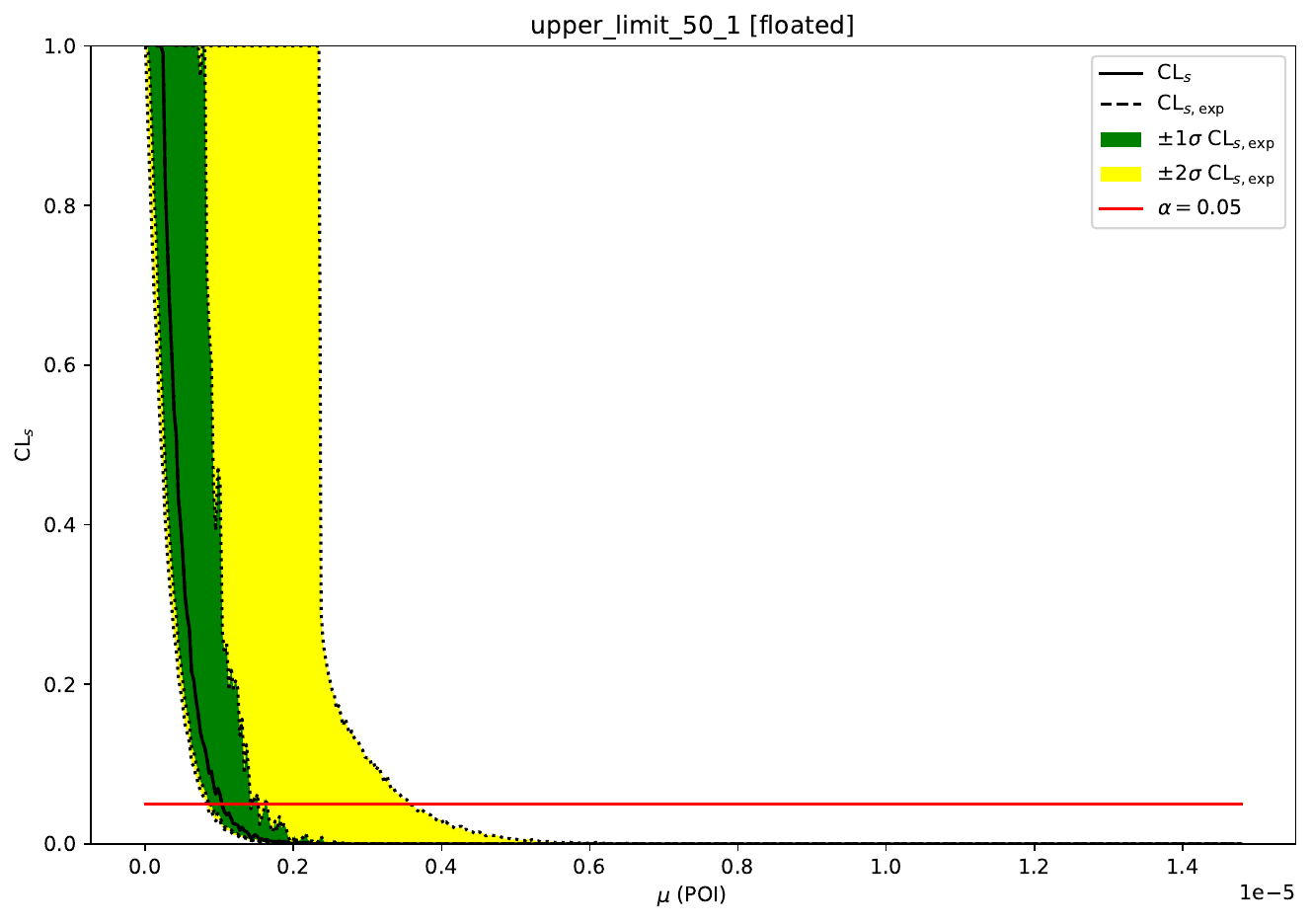}
 \\
 \includegraphics[width=0.30\linewidth ]{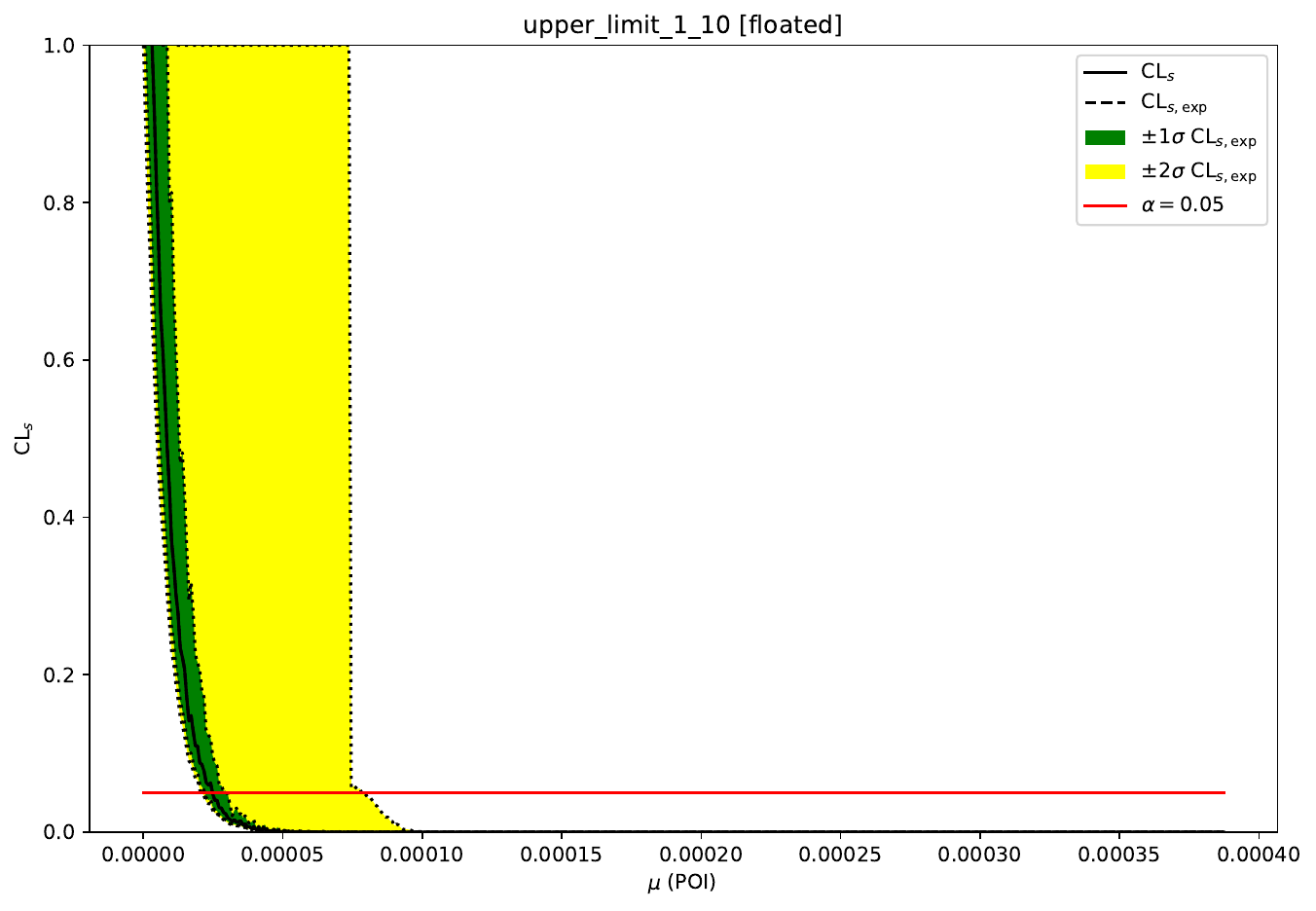}
 \includegraphics[width=0.30\linewidth ]{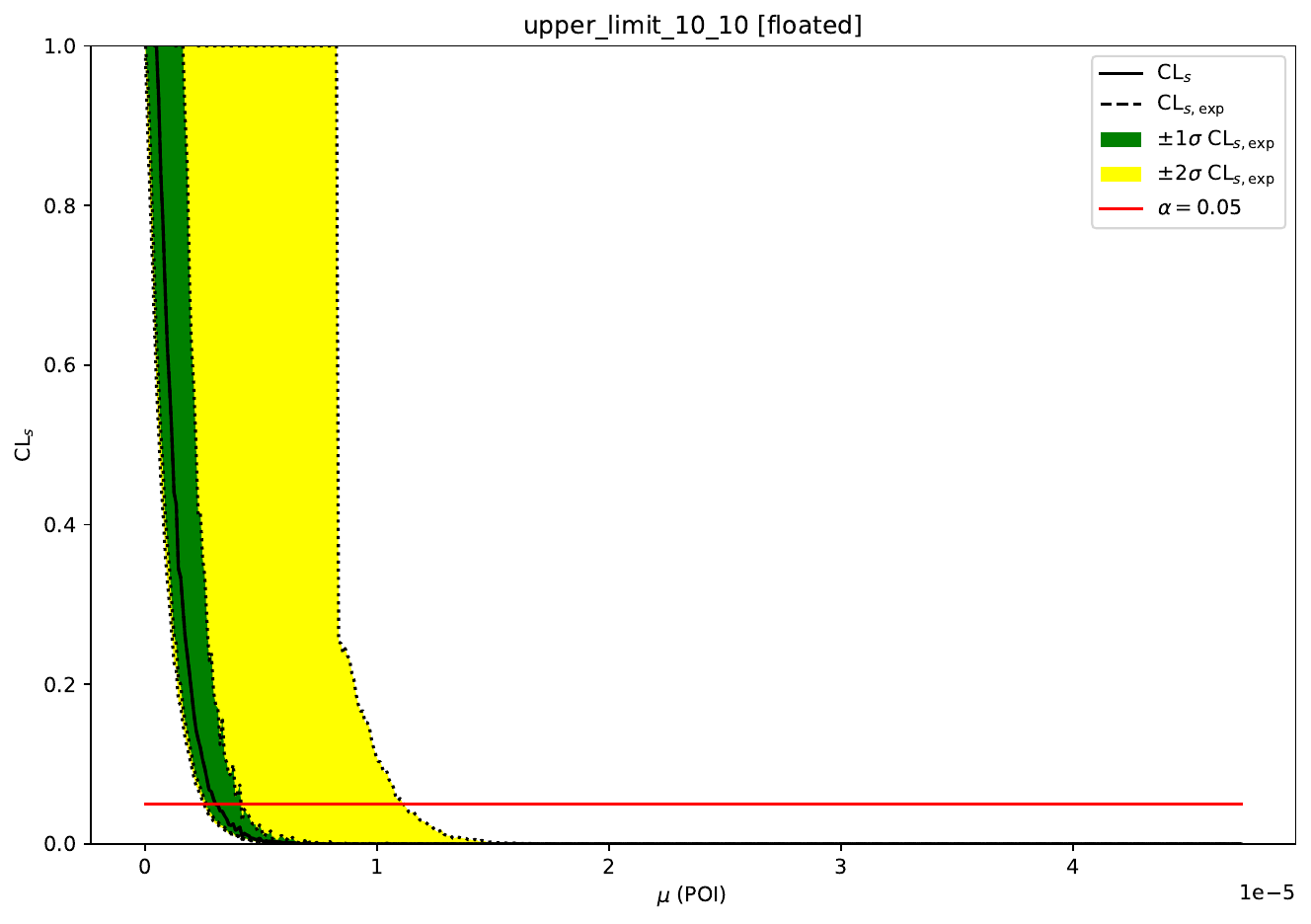}
 \includegraphics[width=0.30\linewidth ]{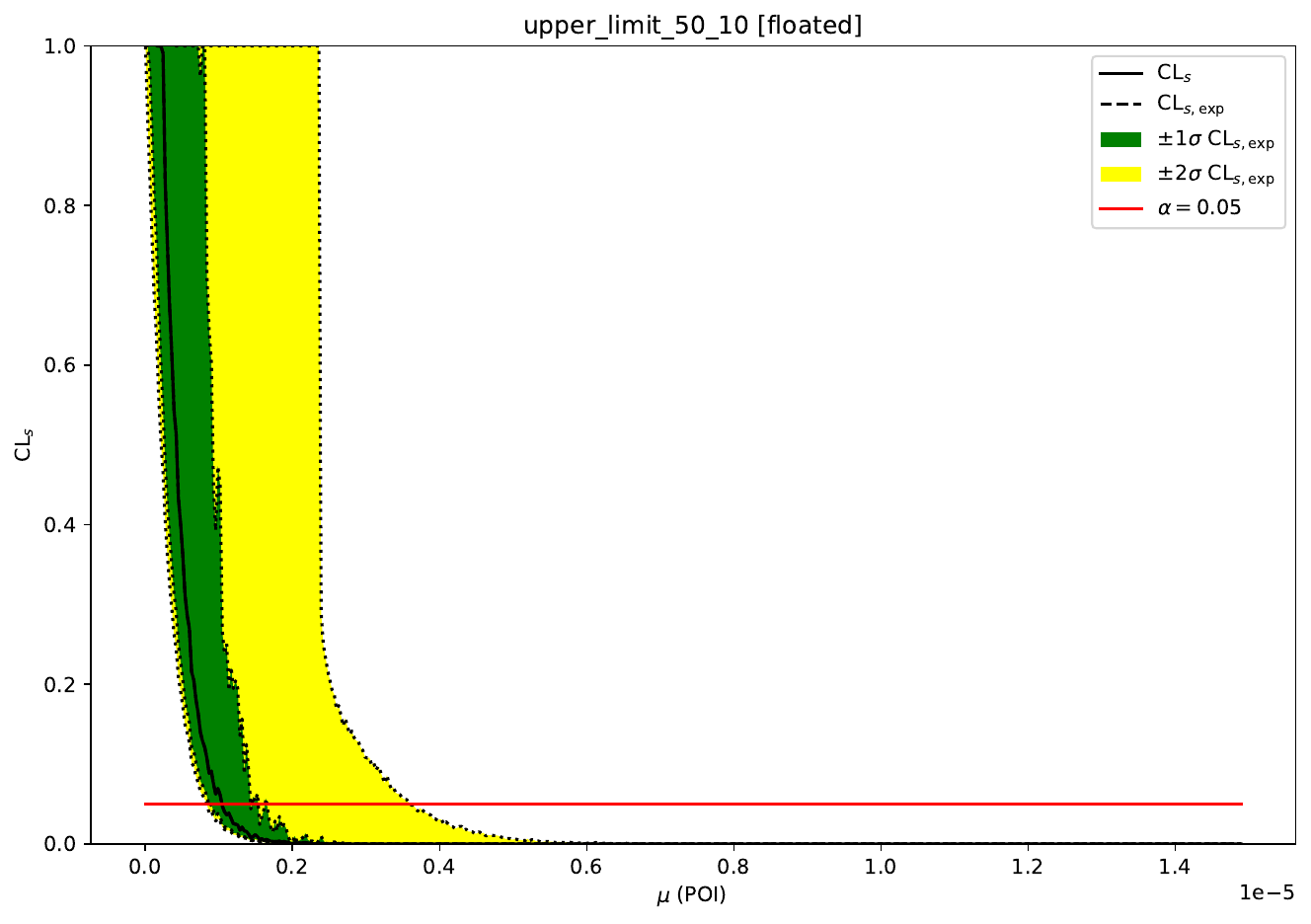}
 \\
 \includegraphics[width=0.30\linewidth ]{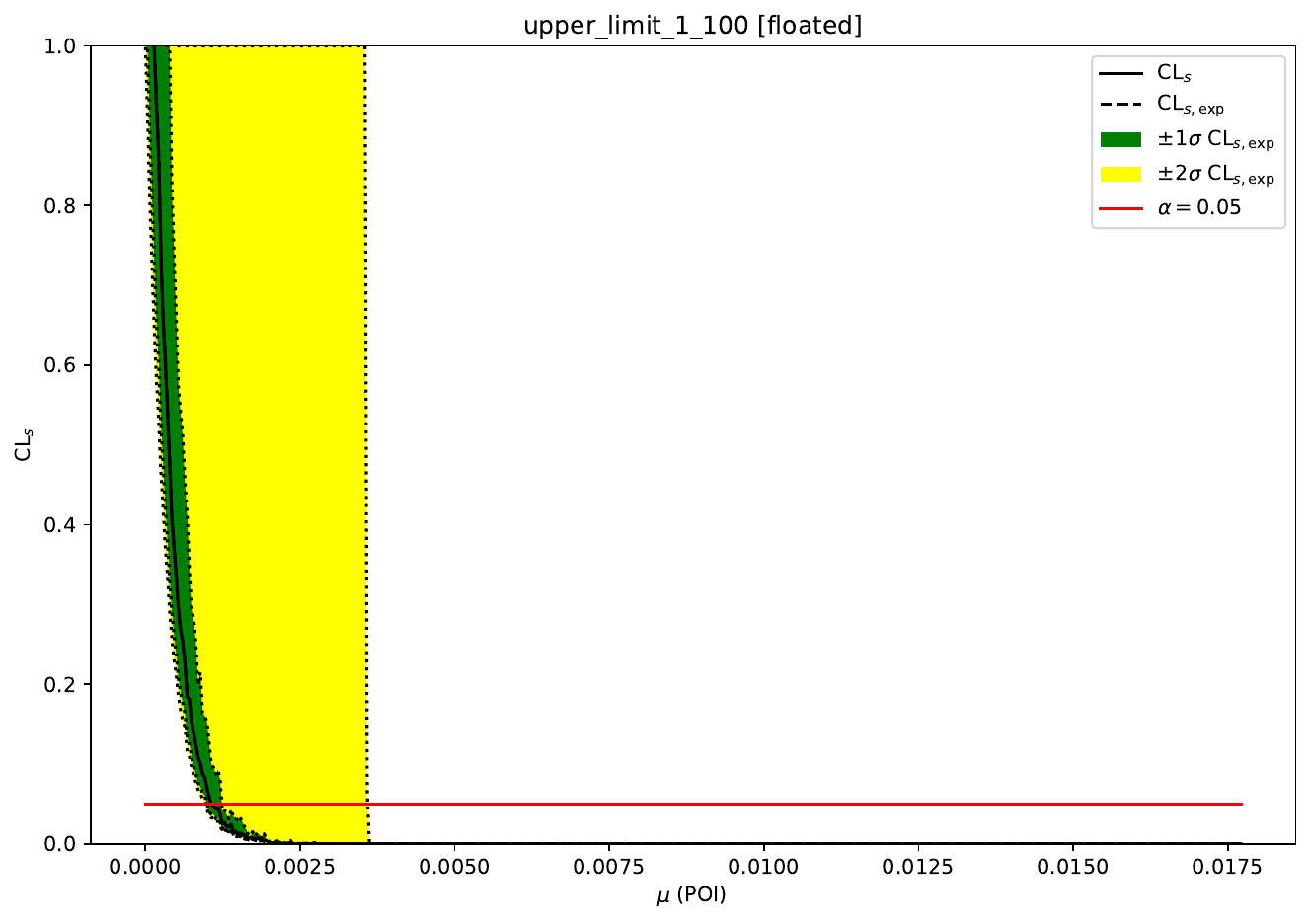}
 \includegraphics[width=0.30\linewidth ]{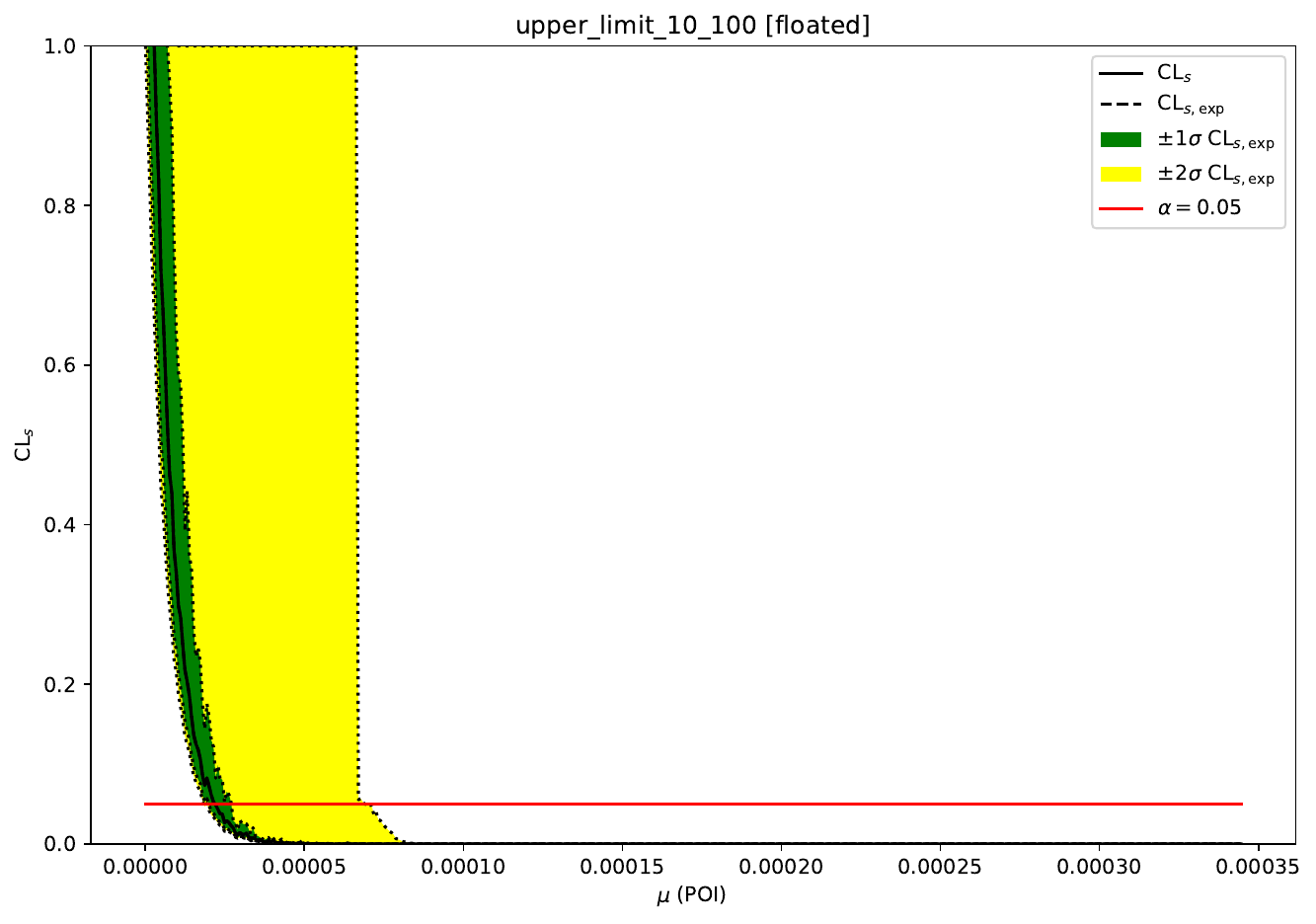}
 \includegraphics[width=0.30\linewidth ]{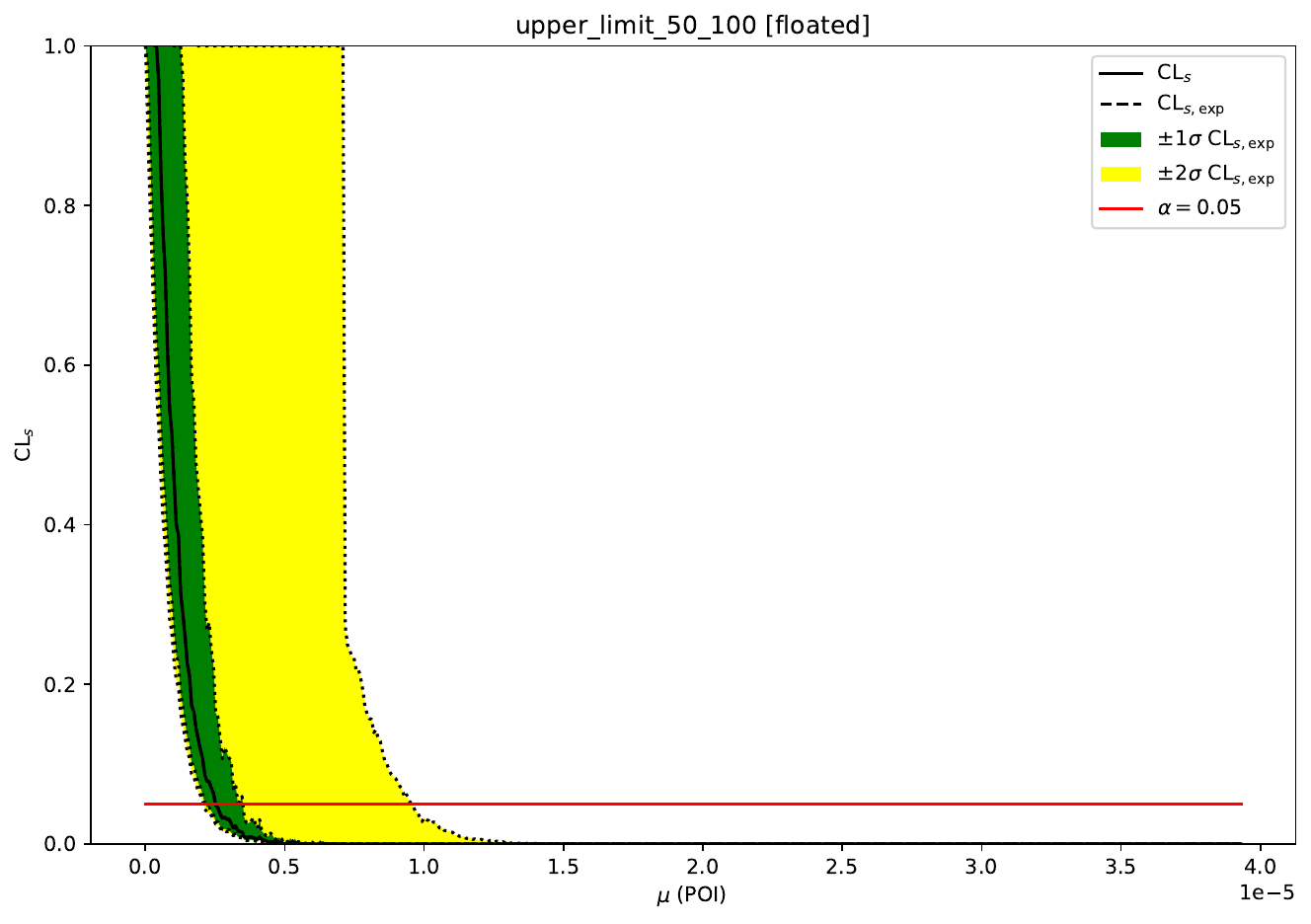}
\caption{Confidence limits obtained from pseudo-experiments under the null hypothesis, as a function of the signal strength ($\mu$), with the ratio $\epsilon_{V}= \frac{BR(X \to \nu\bar{\nu})}{BR(X \to q\bar{q})}$ allowed to float during fitting. Columns correspond to LLP masses of 50, 10, and 1 GeV, respectively (left to right), and rows correspond to LLP lifetimes of 0.0001, 0.1, 1.0, 10, and 100 nanoseconds (top to bottom).}
 \label{fig:cl_limit_free}
 \end{figure*}

Figures~\ref{fig:cl_limit_fix} and~\ref{fig:cl_limit_free} present the confidence limits derived from pseudo-experiments generated under the null hypothesis, as functions of the signal strength parameter $\mu$. The parameter $\epsilon_{V}$, defined as the ratio of branching fractions $\epsilon_{V}= \frac{BR(X \to \nu\bar{\nu})}{BR(X \to q\bar{q})}$, is either fixed to 0.2 (Figure~\ref{fig:cl_limit_fix}) or allowed to float freely during the fitting process (Figure~\ref{fig:cl_limit_free}).

  In both figures, the LLP mass scenarios vary horizontally (from left to right: 50, 10, and 1 GeV), while LLP lifetimes vary vertically (from top to bottom: 0.0001, 0.1, 1.0, 10, and 100 nanoseconds).

%\clearpage
\section{Two dimensional upper limits on ($\mathcal{B}_{\textrm{2-jet}}$, $\mathcal{B}_{\textrm{4-jet}}$)}
\label{app:2d_limit}

\begin{figure*}[!htbp]\label{sec_2D_limits}
  \includegraphics[width=0.29\linewidth]{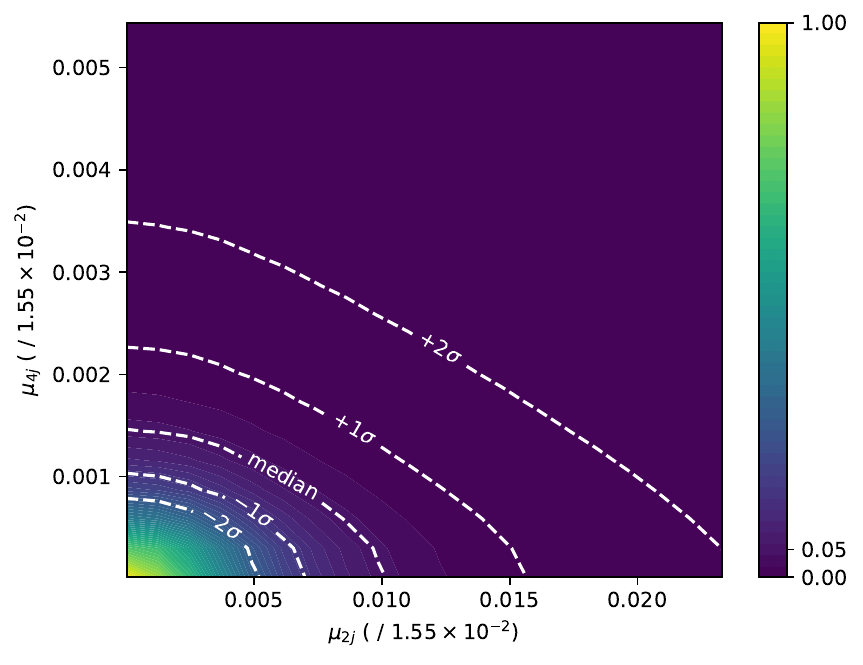}
  \includegraphics[width=0.29\linewidth]{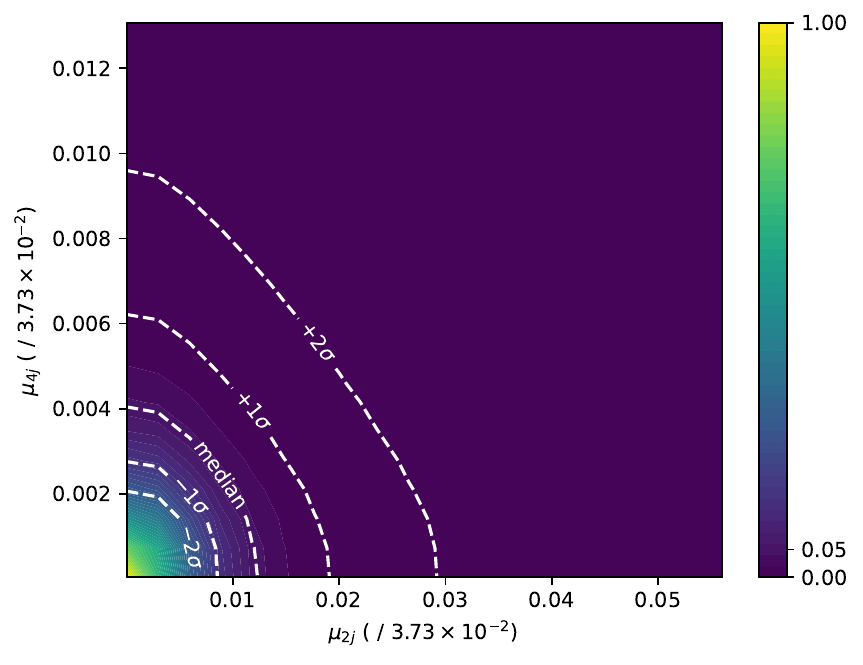}
  \includegraphics[width=0.29\linewidth]{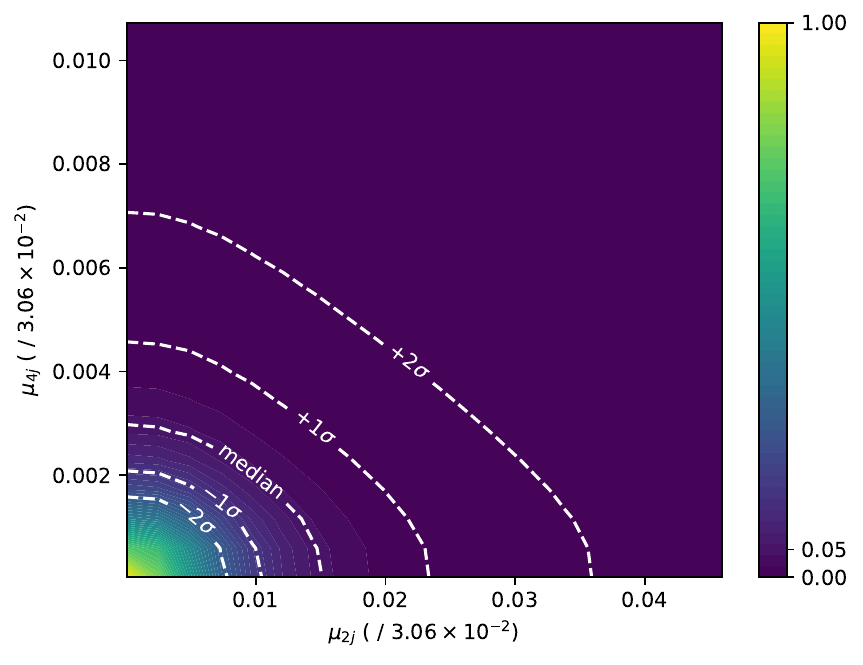} \\
  
  \includegraphics[width=0.29\linewidth]{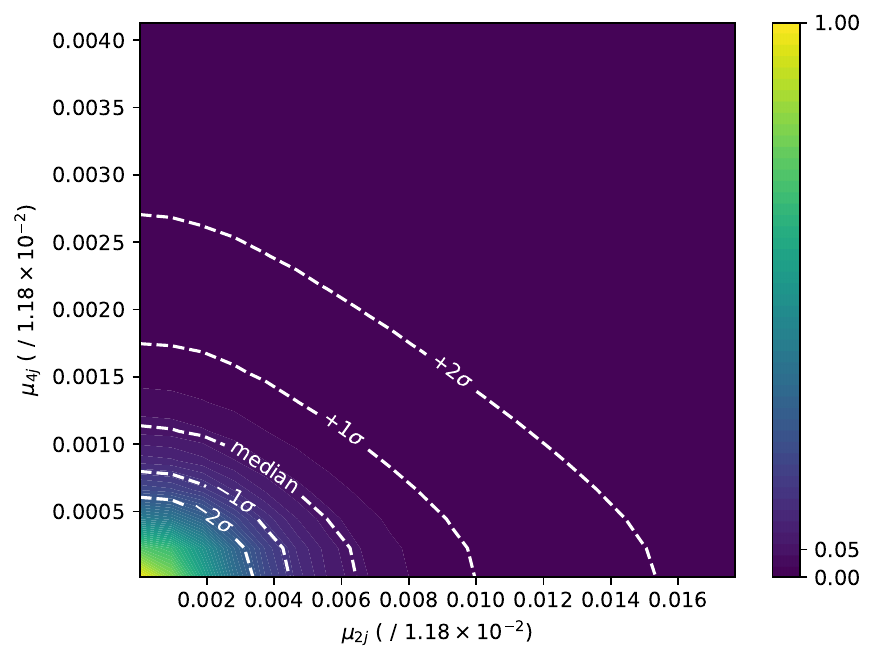}
  \includegraphics[width=0.29\linewidth]{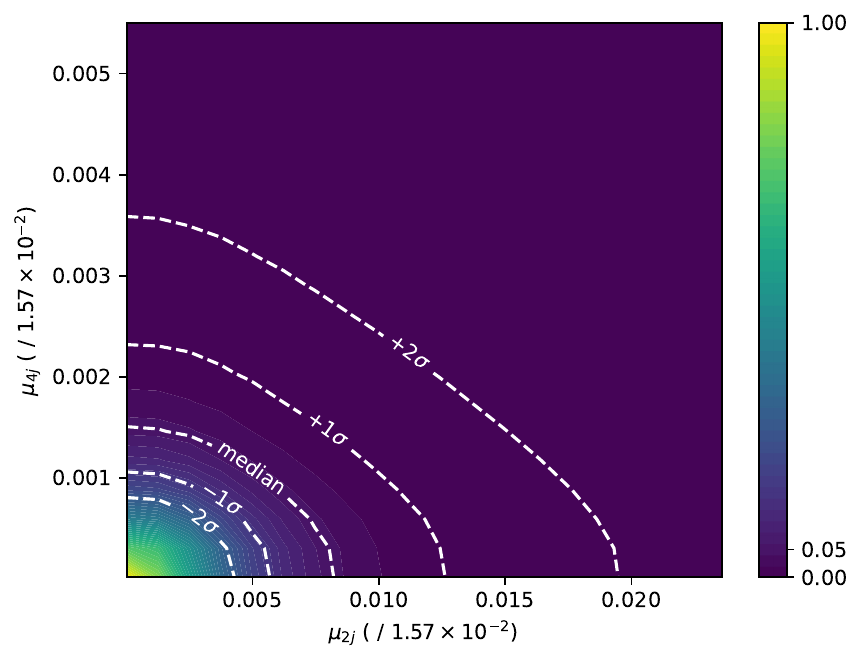}
  \includegraphics[width=0.29\linewidth]{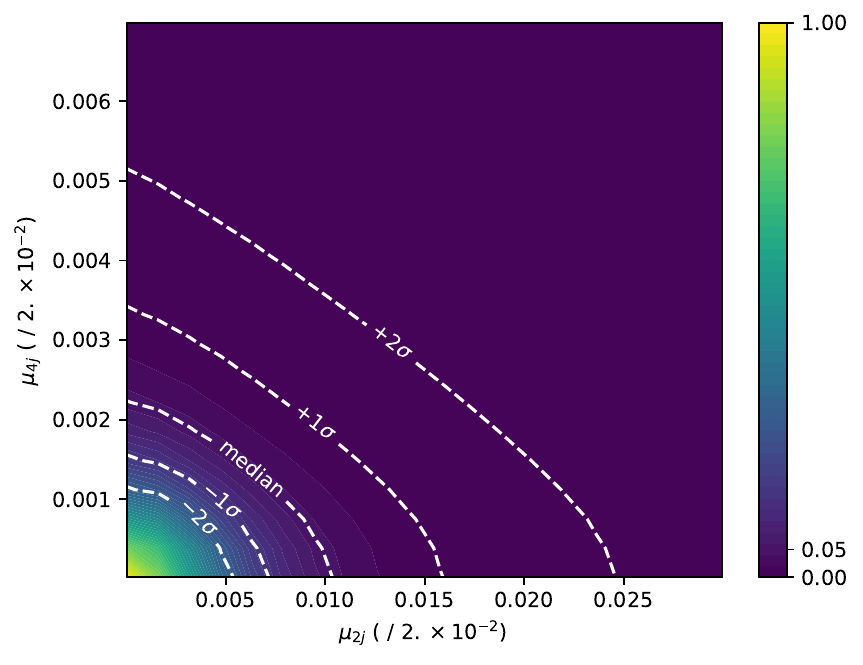} \\

  \includegraphics[width=0.29\linewidth]{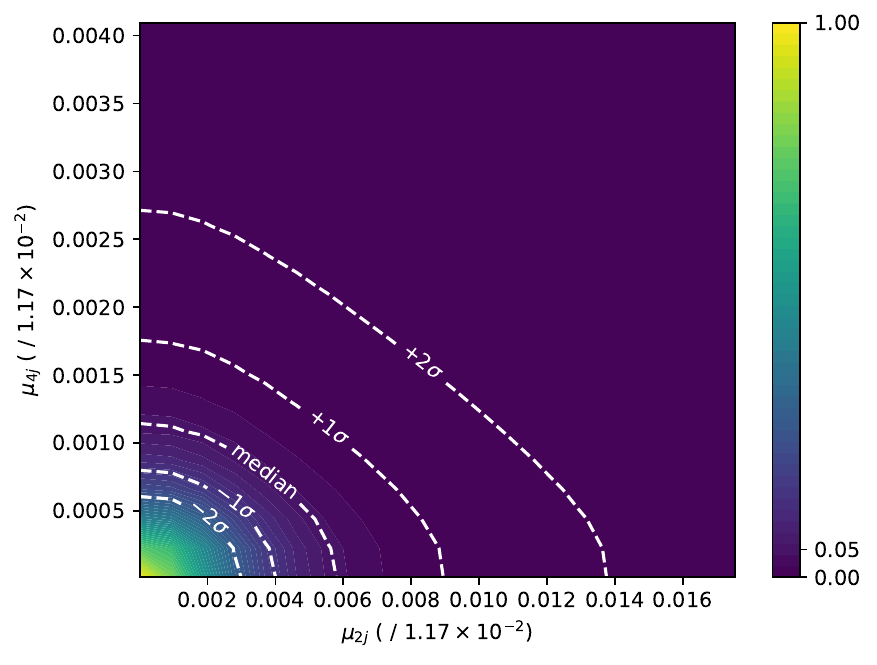}
  \includegraphics[width=0.29\linewidth]{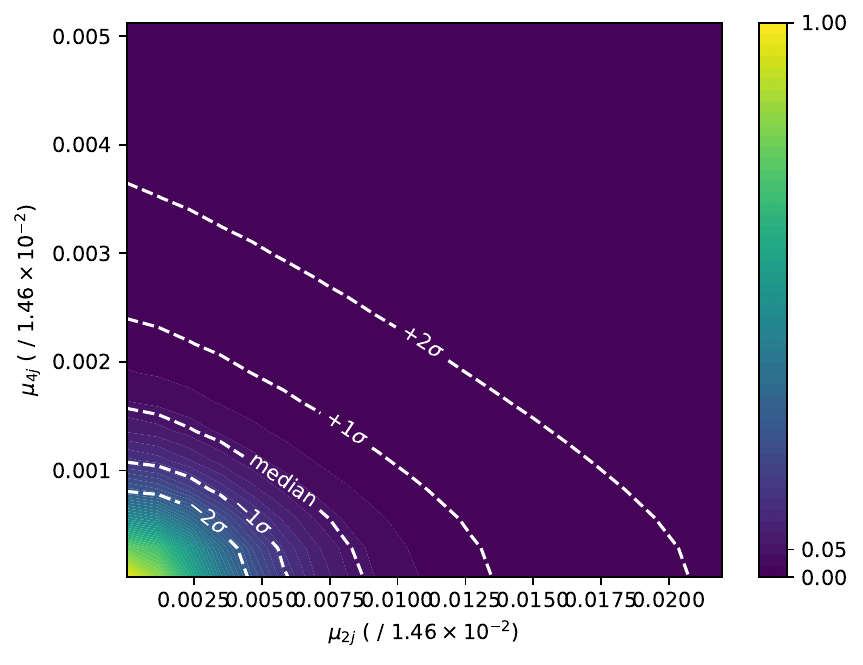}
  \includegraphics[width=0.29\linewidth]{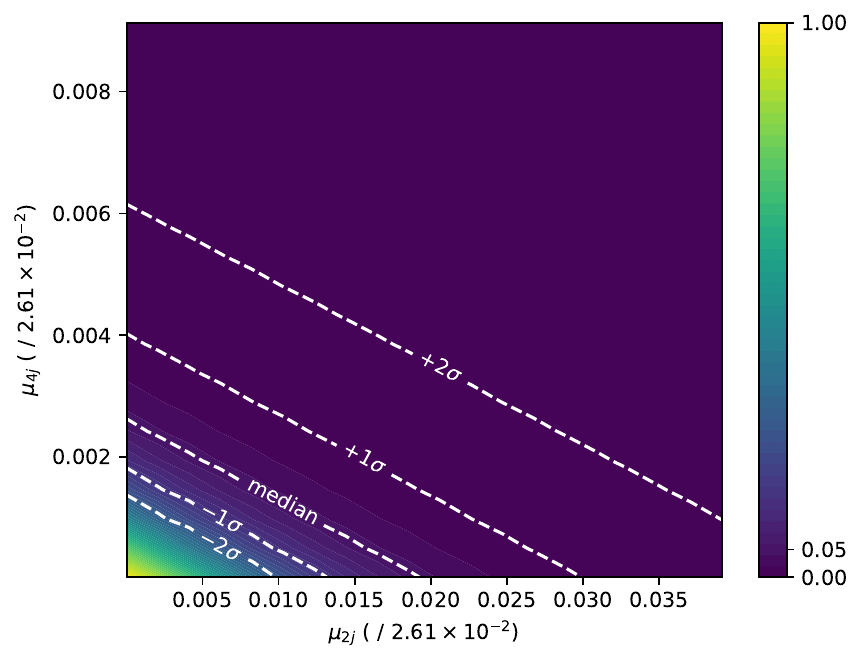} \\

  \includegraphics[width=0.29\linewidth]{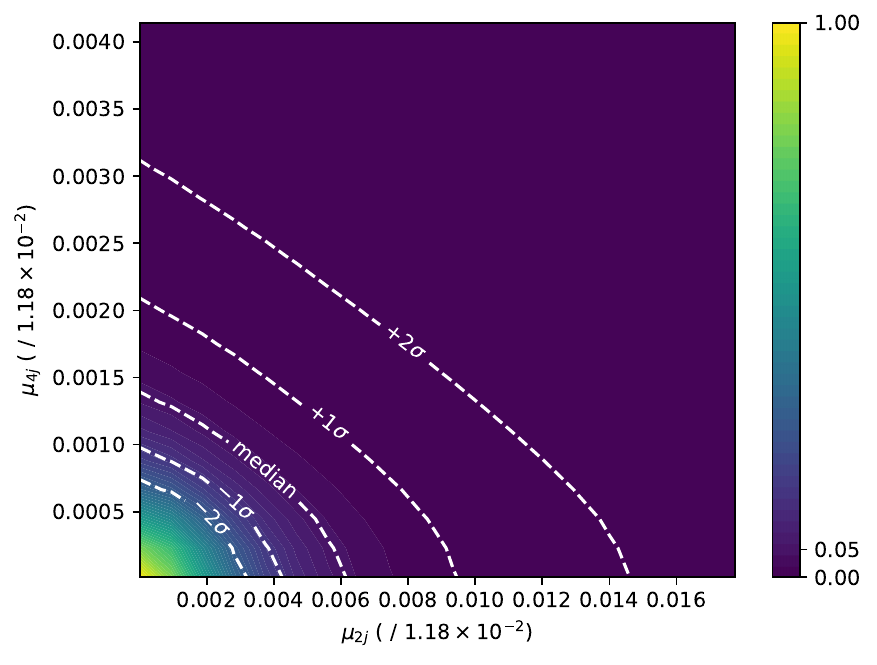}
  \includegraphics[width=0.29\linewidth]{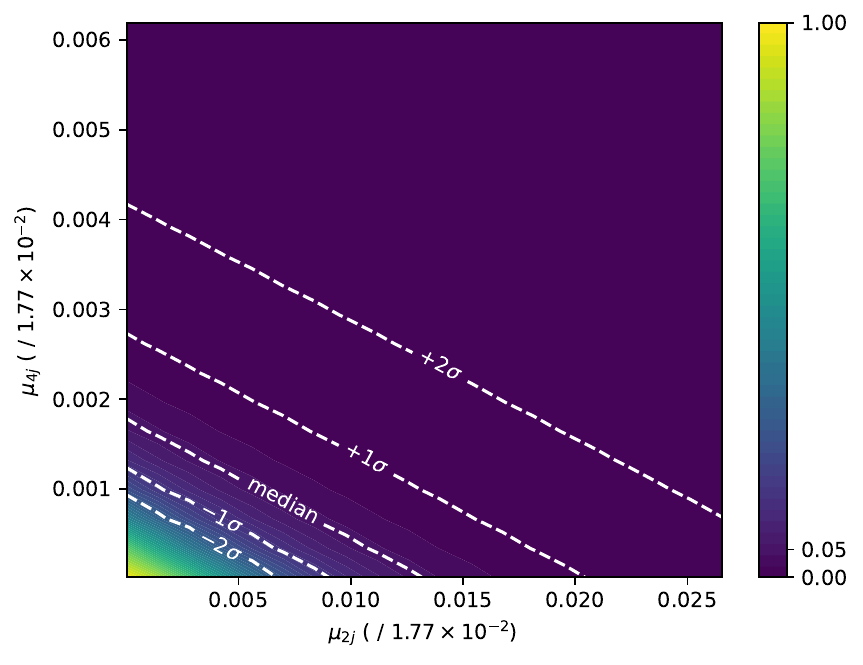}
  \includegraphics[width=0.29\linewidth]{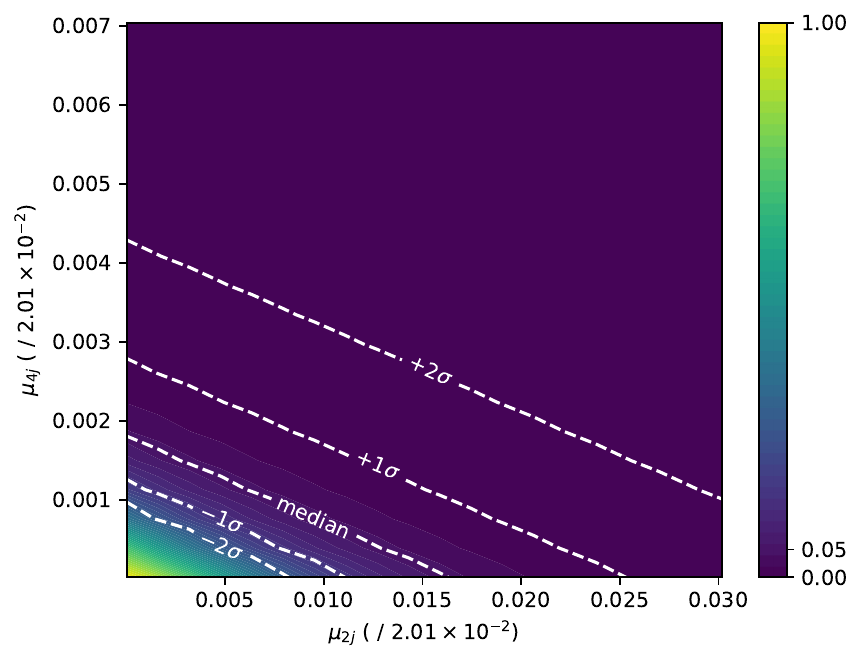} \\

  \includegraphics[width=0.29\linewidth]{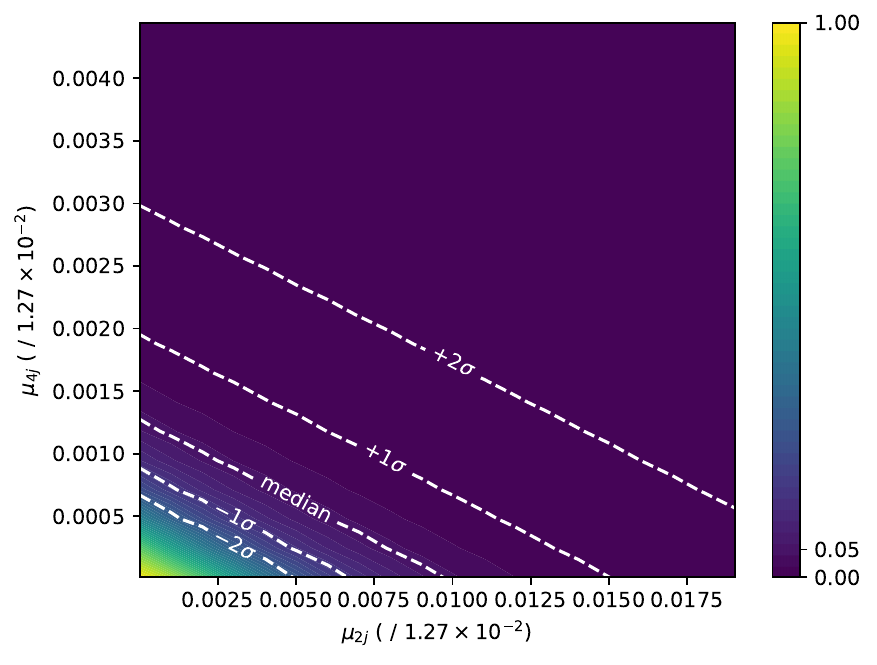}
  \includegraphics[width=0.29\linewidth]{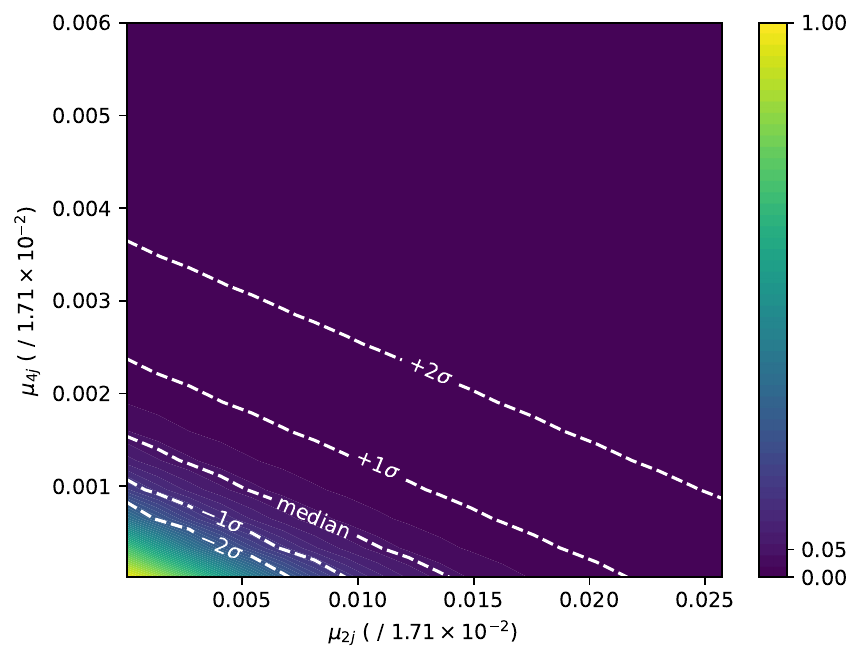}
  \includegraphics[width=0.29\linewidth]{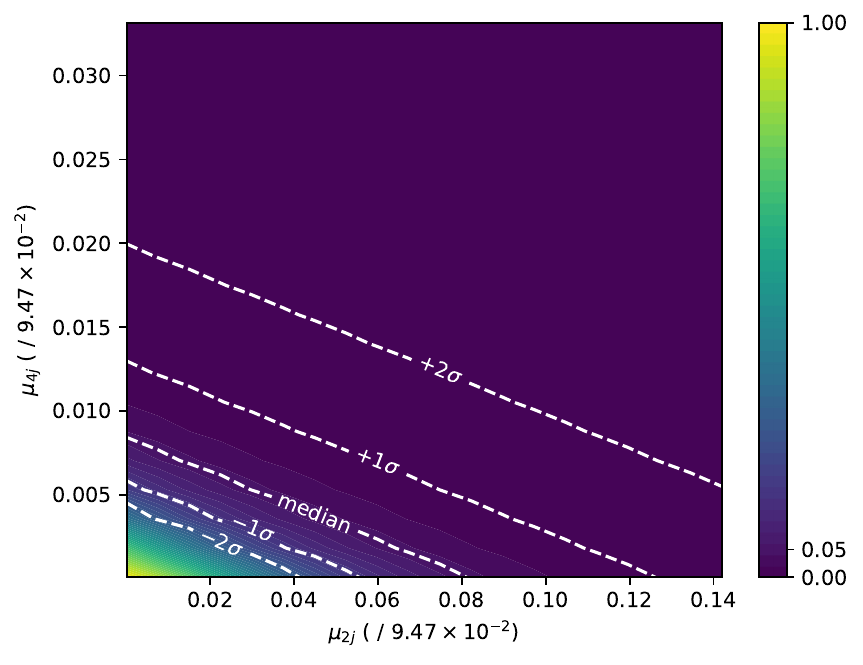}

  \caption{The CLs 2D fitting for ($\mathcal{B}_{\textrm{2-jet}}$, $\mathcal{B}_{\textrm{4-jet}}$). From left to right, the LLP's mass in corresponding samples are set to 50, 10, 1 GeV; From top to bottom, the LLP's lifetime are set to 0.0001, 0.1, 1.0, 10, 100 nanoseconds.}
  \label{fig:limit_2d_all}
\end{figure*}

Figure~\ref{fig:limit_2d_all} illustrates two dimensional upper limits for LLPs assuming a statistics of $4\times 10^6$ Higgs boson.

\section{Sensitivity with different background estimation}
\label{sec:sensitivity_2bkg}
Both 1-D and 2-D exclusion limits are summarized in Table~\ref{tab:limit_combined_2bkg}, but with 2 backgrounds.
\begin{table}[htbp] 
  \centering
  \caption{ 
  The 95\% C.L. exclusion limit on BR($h \to XX$) for all signal channels with both fixed and floating $\epsilon_{V}$ with 2 backgrounds assumption. 
  The limits include $\pm 1 \sigma$ uncertainties after taking into account both statistical and systematic contributions.}
  
  \resizebox{0.9\textwidth}{!}{
  \begin{tabular}{ccccccc}
  \toprule
   \multirow{2}{*}{Scenario} & $\mathcal{B}$  ($\times 10^{-6}$) & \multicolumn{5}{c}{Lifetime [ns]} \\
  \cmidrule{2-7}
   & Mass [GeV] & 0.001 & 0.1 & 1 & 10 & 100 \\ 
   \midrule
   \multirow{3}{*}{Fixed} & 1 & \(2.78^{+1.21}_{-0.54}\) & \(2.40^{+0.95}_{-0.55}\) & \(5.21^{+1.96}_{-1.47}\) & \(34.60^{+12.43}_{-10.17}\) & \(1285.89^{+496.57}_{-301.57}\) \\
   & 10  & \(3.58^{+1.44}_{-0.87}\) & \(1.74^{+0.75}_{-0.42}\) & \(2.07^{+0.79}_{-0.50}\) & \(4.42^{+1.68}_{-1.06}\) & \(31.94^{+11.97}_{-8.64}\) \\
   & 50  & \(1.84^{+0.74}_{-0.44}\) & \(1.54^{+0.59}_{-0.35}\) & \(1.68^{+0.64}_{-0.40}\) & \(1.69^{+0.64}_{-0.40}\) & \(4.04^{+1.54}_{-1.02}\)  \\
  \midrule
   \multirow{3}{*}{Floating} & 1 & \(5.62^{+15.39}_{-4.07}\) & \(2.63^{+4.82}_{-1.40}\) & \(2.80^{+11.86}_{-0.87}\) & \(17.54^{+82.75}_{-5.21}\) & \(675.25^{+3739.15}_{-195.72}\)\\
   & 10 & \(3.75^{+11.82}_{-1.86}\) & \(4.04^{+10.78}_{-3.08}\) & \(1.08^{+4.85}_{-0.31}\) & \(2.30^{+10.34}_{-0.65}\) & \(16.74^{+76.53}_{-4.81}\)\\
   & 50 & \(3.58^{+11.01}_{-2.59}\) & \(2.78^{+8.31}_{-1.93}\) & \(1.96^{+3.24}_{-1.10}\) & \(1.97^{+3.27}_{-1.10}\) & \(2.15^{+9.64}_{-0.71}\) \\
  \bottomrule
  \end{tabular}
  }
  \label{tab:limit_combined_2bkg}
\end{table}

\section{Comparison with selection-based analyses}
\label{app:sec_cut_based_ana}

LLPs searches with the selection-based method has also been carried out. For simplicity, we have investigated a special case that LLPs have long displaced vertices and
decay inside the muon detector.
In this case, shower hits and energy bursts from LLPs decay products are expected in the muon detector while SM backgrounds deposit most energy and hits in the inner detectors. Almost all SM backgrounds are rejected after applying the following selections,
\begin{itemize}
        \setlength{\itemsep}{4pt}
	\item $E_{2jets} >$ 30 GeV: the total energy deposition in the muon detector,
        \item $\Delta T_{j} = \min \left( t_{\textrm{hit},i} - r_{\textrm{hit},i}/c \right) \ge 3$~ns: the minimal time difference, where $t_{\textrm{hit},i}$ represents the hitting time of the $i^\textrm{th}$ component in the jet cluster measured by the muon spectrometer and $r_{\textrm{hit},i}$ is the $i^\textrm{th}$ Euclidean distance to IP, and c the light speed in vacuum.
   \item $N_\textrm{PFOs}$: number of reconstructed particle flow objects,
    \item $\slashed{E}$: the missing energy determined with reconstructed energy in detector subtraction from initial electron-positron c.m. energy.
\end{itemize}
Table~\ref{tab:cutflow_type1} summarizes number of events after imposing a selection chain on both simulated signal and background samples. 

Traditional methods generally
faces various difficulties in reconstructing and identifying both Z bosons and LLPs, and thus require complicated case-by-case analyses,
while the machine learning approach does not.
The latter approach uniformly makes use of all available information of event data,
and not only exhibits a technical superiority but also systematically improves signal efficiencies. Comparing Table~\ref{tab:combined_eff} and Table~\ref{tab:cutflow_type1},  
the ML-based signal efficiency is about 91\%-95\% with a LLP lifetime of 10~ns and mass of 50~GeV, significantly higher than 50.8\% obtained with the selection-based approach.

We have also compared our results with a previous LLP search~\cite{LLP_2019} conducted at lepton colliders, 
where the selection-based method is implemented with fine-tuning and optimization for varying scenarios, 
achieving a maximum selection efficiency~\footnote{"Selection efficiency" specifically denotes the product of signal acceptance and signal efficiency.} of 76\% for a 50 GeV particle with a 1 ns lifetime. 
In contrast, our ML approach significantly outperforms this result and achieves a maximum efficiency of 95\%. 
Notably, our method maintained high selection efficiencies (over 0.9) for lifetimes under 1 ns and masses above 10 GeV, 
and reasonable efficiencies (around 0.4) for masses as low as 1 GeV. 
Furthermore, while the selection-based analysis is limited to lifetimes of up to 1 ns, 
our ML approach can reach up to 100 ns, 
demonstrating its superior capability in performing a broader spectrum of LLP searches.

In summary, we have developed an ML-based approach for LLP searches that outperforms traditional selection-based methods on almost all fronts.

\begin{table}[!htbp]
\centering
\footnotesize
\caption{Number of signal and background events after various selection criteria. MC simulation samples are produced corresponding to the integrated luminosity 5.6 $ab^{-1}$. Signal events have an LLP lifetime of 20~ns and a mass of 50~GeV. }
\label{tab:cutflow_type1}
\begin{tabular}{cccc}
\toprule
 Selections & LLPs Signal with $Z \to j \bar{j}$ & $ee \to q\bar{q}$ & $ee \to ZH$  \\
 generated & $1.0 \times 10^6$ & $2.5 \times 10^8$ & $0.99 \times 10^7$  \\ 
\midrule
decay in muon detector & 134559 & 6516657 & 796596 \\
$|m_{q\bar{q}}-m_Z|<15 GeV$ & 113723 & 4013875 & 39631 \\
$|m_{q\bar{q}}-m_H|<15 GeV$ & 104942 & 229703 & 26862 \\
$0.23 <y_{12}<0.72$  & 93,517 & 129,546 & 20,041  \\
$E_{2jets} > 30 \textrm{GeV}$ & 69,468 & 72 & 16  \\
$min(\Delta T_{j_1}, \Delta T_{j_2}) > 3 \textrm{ns}$ & 68,368 & 50 & 11  \\ 
\midrule
Efficiency &\textbf{50.80\%} & $7.7\times 10^{-6}$ & $1.4\times10^{-5}$\\ 
\bottomrule
\end{tabular}
\end{table}

\section{Searching for LLPs with an external detector}
\label{app:external_detector}

We investigate the possibility of enhancing the sensitivity for detecting LLPs by incorporating an external detector placed significantly farther from the interaction point than the baseline detector. For example, such an external detector may consist of stacked, multi-layer scintillator arrays in the barrel region outside the baseline detector, named as the Far Barrel Detector (FBD), designed to detect LLP decays occurring outside the primary detector. To quantify the sensitivity gain by placing an external detector, we define a gain factor \(F_\textrm{gain}\) as follows:
\begin{equation}
    F_\textrm{gain} = \frac{N_\textrm{obs}}{N_\textrm{gen}} 
    = \frac{\Delta\Omega}{4\pi}\left(\frac{e^{-R_\textrm{max}/d}-e^{-(R_\textrm{max}+\Delta L)/d}}{1 - e^{-R_\textrm{max}/d}}\right) + 1,
\end{equation}

where \(N_\textrm{obs}\) denotes the number of LLP events observed in the FBD, \(N_\textrm{gen}\) is the total number of LLP events generated, and \(\Delta \Omega\) is the geometric acceptance of the external detector relative to the full \(4\pi\) solid angle. Here, \(R_\textrm{max}\) is the outer radius of the baseline detector (beyond which the LLPs enter the FBD), \(d\) is the decay length of the LLPs in the laboratory frame, and \(\Delta L\) is the gap between the FBD and the outer boundary of the baseline detector.

\begin{figure}[h]
    \centering
    \includegraphics[width=0.7\linewidth]{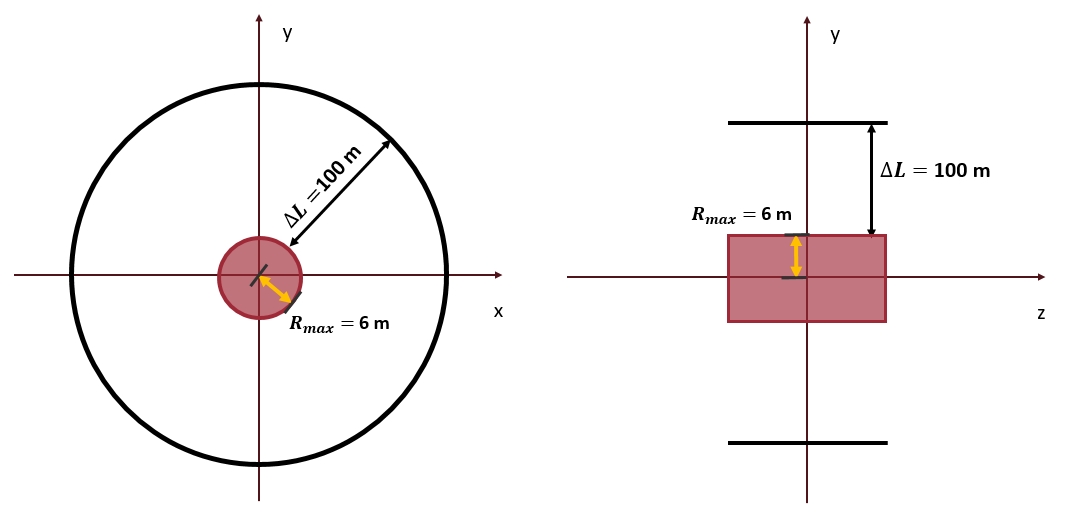}
    \caption{Layout of the Far Barrel Detector (solid lines) relative to the CEPC baseline detector (red) in the $x$-$y$ (left) and $y$-$z$ (right) planes.}
    \label{fig:sup_fbd}
\end{figure}

\begin{table}[hp]
  \footnotesize
  \centering
  \caption{Sensitivity gain factor \(F_\textrm{gain}\) estimated for different LLP masses and lifetimes.}
  \begin{tabular}{ccccccc}
  \toprule
    \toprule
   \multirow{2}{*} & $F_\textrm{gain}$  & \multicolumn{5}{c}{Lifetime [ns]} \\
  \cmidrule{2-7}
   & Mass [GeV] & 0.001 & 0.1 & 1 & 10 & 100 \\ 
   \midrule
   \multirow{2}{*} & 1 & 1     & 1   & 2.8 & 9.9 & 13.7 \\
     & 10 & 1     & 1   & 1   & 2.9  & 10.1 \\ 
     & 50 & 1     & 1   & 1   & 1.1  & 3.3 \\ 
    \bottomrule
  \end{tabular}
  \label{tab:f_gain}
\end{table}

Figure~\ref{fig:sup_fbd} illustrates the relative positions of the external and baseline detectors, where \(R_\textrm{max}\) is 6 meters and \(\Delta L\) is 100 meters.
The estimated sensitivity gain factor \(F_\textrm{gain}\) obtained by adding the external detector in the barrel region with a geometry acceptance factor of about 0.7 is summarized in Table~\ref{tab:f_gain}, highlighting significant improvements, especially for lighter LLPs and longer lifetimes where a maximum gain factor of 13.7 can be achieved. Such an external detector setup thus offers valuable complementary sensitivity to LLP searches at future collider experiments.

\end{document}